\documentclass[a4,twoside]{article}

\usepackage{graphicx}

\usepackage{color}

\begin{document}



\def\la{\,\raise 0.3 ex\hbox{$ < $}\kern -0.75 em
 \lower 0.7 ex\hbox{$\sim$}\,}
\def\ga{\,\raise 0.3 ex\hbox{$ > $}\kern -0.75 em
 \lower 0.7 ex\hbox{$\sim$}\,}

\begin{titlepage}

\vspace*{0.1cm}

to appear in Annual Reviews of Astronomy and Astrophysics, Vol. 45, 2007

\begin{center}

{\huge \bf Toward Understanding\\Massive Star Formation}

\vspace{0.5cm}

{\large Hans Zinnecker\footnote{Astrophysikalisches Institut Potsdam,
An der Sternwarte 16, D-14482 Potsdam, Germany;
email: hzinnecker@aip.de} \hspace{1.0cm}
Harold W. Yorke\footnote{Jet Propulsion Laboratory,
California Institute
of Technology, 4800 Oak Grove Drive, Pasadena, CA 91109 USA;
email: Harold.Yorke@jpl.nasa.gov}}

\end{center}

\vspace{0.5cm}

\begin{abstract}
Although fundamental for astrophysics, the processes that
produce massive stars are not
well understood. Large distances, high extinction, and short
timescales of critical evolutionary phases
make observations of these processes challenging. Lacking good
observational guidance, theoretical
models have remained controversial. This review offers a basic
description of the collapse of a
massive molecular core and a critical discussion of the three
competing concepts of massive star
formation:
\begin{itemize}
\item monolithic collapse in isolated cores
\item competitive accretion in a protocluster environment
\item stellar collisions and mergers in very dense systems
\end{itemize}
We also review the observed outflows, multiplicity, and
clustering properties of massive stars, the
upper initial mass function and the upper mass limit.
We conclude that high-mass star formation is
not merely a scaled-up version of low-mass star formation
with higher accretion rates, but partly a
mechanism of its own, primarily owing to the role of stellar
mass and radiation pressure in
controlling the dynamics.
\end{abstract}

\vspace{0.3cm}

\end{titlepage}

\newpage

\tableofcontents

\newpage

\section{INTRODUCTION}

\subsection{Basic Issues}

Massive stars play a key role in the evolution of
the Universe. They are the principal source of
heavy elements and UV radiation. Through a
combination of winds, massive outflows,
expanding HII regions, and supernova
explosions they provide an important source of
mixing and turbulence in the interstellar
medium (ISM) of galaxies. Turbulence in
combination with differential rotation drives
galactic dynamos. Galactic magnetic fields are
generated, interacting with supernova shock
fronts that accelerate cosmic rays. Cosmic rays,
UV radiation, and dissipation of turbulence are
the principal sources of heating in the ISM,
whereas heavy elements found in dust,
molecules, and in atomic/ionic form ultimately
are responsible for its cooling. Massive stars
thus profoundly affect the star- and planet-formation
process (Bally, Moeckel \& Throop
2005) as well as the physical, chemical, and
morphological structure of galaxies (e.g.,
Kennicutt 1998, 2005).

In spite of the dominant role that massive
stars play in shaping galactic structure and
evolution, our understanding of their formation
and early evolution is still sketchy. There are
many reasons. High dust extinction makes it
difficult to observe high-mass stars during
critical early formation phases. They are rare.
They evolve quickly and important
evolutionary phases are short-lived. The
theoretical problem is extremely
complex. Finally, massive stars are seldomly (if
at all) formed in isolation; the proximity of
other high-mass stars compounds the complex
influence of the forming star on its local
environment via gravitational interactions,
powerful outflows and winds, ionizing
radiation, and supernovae.

The low number statistics of young or
forming high-mass stars is only partially offset
by their higher luminosities, which allow us to
study them at greater distances than their low-mass
counterparts. However, insufficient spatial
resolution is an issue -- an entire OB-star cluster
is often contained in a single observing pixel
(e.g., Henning \& Stecklum 2002).

\subsection{Definitions}

Star formation typically starts with a collapsing
gas condensation (core)  inside a larger subunit
(clump) of a molecular cloud (cf. Williams,
Blitz \& McKee 2000). A protostar forms that
increases its mass by accretion (accumulation)
of neighboring gas, while at the same time
some mass loss occurs through a bipolar
outflow and/or a collimated jet. Let us define
some of the terminology adopted here.

One of the most misused terms in papers
dealing with star formation is protostar, which
is considered the Holy Grail (Wynn-Williams
1982) of IR astronomy. Here, we reserve the
term protostar or protostellar object for a
gaseous object in hydrostatic equilibrium (gas
pressure forces balance the gas' self-gravity),
which has not yet begun hydrogen burning but
which will, given time, burn hydrogen. At the
point hydrogen burning commences, we shall
speak of a zero-age main-sequence (ZAMS)
star; as long as hydrogen burning occurs in the
center, we shall speak of a main-sequence star.
Note that the size scale of a protostar is at most
a few tens of solar radii.

We use the terms massive star and high-mass
star interchangeably to denote an OB star
sufficiently massive to produce a type II
supernova (${\rm M}_*/{\rm M}_\odot \ga$\,8
for solar abundances).
With these definitions in mind, the term high-mass
protostar denotes a $\ga$\,8\,M$_\odot$ hydrostatic
object that has not yet begun hydrogen burning.
As we shall see in the following, such objects
exist only briefly during a transitory stage
between "accreting intermediate-mass
protostar" and "accreting high-mass star."
Because it will be impossible to distinguish
observationally when an accreting object begins
burning hydrogen, we suggest that the terms
massive protostar and high-mass protostar
generally be avoided.

In {\bf Table~1}, we give a crude classification of
massive stars in terms of logarithmic mass
intervals and the corresponding main sequence
spectral types.

\begin{table}
\begin{center}
\caption{Main Sequence massive star definition 
        (logarithmic mass ranges)}
\vspace{0.5cm}
\begin{tabular}{@{}ccccccc@{}}
\hline\hline
{\bf Mass}    & \hspace{0.3cm} & {\bf Designation} & \hspace{0.3cm} & {\bf Sp. type}\\
\hline
8\,--\,16\,M$_\odot$      & & Early B-type massive stars & & B3V to B0V\\
16\,--\,32\,M$_\odot$     & & Late  O-type massive stars & & O9V to O6V\\
32\,--\,64\,M$_\odot$     & & Early O-type massive stars & & 05V to O2V$^a$\\
64\,--\,128\,M$_\odot$    & & O/WR-type massive stars    & & WNL-H$^b$\\
\hline
\end{tabular}
\label{tab:defs}
\end{center}
\vspace{-1.0cm}
\begin{tabular}{@{}ll@{}}
\hspace{0.8cm} & \\
& $^a$O2V main sequence stars have been identified by Walborn et
al. (2002).\\
& $^b$WNL-H: N-rich late-type Wolf-Rayet (WR) stars, still on the\\
& Main Sequence (H-burning) -- see Crowther (2007).\\ 
\end{tabular}
\end{table}

We reserve the terms very massive star
(VMS) and supermassive star (SMS) for stars
in the mass ranges of 
$100 \la {\rm M}_*/{\rm M}_\odot \la 1000$ and
$10^4 \la {\rm M}_*/{\rm M}_\odot \la 10^8$ , 
respectively, and introduce
the term ultramassive star (UMS) for stars in
the mass range of $10^3 \la {\rm M}_*/{\rm M}_\odot \la 10^4$. 
SMSs are
equilibrium configurations that are dominated
by radiation pressure --- baryons and electron-positron
pairs provide only a minor contribution
to the equation of state. At some point  during
their evolution SMSs collapse owing to a
general relativistic gravitational instability.
Whereas in the present epoch VMSs, UMSs,
and SMSs are unlikely to be formed except
under very special conditions, stars with masses
in excess of 100\,M$_\odot$ are expected during the
first epoch of star formation (Bromm \& Larson
2004; Abel, Bryan \& Norman 2000). VMSs,
UMSs, and SMSs are not discussed in this
review. A recent discussion of the formation
and evolution of VMSs is given in Portegies
Zwart et al. (2006) and in Belkus, van Bever \&
Vanbeveren (2007).

What is accretion? The term accretion is used
in a variety of senses. Measured accretion rates
often refer to the rate of mass inflow toward
star-forming sites -- not the rate at which a star
or protostar gains mass. Originating from a 0.1\,pc
scale, this material cannot possibly fall into a
sub-10$^{-6}$\,pc region without carrying significant
angular momentum. Instead, it either forms a
disk or hits and is mixed with prior existing
disk material (see {\bf Figure~1}). Thus, we
distinguish between the accretion of cloud core
material onto a disk (${\rm \dot M_{D-acc}}$) 
and the accretion
onto a (proto-)star (${\rm \dot M_{S-acc}}$).

Analogous to accretion, mass loss is used in
a variety of senses. Measured mass loss rates
from jets and outflows do not necessarily
reflect the mass loss from an isolated young
star. We thus distinguish between the mass loss
from the (proto-)star via a wind 
(${\rm \dot M_{S-wind}}$), the
mass loss launched from the accretion disk
(${\rm \dot M_{D-wind}}$), 
which never reaches the (proto-)star,
and the material swept up into the outflow from
the surrounding molecular cloud 
(${\rm \dot M_{load}}$). The
measured outflow could have contributions
from several stars and several disks.

The interrelation between disk accretion and
disk winds is a fascinating aspect of massive
star formation (see, e.g., the recent
magneto-hydrodynamic models of Banerjee \& Pudritz
2007) -- and is at the focus of the frequently
asked question: Is high-mass star formation a
scaled-up version of low-mass star formation?
The answer to this question will be better
defined by the time we reach the end of this
review.

\subsection{Recommended Reading}

The study of the origin of massive stars is a
relatively new field of astrophysical research.
There is no comprehensive monograph on the
subject, but there are several conference
proceedings over the past few years dedicated
to the topic, of which we recommend the
following:
\begin{itemize}
  \item {\it Massive Stars: Their Lives in the interstellar Medium}
        (Cassinelli \& Churchwell 1993)
  \item {\it Hot Star Workshop III: The Earliest Stages of Massive Star Birth}
        (Crowther 2002)
  \item {\it Massive Star Birth: A Crossroads of Astrophysics}
        (Cesaroni et al. 2005a)
\end{itemize}

The reviews on {\it Environment and Formation
of Massive Stars} (Garay \& Lizano 1999),
{\it Control of Star Formation by Supersonic
Turbulence} (Mac Low \& Klessen 2004), {\it The
Formation of the First Stars in the Universe}
(Glover 2005),  {\it The Birth of Massive Stars and
Star Clusters} (Tan 2005), {\it High Mass Star
Formation by Gravitational Collapse of
Massive Cores} (Krumholz 2007),  and {\it The
Critical Role of Disks in the Formation of
High-Mass Stars} (Cesaroni et al. 2007) are also
recommended. Among the reviews in the
Proceedings of Protostars and Planets V, we
particularly recommend {\it The Formation of
Massive Stars} (Beuther et al. 2007).

There are also a few related Annual Reviews
articles:
\begin{itemize}
  \item {\it Compact HII Regions and OB Star Formation}
        (Habing \& Israel 1979)
  \item {\it The Search for Infrared Protostars}
        (Wynn-Williams 1982)
  \item {\it The Dynamic Evolution of HII Regions -- Recent Theoretical
        Developments} (Yorke 1986)
  \item {\it The Orion Molecular Cloud and Star-Forming Region}
        (Genzel \& Stutzki 1989)
  \item {\it Physical Conditions in Regions of Star Formation}
        (Evans 1999)
  \item {\it Ultra-Compact HII Regions and Massive Star Formation}
        (Churchwell 2002)
  \item {\it Massive Stars in the Local Group:  Implications for Stellar
       Evolution and Star Formation} (Massey 2003)
  \item {\it The First Stars} (Bromm \& Larson 2004)
\end{itemize}

Except for the last, none of these earlier
reviews had its focus on the formation aspect
of massive stars but rather provided a
descriptive observational summary of the
properties of young OB stars and their HII
regions. The present Annual Reviews article
will be accompanied in the same volume by
an article on 'Physical properties of Wolf-Rayet
stars' (Crowther 2007) and an article
on 'Theory of Star Formation' (McKee \&
Ostriker 2007) which mostly addresses low-mass
star formation, but also includes an
important section on high-mass star
formation.

\subsection{The Focus of This Review}

The major questions we wish to address in this review are:
\begin{enumerate}
\item What is the sequence of observable states
      leading from molecular clouds to young
      high-mass stars?
\item What are the initial conditions of massive
      star formation (gas densities, temperatures,
      clump masses, etc.) and how do they come
      about?
\item Do massive stars always form in dense stellar
      clusters or can they form in isolation? What
      special conditions are necessary to allow
      coalescence, i.e., mergers of stars?
\item Which clues to the origin can be gleaned
      from multiplicity observations? How do we
      explain the very tight massive spectroscopic
      binaries and OB runaway stars?
\item How does the forming massive star influence
      its immediate surroundings, possibly limiting
      its final mass and/or the final mass of its
      neighbors?
\item How do young massive stars influence their
      global environment, either by inhibiting or
      by triggering further star formation? How do
      we get a starburst?
\end{enumerate}

To tackle these questions, we first discuss some
key observations related to massive star
formation.

\newpage

\section{MASSIVE STAR FORMATION:\\KEY OBSERVATIONS}

\subsection{Observable Stages}

The optically visible main-sequence life of OB-type
stars is preceded by an embedded phase
that lasts about 15\,\% of their lifetime
(Churchwell 2002). As summarized by Menten,
Pillai \& Wyrowski (2005) and van der Tak \&
Menten (2005), observations at mid-IR through
radio wavelengths have shown that this
embedded phase can be subdivided into several
groups of objects:

\begin{itemize}
\item IR dark clouds (Perault et al. 1996,
      ISOCAM; Egan et al. 1998, MSX; Benjamin
      et al. 2003, {\it Spitzer}). Their internal density
      maxima and temperature minima likely
      represent the initial conditions of high-mass
      star formation; a compilation of several
      dozens of such high-mass starless cores has
      been given by Sridharan, Williams \& Fuller
      (2005). Some of these cores probably contain
      low-mass and intermediate-mass accreting
      protostars, which are faint and hard to detect.
      Protostellar outflow activity has been
      detected in one of them (Beuther, Sridharan
      \& Saito 2005).
\item Hot molecular cores (Kurtz et al. 2000,
      Cesaroni 2005). These have large masses of
      warm and dense gas, and large abundances
      of complex organic molecules evaporated off
      dust grains; they are signposted by methanol
      maser emission (Menten 1991, Walsh et al.
      1998, Hill et al. 2005); ground-based
      detectability on the Wien part of the spectral
      energy distribution with sufficient spatial
      resolution is difficult (Stecklum et al. 2002),
      but comes into reach with dedicated 8-m
      class telescope observations (De Buizer \&
      Minier 2005, Linz et al. 2005).
\item Hypercompact and ultracompact HII regions
      (Kurtz 2005, Hoare et al. 2007). In these
      regions, small but growing pockets of
      ionized gas have developed that stay
      confined to the stellar vicinity.  Whereas
      hypercompact HII regions probably represent
      individual photoevaporating disks (Keto 2007;
      see also the example in Nielbock
      et al. 2007), ultracompact HII regions
      probably represent disk-less stars
      photoionizing their own cocoons and
      massive envelopes.
\item Compact and classical HII regions (Mezger
      et al. 1967, Yorke 1986). Their gas is ionized
      globally, often by several ionizing sources. It
      expands hydrodynamically as a whole and
      disrupts the parent molecular cloud,
      revealing both the embedded high-mass and
      lower  mass stellar population for optical and
      near-IR observations (Carpenter et al. 1993;
      Zinnecker, McCaughrean \& Wilking 1993).
\end{itemize}

\subsection{Initial Conditions}

Massive star formation occurs inside dense,
compact clumps in giant molecular clouds (H$_2$
column densities are 
10$^{23}$\,--\,10$^{24}$\,cm$^{-2}$). Smaller
mass clumps with lower peak H$_2$ column
densities do not form massive stars. Several
types of molecular cloud surveys,
predominantly near HII regions, have been
carried out: CS-molecule surveys for dense
molecular gas, 1.2-mm dust continuum surveys
for massive cold dust (and hence gas)
condensations, as well as OH, H$_2$O, and
methanol maser emission surveys for shock-excited
compact regions as signposts for
massive star formation. (Note that methanol
maser and OH maser emission is exclusively
associated with high-mass star formation,
whereas H$_2$O masers  may also be found in
low-mass star-forming regions.  This is because
methanol and OH masers are radiatively
pumped and need an intense far-IR source in
their vicinity; H$_2$O masers, in contrast, are
collisionally pumped in gas shocked by
outflows.) These gas and dust surveys have
revealed dense cold clumps (molecular
hydrogen density n$_{\rm H_2}$\,=\,10$^5$\,cm$^{-3}$, gas
temperature T\,=\,10\,--\,20\,K, diameter $\sim$\,0.5\,pc)
with gas masses ranging from a few hundred to
a few thousand solar masses 
(Plume et al. 1997, Shirley et al. 2003, 
Garay et al. 2004, Motte
et al. 2005, Evans 2005).
The methanol maser surveys at 6.7\,GHz
point to hot molecular cores with internal heat
sources and outflows, as well as protoclusters
(Burton et al. 2005, De Buizer 2003, Minier et
al. 2005).

As mentioned before, large-scale observations
with the ISO, MSX, and {\it Spitzer} satellites have
revealed a new class of clouds, the so-called IR
dark clouds or IRDCs, and Simon et al. (2006)
have identified more than 10,000 such IRDCs
from the MSX data base. Many of these appear
to be located in the 4\,--\,5\,kpc Galactic molecular
ring (Jackson, private communication; see also
Bronfman et al. 2000). The IRDCs are dense
clouds seen in absorption against mid-IR
background emission. They are mostly
filamentary structures that contain
condensations of cold massive cores where
massive stars or even star clusters seem to form
(Rathborne, Jackson \& Simon 2006). Recent
mid-IR and millimeter-continuum observations
show different evolutionary stages of massive
star formation in adjacent cores: dense
millimeter-continuum sources with and without
mid-IR emission (Garay et al. 2004, their figure
4).

The origin of these structures appears to derive
from supersonic turbulence in  giant molecular
clouds, that is, shock compression from convergent
turbulent gas streams. Depending on the direction of
the compression with respect to the direction of the
magnetic field lines, the magnetic field will be
boosted through flux freezing and hence the
resulting clump will be stabilized by magnetic forces
against gravitational collapse (subcritical
compression). If not, the compressed clump is
quickly set up for collapse (supercritical
compression), in fact so quickly that the set-up time
is shorter than the free-fall time. Magnetically
stabilized clumps take much longer for collapse to
begin, and their internal turbulent structure may
make the clump prone to subfragmentation. If this is
true, only supercritical compression leads to massive
star formation (cf. Shu, Adams \& Lizano 1987). An
interesting speculation would be that clouds without
sufficiently strong magnetic fields can form lots of
massive stars quasi-simultaneously, giving rise to
gigantic starbursts.

\subsection{Endproducts}

\subsubsection{OB Clusters} The endproducts of massive
star formation are either dense gravitationally bound
OB star clusters or loose unbound OB associations
(Lada \& Lada 2003, Brice\~{n}o et al. 2007). Classical
examples of OB star clusters include the Orion
Nebula Cluster (ONC), the dense compact cluster
associated with the giant galactic HII region NGC
3603, and the R136 cluster in the 30 Dor region in
the Large Magellanic Cloud (LMC; see {\bf Figure~2}).
These clusters roughly define a richness sequence in
powers of 10: they contain 1, 10 (21), and $\sim$\,100
massive O-type stars per cluster, with estimated total
cluster masses of 10$^3$, 10$^4$, and 10$^5$\,M$_\odot$,  
respectively
(e.g., ONC: Hillenbrand 1997, Hillenbrand \&
Hartmann 1998; NGC 3603: Moffat, Drissen \&
Shara 1994, Drissen et al. 1995; R136: Parker \&
Garmany 1993,  Massey \& Hunter 1998). Although
R136 can probably be considered a small young
globular cluster (M. Andersen et al., submitted),
there are more massive young clusters in the nearby
universe, such as the very massive embedded super
star cluster in the center of the NGC 5253 dwarf
galaxy (Turner et al. 2003). This cluster has an
ionizing flux equivalent to the presence of 4000\,--\,6000
O7V stars that comes from a very compact
region about 1\,pc in size, measured with the VLA.
How can the formation of so many massive stars in
such a small volume be possible? This question is at
the heart of understanding the origin of globular
clusters.

\subsubsection{OB Associations} The classical examples of
OB associations include the nearby Scorpius OB2
and Orion OB1 associations (Blaauw 1964, 1991).
Another fine example of what may ultimately
become an OB association is the Carina star
formation complex at 2.3\,kpc (Smith \&
Brooks 2007). One of the best studied
extragalactic OB associations is NGC 604 in M33
(Ma\'{i}z-Apell\'{a}niz, P\'{e}rez \& Mas-Hesse 2004). In all
these cases, the OB stars are spread over the whole
face of the parent giant molecular cloud and are not
densely packed at all, with distances between
massive stars ranging from 1 to 10\,pc. This then
appears to be a completely different mode of
massive star formation, although it must be noted
that OB associations often contain dense clusters,
too (e.g., the Carina complex harbors the well-known
Trumpler 14 and 16 clusters). The question is
whether OB associations are superpositions of
expanded young clusters (e.g., Kroupa, Aarseth \&
Hurley 2001, Bastian \& Goodwin 2006).

\subsubsection{Field OB Stars} Do massive stars always
occur in young star clusters or OB associations, or
can massive stars also be found outside these
regions, i.e., in the field? The answer is they can.  It
has long been realized that there exists a class of
massive stars, the so-called runaway OB stars
(Blaauw 1961, Poveda, Ruiz \& Allen 1967, Gies \&
Bolton 1986) that are ejected from their birthplaces --
clusters and associations -- with velocities in excess
of 40\,kms$^{-1}$. About 10\,--\,25\,\% of all O stars and about
2\,\% of all B stars belong to this class of massive field
stars. The question whether -- these runaway stars
aside -- other massive stars occur in the field
(implying that they would be formed in isolation)
has been studied by de Wit et al. (2004, 2005),
following Mason et al. (1998). These authors
established that 43 among the 227 O stars brighter
than eighth V-magnitude are in the field. Of these,
about half can be traced back to a cluster or
association origin, but about 10\,--\,20 stars (i.e.,
$\sim$\,5\,--\,10\,\%) could not be assigned to any group of origin
and therefore might be true field stars, born outside
clusters and associations, an issue already raised in a
pioneering paper by Roberts (1957). A case in point
is HD93521, a high-latitude O9.5V star, more than
1\,kpc above the Galactic plane, which must have
formed locally in the halo (Irvine 1989)! These
examples suggest that the question of the birthplaces
of massive stars is not yet completely settled. The
forthcoming {\it Spitzer} 8-micron imaging survey of the
Magellanic Clouds (see Meixner et al. 2006) can
shed new light upon the question, and in particular
can pinpoint isolated massive stars in the
LMC/SMC, should these objects indeed exist.

\subsection{Clues from multiplicity}

The multiplicity of
massive stars is believed to be higher than that of
young low-mass premain-sequence stars (Preibisch,
Weigelt \& Zinnecker 2001, Duch\^{e}ne et al. 2001).
This means massive stars have more physical
companions than low-mass stars on average. For
reference, the multiplicity or, more precisely, the
companion star fraction (csf) of a stellar population
has been defined by Reipurth \& Zinnecker (1993) to
be
\[
{\rm csf\,=\,(B\,+\,2T\,+\,3Q\,+\,...)/(S\,+\,B\,+\,T\,+\,Q\,+\,...)},
\] 
where S is
the number of single, B the number of binary, T the
number of triple, and Q the number of quadruple
systems, etc. (i.e., triple systems contribute two
companions, quadruple systems three companions,
etc.). For example, the multiplicity of the four OB
stars in the Trapezium Cluster in Orion is as follows:
the most massive star $\theta^1$~C is double, the next
massive star $\theta^1$~A is triple (a hierarchical system
with a close spectroscopic binary and a wider
companion), $\theta^1$~B is at least quadruple (kind of a
Trapezium system within the Trapezium Cluster),
and $\theta^1$~D is apparently single. [Kraus et al. (2007)
find some indications that $\theta^1$~D appears extended
in their speckle images. There is a fifth star in the
Trapezium Cluster, $\theta^1$~E, which has recently been
found to be a double-lined intermediate-mass
spetroscopic binary (Herbig \& Griffin 2006, Costero
et al. 2006).] Putting the number of companions of
$\theta^1$~A, B, C, D into the above formula, we get
csf\,=\,1.5 at face value (probably a lower limit). This
should be compared to the multiplicity of the low-mass
stellar members in the Orion Nebula cluster,
which is csf\,=\,0.5 (Padgett, Strom \& Ghez 1997; B.
Reipurth et al., submitted), i.e., significantly lower.

The multiplicity statistics of other OB
clusters, including clusters rich in O stars (N\,$>$\,5)
and poor in O stars (N\,$<$\,3) has been studied by
Mermilliod \& Garc\'{i}a (2001) and Garc\'{i}a \&
Mermilliod (2001), with interesting results: The
spectroscopic binary frequency in O-star rich
clusters can vary enormously in different
clusters, from 15\,\% to 80\,\%, with no apparent
correlation. If anything, there is an
anticorrelation of the binary frequency and the
cluster density, but this needs to be
reinvestigated and confirmed. The above
statistics often rely on relatively poor data; and
the sampling is not complete. Recent work on
higher quality data, but on a more limited
number of clusters, tends to obtain a lower
binary fraction in the range of 20\,\% to 60\,\%
(Sana, Rauw \& Gosset 2005). The most
dramatic example is the IC 1805 cluster where
the binary frequency went down from 80\,\% to
20\,\% based on better data (De Becker et al.
2006).

In the O-star poor clusters almost all O stars
are spectroscopic binaries, often double-lined
and even eclipsing. These massive binaries are
usually members of hierarchical triple or
quadruple systems, or of trapezia, and are often
located at the cluster center. The exciting star of
the Orion Nebula Cluster, $\theta^1$~Ori C belongs in
this category, although $\theta^1$~Ori C is not a
massive close spectroscopic binary but a very
eccentric visual binary, with masses of 34\,M$_\odot$
(O5.5V) and 15\,M$_\odot$ (O9.5V) and an orbital
period of about 11\,yr (Kraus et al. 2007).
The orbital periods of the spectroscopic binaries
in the O-star rich clusters are concentrated in
the range of 4\,--\,5 days, whereas in the O-star
poor clusters there is a pile-up of orbital periods
around 3\,$\pm$\,1 days. In NGC 6231, for example,
according to Sana (private communication), 10
out of the 16 O stars are double-lined
spectroscopic binaries: 6 with periods under 10
days (4 below 5 days); 2 with periods between
3 and 9 months, and 2 with periods of the order
of a year or greater (in addition, 1 star is
probably a triple-lined spectroscopic binary).
The luminosity ratios are all in the range 1\,--\,10,
otherwise one would not detect them as double-lined
spectroscopic binaries. This implies that
the secondaries are probably early B stars, and
the primary-to-secondary mass ratios must be
about 3 (at maximum) or lower. It is also worth
mentioning that there is no very short-period
highly eccentric O+O binary known at this time
(Sana, private communication). These
fascinating and surprising facts challenge our
views of massive star formation and provide
clues to their origin, clues too complex to fully
decipher yet but hinting at gravitational
dynamics playing a role -- beyond mere disk or
filament fragmentation (see {\bf Section~5}, where
the formation of binary and multiple systems is
extensively discussed; see also Zinnecker
2003).

Of course, it is equally important to study the
multiplicity of massive stars in OB associations
where the stellar density is much lower than in
OB clusters, and dynamical interactions
between the forming massive stars should be
less of an issue. For example, the Orion OB1
association contains $\sim$\,70 massive stars in its
three subgroups 1a,b,c (subgroup 1d is the
Orion Nebula Cluster). Of these 70 OB stars,
20\,\% are spectroscopic binaries with periods
less than 10 days (Morrell \& Levato 1991). The
three subgroups show variations in their
spectroscopic binary fractions: subgroup 1a is
average, subgroup 1b is a factor 1.5 above the
average, and subgroup 1c is a factor 1.5 below
the average. The situation in the nearby
Scorpius-Centaurus OB2 association is as
follows: Among the 48 early B stars (B05 to
B3V) there are 25 binaries, and 20 of them are
spectroscopic binaries with known periods in
the range of 0.9\,--\,34.2 days, with a median of
5.7 days (Brown 2001).

In conclusion, it seems that the spectroscopic
binary fraction among massive stars in OB
associations is surprisingly similar to that in OB
clusters rich in O stars (about 40\,\% on average),
and the fraction of very close spectroscopic and
eclipsing massive binaries in OB associations
with orbital periods below 5\,--\,10 days (about
20\,\%, judging from the Orion and Scorpius-Centaurus
regions) is a factor of two lower than
in OB clusters. Thus OB clusters appear to
contain more of a population of very tight
(hard) binaries, possibly an effect owing to
dynamical encounters after birth; this is an
effect that is absent in OB associations.

The future of spectroscopic massive binary
research lies in the near-IR and in multiepoch
radial velocity surveys of embedded massive
stars. First results (and successes) have been
reported by Apai et al. (2007), indicating that
massive close binaries indeed form at a very
early stage.

\subsection{Upper Initial Mass Function and Upper Mass
Limit}

The mass distribution function of massive stars
at birth (the so-called Initial Mass Function or
IMF for short) is a complicated matter, because
({\it a}) massive stars quickly lose some of their
initial mass through stellar winds, ({\it b}) many of
these massive stars are unresolved binaries, and
({\it c}) massive stars tend to be born in the centers
of OB clusters or tend to sink preferentially
toward the cluster center, leaving behind their
lower mass siblings that live in the cluster
outskirts. This introduces a bias into the mass
distribution, flattening a power-law slope. The
upper IMF from about 10 to 100\,M$_\odot$ is usually
found to be a universal power law, with
logarithmic slope -1.35, first found by Salpeter
(1955) for a range of masses below 10\,M$_\odot$.  We
refer here to the early observational work of
Garmany, Conti \& Chiosi (1982) and the
summary of Massey (1998). The implication of
a Salpeter slope or other similar slopes of the
IMF for the number of stars born in different
mass intervals (for convenience spaced by a
factor of two) can be seen in {\bf Table~2}, which
has been normalized to contain exactly one
object in the highest mass interval. This is
instructive, because it shows dramatically how
rare the O/WR-type massive stars (interval
64\,--\,128\,M$_\odot$)  are compared with the early B-type
massive stars (8\,--\,16\,M$_\odot$),  or with the
solar-type low-mass stars (1\,--\,2\,M$_\odot$);
see Zinnecker (1996).

\begin{table}
\begin{center}
\caption{Initial Mass Function (dN/dlogM\,$\sim$\,M$^{\rm -x}$)
         examples}
\vspace{0.5cm}
\begin{tabular}{@{}ccccc@{}}
\hline\hline
{\bf Mass range} & \hspace{1cm} & & {\ Logarithmic slope} & \\
\hline
&                           & x\,=\,1 & x\,=\,1.35 & x\,=\,1.7\\
\hline
0.5\,--\,1\,M$_\odot$     & & 128     & 700        & 3822 \\
1\,--\,2\,M$_\odot$       & & 64      & 275        & 1176 \\
2\,--\,4\,M$_\odot$       & & 32      & 108        & 362 \\
4\,--\,8\,M$_\odot$       & & 16      & 42         & 111 \\
8\,--\,16\,M$_\odot$      & & 8       & 16.6       & 34.3 \\
16\,--\,32\,M$_\odot$     & & 4       & 6.5        & 10.6 \\
32\,--\,64\,M$_\odot$     & & 2       & 2.55       & 3.25 \\
64\,--\,128\,M$_\odot$    & & 1       & 1          & 1 \\
\hline
\end{tabular}
\end{center}
\end{table}

The IMF of massive stars in the
aforementioned OB clusters (Orion Nebula
Cluster, NGC 3603, and R136 in 30 Dor) is
discussed by Pflamm-Altenburg \& Kroupa
(2006), Stolte et al. (2006), and Massey \&
Hunter (1998), respectively. [Other studies of
the Orion Nebula Cluster include Zinnecker,
McCaughrean \& Wilking (1993); Hillenbrand
\& Hartmann (1998). The stellar content in the
biggest galactic HII region NGC 3603 was also
investigated by Drissen et al. (1995); Hofmann,
Seggewiss \& Weigelt (1995); Eisenhauer et al.
(1998); Brandl et al. (1999); Moffat et al.
(2004); and Sung \& Bessell (2004). An
important investigation of the IMF of R136 is
that of Sirianni et al. (2000).]  Whereas
Pflamm-Altenburg \& Kroupa (2006) find a
deficit of high-mass stars in the Orion cluster,
Stolte et al. (2006) derive an excess of massive
stars in the NGC 3603 cluster core, reflected by
a power-law slope of -0.9 (probably due to
mass segregation). It is only in R136 that the
power-law slope is almost exactly the same as
the Salpeter value. It is worth noting here that
the slope of the mass function of high-mass
stars in the range of 8\,--\,40\,M$_\odot$ in the wider
field in the 30 Dor region is apparently the
same as in the R136/NGC 2070 cluster (Selman
\& Melnick 2005). The latter authors do not find
the much steeper slope of the IMF (for the
range of 25\,M$_\odot$ to 120\,M$_\odot$) derived by Massey
(2002) for the global OB field population in the
LMC. They suspect that selective
incompleteness at V\,=\,12 owing to detector
saturation and Be star contamination lies at the
origin of this discrepancy.

The question of the IMF of massive stars in
OB associations was discussed long ago by
Garmany, Conti \& Massey (1980) and later by
Massey, Johnson \& Degioia-Eastwood (1995).
They concluded that the IMF is normal, i.e.,
consistent with a Salpeter power-law. A fine
discussion of upper IMF slopes in various
young clusters and associations including all
the caveats and selection effects was given by
Scalo (1998). He noted that individual
realizations of IMF slopes can vary, but the
average slope is indeed close to the Salpeter
value -1.35.

The question of whether there is a physical
(rather than statistical) upper mass end to the
IMF is of great interest to anyone interested in
population synthesis, galactic evolution, and
cosmology. For example, if the stellar upper
mass limit were 120\,M$_\odot$,  pair instability
supernovae requiring stellar masses at the time
of explosion between 140 and 260\,M$_\odot$ (Heger
et al. 2003) could not happen. Larson (1982)
originally asked the observational question
whether there was a correlation between the
mass of a molecular cloud and the maximum
mass of a star that could form in it. His result
was that indeed the maximum stellar mass
scaled with the mass of the parent cloud,
roughly with its square root. It takes a giant
molecular cloud of 10$^5$\,M$_\odot$ to form a 50\,M$_\odot$
star; a cloud of 10$^3$\,M$_\odot$ can only spawn a
maximum stellar mass of 8\,M$_\odot$.  The
implication is that massive stars form in clouds
of mass between 10$^3$ and 10$^5$\,M$_\odot$ or more,
probably because only these have sufficiently
massive substructure (clumps).

Weidner \& Kroupa (2004) and Figer (2005)
discussed the upper limit to the masses of stars,
based on observations of the R136 cluster in the
LMC and the Arches cluster near the Galactic
Center, respectively. They pointed out that
these clusters are so massive that given a
Salpeter IMF one would expect to find stars as
massive as 750\,M$_\odot$ and 500\,M$_\odot$,  respectively,
whereas the most massive stars seen do not
exceed 140\,M$_\odot$ and 130\,M$_\odot$,  respectively.
This suggests a firm upper mass limit of 150\,M$_\odot$.
Otherwise, a sharp down-turn of the IMF
near 150\,M$_\odot$ would be required (see the
extensive discussion in Elmegreen 2000). Or
the very massive stars have already
exploded/imploded during the dust-obscured,
hidden, early evolutionary stages -- an unlikely
scenario. Oey \& Clarke (2005) also gave a
statistical confirmation of a stellar upper mass
limit around 120\,--\,200\,M$_\odot$,  if the IMF is
Salpeter-like. Koen (2006) further analyzed the
upper IMF in the R136 cluster with two
different statistical techniques and suggested an
upper mass limit of 140\,--\,160\,M$_\odot$.  Thus all four
studies agree on the existence of a physical
upper limit in the stellar mass distribution.

\subsection{Feedback and Triggering}

This topic deserves its own review. The
question we ask here is the following: What
does the energy and momentum input of
massive stars in terms of expanding HII
regions, stellar winds, or supernova shock
waves do to the parent clouds? Is the cloud
primarily disrupted or is new star formation
triggered? Which of the above agents (HII
regions, stellar winds, or supernova shock
waves) provides the best trigger for new OB
star formation and for new low-mass star
formation? We are only beginning to answer
these questions.

A key observation in this context is the fact
that the high-mass and low-mass stellar
populations in the subgroups of OB
associations appear to be coeval, i.e., the
nuclear age of the massive stars is the same as
that of the lower mass premain-sequence
objects (Preibisch \& Zinnecker 1999, 2007;
Brice\~{n}o et al. 2007). This would appear to
require a fast, coherent trigger, such as a
supernova shock wave. Indeed, in the
Scorpius-Centaurus association there is evidence that the
shock wave of a supernova in one OB subgroup
triggered the formation of another subgroup (de
Geus 1992). However, there is also other
evidence that radiation from massive stars [by a
process called radiative implosion (e.g., Kessel-Deynet
\& Burkert 2003)] can only trigger the
formation of low- and intermediate-mass
objects (Lee \& Chen 2007 and references
therein).

The classic theory of triggered massive star
formation is the one by Elmegreen \& Lada
(1977). In their theory, an ionization shock
front provides the pressure on an adjacent layer
of molecular gas to compress it and heat it, thus
stimulating gravitational instability of massive
gas layers, hence the formation of massive
stars -- and perhaps only massive stars. Low-mass
star formation may not be triggered with
this mechanism. If true, this would produce an
anomalous IMF with only massive stars (known
as bimodal star formation, cf. G\"usten \& Mezger
1982). It is worth noting that such
characteristics -- small groups of massive stars
only -- may be needed for the decay of small
N-body systems that give rise to the dynamical
ejection of runaway OB stars (Clarke \& Pringle
1992).

\newpage

\section{MASSIVE STAR FORMATION:\\BASIC THEORY}

\subsection{Sequence of Events}
Starting from a pre-existing giant molecular
cloud, the sequence of events is likely as
follows:
\begin{enumerate}
\item Formation of cold dense molecular cores or
      filaments, induced by gravo-turbulent cloud
      fragmentation (Mac Low \& Klessen 2004).
      This means that supersonic turbulence
      rapidly produces localized compressed
      pockets of gas, some of which remain
      gravitationally bound and provide the initial
      conditions for collapse (Padoan \& Nordlund
      2002; Klessen et al. 2005). A characteristic
      density of about 10$^5$\,cm$^{-3}$ and temperature
      of 10\,--\,15\,K results from the equation of
      state of dusty molecular gas of solar
      abundance (Jappsen et al. 2005, Larson
      2005).
\item Nonhomologous gravitational collapse of
      portions of the cores into optically thick,
      pressure-supported protostellar embryos
      with initial masses of the order of 10$^{-3}$\,M$_\odot$
      (Larson 1969, Bate 2000). The term
      nonhomologous collapse refers to the fact
      that the relative distribution of material
      changes, as opposed to a homologous or
      self-similar collapse. (This is different from
      Shu's (1977) inside-out collapse of a self-similar
      isothermal sphere).
\item Accretion of material onto protostellar
      objects as they evolve toward the main
      sequence. For low-mass objects the
      accretion stops well before hydrogen
      burning commences. These premain-sequence
      objects of fixed mass then slowly
      and quasi-hydrostatically contract to the
      main sequence (Palla \& Stahler 1993,
      Baraffe et al. 2002). However, high-mass
      objects eventually start burning hydrogen
      and develop radiation-driven winds as they
      continue to accrete and evolve up the main
      sequence to hotter and more luminous states
      (Kudritzki 2002).
\item Disruption of the birth cloud, as the first
      high-mass stars strongly influence their
      environment by their winds, outflows, and
      UV radiation, and eventually become
      supernovae. The most massive stars go
      supernova after $\sim$\,3\,Myr. When the remnant
      molecular cloud has been dissipated, the
      result is mostly a cluster of OB stars or an
      OB association, with an associated cospatial
      population of lower  mass stars (Zinnecker,
      McCaughrean \& Wilking 1993). Often,
      several evolutionary stages of star
      formation can be found side by side; the 30
      Dor region in the LMC is a good example
      (Walborn et al. 1999).
\end{enumerate}

We shall denote these four phases as the
compression, collapse, accretion, and disruption
phases of high-mass star formation,
respectively. All phases can occur
simultaneously and side-by-side in a molecular
cloud. Below, we give a fairly detailed yet still
greatly simplified discussion of these phases,
including protostellar luminosity and ionization
evolution as a function of growing stellar mass.

\subsection{The Compression Phase}

This first step toward high-mass star formation
is either a starless core ($\sim$\,100\,M$_\odot$)  or a starless
clump ($\sim$\,1000$_\odot$)  of molecular gas in a giant
molecular cloud. McKee \& Tan (2003)
envisage that these cores are molecular
condensations in a turbulence-supported
quasi-equilibrium that ultimately form single or
gravitationally bound multiple massive
protostars. As these authors argue, turbulent
and pressurized clouds permit sufficient
material to be available in the cores of giant
molecular clouds for high-mass star formation.
Mechanical energy must be continuously
injected into the clump in order to maintain this
quasi-equilibrium between turbulence and
gravity. The assumption is that this energy is
either injected from within the cores from the
kinetic energy of outflows and accretion
shocks, or it comes from the outside and
cascades down to smaller scale sizes. It is
furthermore assumed that (small-scale)
turbulence acts as an isotropic pressure.

In an alternate scenario by Bonnell et al.
(1997, 2001a), the compression phase is more
of a transient phase due to the random motions
in the self-gravitating cloud. Smoothed particle
hydrodynamics (SPH, see the sidebar Smoothed
Particle Hydrodynamics with Sink Particles)
simulations of this phase show sheets,
filaments, and cores forming (see {\bf Figure~3}).
Some cores collapse and fragment, marking the
beginning of the collapse phase for these
objects, while in other parts of the cloud
compression is still occurring.

The philosophical difference between the
approach of McKee \& Tan (2003) and that of
Bonnell et al. (1997, 2001a) can be summarized
as "monolithic collapse" versus "competitive
accretion." For the former, the mass necessary
for massive star formation is intimately
associated with the final product. If there are
bulk motions of the embryo star, its protostellar
core participates in those motions. The only
competition for the infalling material is
between close members of a multiple system.
For competitive accretion, the material that
makes up a particular star can come from
various parts of the parent cloud. Protostars
move relative to the molecular gas; the only gas
that is intimately associated with any particular
protostar is in its circumstellar disk and
envelope. Because multiple protostars are often
formed together at the same time, each
protostar competes for the available molecular
material (see {\bf Section~4.2}).

Crutcher's (1999, 2005) summary of the
available Zeeman measurements of magnetic
field strengths in molecular clouds suggests that
magnetic fields likely play an important role in
molecular cloud dynamics. V\'{a}zquez-Semadeni
et al. (2005) studied the evolution of clumps
and cores formed as turbulent density
fluctuations in nearly isothermal molecular
clouds, considering both the magnetic and
nonmagnetic cases with driven turbulence. In
the nonmagnetic case the authors find that the
cores are unlikely to reach a hydrostatic state --
necessary for monolithic collapse -- if the
molecular clouds have an effective polytropic
exponent less than 4/3. In this case, cores are
transient, either proceeding directly to collapse
or re-expanding on a dynamical timescale.

The magnetically subcritical clouds simulated by
V\'{a}zquez-Semadeni et al. (2005) do not produce
magnetostatic clumps, but rather a few marginally
bound clumps that are subsequently dispersed.
Ambipolar diffusion -- had it been included in the
simulations -- could have increased the clumps'
likelihood to become bound and subsequently to
collapse. For clouds with weaker magnetic fields a
few cores form and collapse on a timescale slightly
larger than the cloud's free-fall timescale. In their
most supercritical simulation, fewer clumps and
cores form than in their nonmagnetic counterpart,
and these cores reexpand because they are not Jeans
unstable. The authors thus conclude that not all cores
observed in molecular clouds will necessarily form
stars and that magnetic fields may help reduce the
star-formation efficiency by reducing core formation
rather than by delaying or inhibiting the collapse of
individual cores. 

\noindent
\textcolor{blue}{
{\bf Smoothed Particle
Hydrodynamics with Sink
Particles}}

\noindent
\textcolor{blue}{
Many simulations of star
formation in turbulent
interstellar clouds are
conducted using a Lagrangian
method to solve the equations
of hydrodynamics. This is
called smoothed particle
hydrodynamics (SPH). It is a 3D
method that assumes no
symmetries and uses no
imposed grids. Monaghan
(2005) gives a recent review of
the method and its applications
(see also Monaghan 1992). With
SPH, the gas is represented by
a number of sampling points
("particles"), each associated
with a certain gas mass. The
mass of each particle is
distributed in space according
to a spread function so that
density and other
hydrodynamic quantities are
continuous in space. The
particles are allowed to move in
the computational domain
under the laws of self-gravitating
hydrodynamics
(which are evaluated at the
particle positions). The system
is evolved in time using a
standard integration routine.}

\textcolor{blue}{
The timestep with which the
simulations evolve is chosen
such that all changes to the
system's state are examined at
the appropriate speed. For
example, during the collapse
phase, when infalling particles
approach each other, the
timestep gets progressively
shorter. This may potentially
bring the simulations to an
early halt as the timestep
converges to zero. This
problem is overcome with the
introduction of star (sink)
particles of finite radii that each
replace groups of dense
neighboring gas particles
above a certain density
threshold (see Bate, Bonnell \&
Price 1995). The sink particles
are then evolved as
collisionless matter that can
accrete from the lower density
gas whenever gas particles
approach within the sink
radius. The final state of the
system in such simulations
involves groups of sink
particles in spatial distributions
resembling observed young
clusters. However, all
information on scales smaller
than the sink radius is lost;
such sink particles cannot later
become sources of outflows,
unless specifically instructed to
do so. Note that sink particles
are also usefully employed in
Eulerian collapse calculations,
as shown by Krumholz, McKee
\& Klein (2004).}

\textcolor{blue}{
In order for fragmentation to be
realistically modelled, the
numerical resolution of the SPH
simulations must be such that
they obey the Jeans condition.
This states that "the local
Jeans mass must always
remain resolved," i.e., it must
contain more than a certain
number of particles (Truelove et
al. 1997, Bate \& Burkert 1997,
Kitsionas \& Whitworth 2002,
Martel, Evans \& Shapiro 2006).
The Jeans mass decreases as
the density increases up to the
point at which the gas stops
being approximately isothermal
and begins to heat up
adiabatically. From there on,
the Jeans mass starts
increasing again with density.
The density at which this
occurs defines the minimum
Jeans mass that the simulation
has to resolve and, thus, the
minimum number of particles
required. Present-day computer
capabilities allow the use of
several million SPH particles
per simulation, a number that is
nominally adequate for
resolving the density at which
adiabatic heating switches on.
Using parallel computing
facilities presently available,
such computations take several
months for each evolutionary
model.}

The assumption of driven as
opposed to decaying turbulence is important in these
and similar simulations (cf. Heitsch, Mac Low \&
Klessen 2001, Li \& Nakamura 2006).

Indeed, Li \& Nakamura (2006) using the best
3D MHD simulation technique to date have shown
that initial turbulence in a cluster forming region is
quickly replaced by motions generated by
protostellar outflows. This protostellar outflow-driven
turbulence can keep a protocluster clump
close to virial equilibrium long after the initial
turbulence has decayed away. This may explain the
early molecular cloud observations of Bertoldi \&
McKee (1992) and lend support to the recent
equilibrium cluster formation models of Tan,
Krumholz \& McKee (2006). It could even imply that
the stellar IMF is regulated by outflow feedback
(Silk 1995, Adams \& Fatuzzo 1996). However, the
ultimate test of these predictions has not yet been
made and requires much more observational work.
At the same time, competing, more violent cluster
formation models have been proposed in the
literature (e.g. Bonnell, Bate \& Vine 2003), and the
jury is still out. Essentially, the question is whether
star formation is slow or fast (Ballesteros-Paredes,
Hartmann \& V\'{a}zquez-Semadeni 1999, Glover \&
Mac Low 2007).

\subsection{The Collapse Phase}

Gravity plays the dominant role in star
formation. To form a star, gravity must
overcome pressure, magnetic forces, internal
turbulence, and rotation. In the simplest case of
gravity versus gas pressure, one defines the
Jeans mass,
\[
  {\rm M_{Jeans}} \simeq 1.1\;{\rm M}_\odot
  \left[ { {\rm T}_{\rm gas} \over 10\,{\rm K}} \right]^{3/2}
  \left[ { \rho \over 10^{-19}\,{\rm g\; cm}^{-3}} \right]^{-1/2}
\]
as the smallest mass for which gravity can become
dominant. The normalization is consistent with
typical initial conditions.

Turbulence as a repulsive force will exceed
gas pressure if motions are supersonic. Unless
continually replenished, however, supersonic
turbulence dies out on a dynamical timescale
(see, e.g., Stone, Ostriker \& Gammie 1998;
Clark \& Bonnell 2005; Kritsuk, Norman \&
Padoan 2006; for a summary of earlier work,
see Mac Low \& Klessen 2004).

As Shu, Adams \& Lizano (1987) point out,
once gravity dominates pressure and magnetic
forces in an optically thin gas capable of
radiating compressional heat, it remains
dominant and the gas collapses on a free-fall
timescale
\[
  {\rm t_{ff}} \simeq 2.1 \times 10^5\, {\rm yr}
  \left[ { \rho \over 10^{-19}\,{\rm g\; cm}^{-3}} \right]^{-1/2} \; .
\]

That is, the densest parts collapse the fastest,
and the Jeans mass decreases during collapse.
The gas collapses nonhomologously until the
densest parts become optically thick, allowing
the gas to heat up adiabatically and to increase
the gas pressure dramatically. Rotational
(centrifugal) forces increase during
gravitational collapse owing to conservation of
angular momentum, so flattened structures and
accretion disks (e.g., Black \& Bodenheimer
1975; Terebey, Shu \& Cassen 1984; Yorke \&
Bodenheimer 1999) are expected phenomena of
gravitational collapse.

\subsection{The Accretion Phase}

\subsubsection{Formation of the Hydrostatic Core} The
formation of low-mass stars is explained by the
nonhomologous collapse of a slowly rotating
fragment of molecular material; the collapse is
stopped in the central regions when the object
becomes optically thick (Larson 1969, Woodward
1978, Winkler \& Newman 1980, Masunaga,
Miyama \& Inutsuka 1998; for a more recent review,
see Klein, Fisher \& McKee 2004). There is a
second, inside-out collapse in the center for
${\rm T_{gas}} \simeq 2000$\,K when molecular hydrogen
dissociates. When the second core is optically thick
and thermally ionized, pressure forces are able to
balance gravity on a dynamical timescale, and one
speaks of an accreting quasi-hydrostatic core as
more material rains down. When it forms, the
second hydrostatic core contains somewhat more
than a Jupiter mass and has a radius on the order of
(3\,--\,5)\,R$_\odot$ (Winkler \& Newman 1980, Masunaga \&
Inutsuka 2000).

As long as material continues to flow onto
the quasi-hydrostatic core, the core grows in
mass. Simultaneously, it contracts on a thermal
Kelvin-Helmholtz timescale
${\rm \tau_{KH} \sim GM_*^2/R_*L_*}$
toward hydrogen-burning densities and
temperatures. {\bf Figure~4} shows the Kelvin-Helmholtz
timescale to the ZAMS as a function
of stellar mass. Except for the growth of mass
and angular momentum, the interior regions of
the hydrostatic core are generally ignorant of
and not influenced by the details of the
accretion process. This allows us to separate the
problem of star formation into several distinct
parts: ({\it a}) the evolution of the central core, ({\it b})
the details of transporting material from the
disk onto the core, ({\it c}) transporting material
inward within the disk, and ({\it d}) accretion onto
the disk. Fundamental differences between low-mass
and high-mass star formation can be
attributed to differences in the above processes
and, in particular, to significant differences in
the timescales involved and in the local
radiative environment.

\subsubsection{Evolution of Accreting Cores} Accretion
onto the (proto)star will affect the core's outer
(atmosphere) regions and the overall spectral
appearance. We estimate the thickness ${\rm \Delta R_{dyn}}$ of the
outer stellar regions that are dynamically affected by
accretion to be:
\[
  {\rm \frac{\Delta R_{dyn}}{R_*} \sim
   \frac{\dot M_{S-acc} t_{S-dyn}}{M_*} =
    \frac{t_{S-dyn}}{t_{S-acc}}}
\]
where ${\rm t_{S-dyn} \simeq R_*/v_{esc}}$, the stellar dynamical
timescale, is the time for an acoustic wave to cross
through the core or the orbital period of a body just
above the core's surface (generally less than a day).
Because ${\rm t_{S-dyn} <<< t_{S-acc} =}$
${\rm M_*/\dot M_{S-acc}}$
(the accretion
timescale), ${\rm \Delta R_{dyn}/R_*}$ is extremely small 
($<$\,10$^{-5}$ for
accretion rates 
${\rm \dot M_{S-acc} < 10^{-2}}$\,M$_\odot$\,yr$^{-1}$).

It is not easy to estimate the size of the
region thermally affected by the accretion flow
onto the (proto)star, because the radiation
transfer in an optically thick plasma depends on
details of the complex accretion geometry.
Exactly how material flows from the disk onto
the (proto)star remains an unsolved theoretical
problem.

One can expect a (proto)star's location in the
Hertzsprung-Russell (HR) diagram to be
strongly affected by accretion whenever the
Kelvin-Helmholtz timescale ${\rm \tau_{KH}}$ exceeds the
accretion timescale ${\rm \tau_{acc}}$. However, even when
this criterion is not fulfilled, the Kelvin-Helmholtz
timescale for the outer regions of a
star can locally exceed the accretion timescale,
and accretion can affect the radius, effective
temperature, and luminosity of the star
(Kippenhahn \& Meyer-Hofmeister 1975).

Kippenhahn \& Meyer-Hofmeister (1975)
consider the case of mass transfer within a close
binary system. At high accretion rates
${\rm \dot M_{S-acc} \ga 10^{-3}}$\,M$_\odot$\,yr$^{-1}$,
accreting main-sequence
stars of mass M$_* \ga 5$\,M$_\odot$ bloat up to
radii exceeding 10 times their ZAMS values.
Although not strictly applicable to the case of
accreting premain-sequence stars, one can
expect the same qualitative effect of high
accretion rates in this mass range. These results
demonstrate the importance of accretion for the
appearance of the central hydrostatic cores.

For low-mass stars, it is generally accepted
that material is transported onto the central core
through an accretion disk (e.g., Shu, Adams \&
Lizano 1987). The net gain per unit time of
gravitational potential energy of the accreted
material ${\rm G M_*\dot M_{S-acc}/R_*}$
is partly converted into
rotational energy of the core and disk ($\sim$\,1/4)
and partially converted into heat ($\sim$\,3/4), which
is radiated away (Yorke \& Kr\"ugel 1977). Part
of the gravitational energy is converted into
heat in a series of disk accretion shocks (Yorke
\& Bodenheimer 1999); part of it is converted
into heat within the disk by the same viscous
processes that transport angular momentum
outward and allow the radial flow of material
inward; and part of it is converted into heat in
the accretion flow and shocks/relaxation zones
on the (proto)stellar surface.

As mentioned above, the details of how the
gas is ultimately transported from the disk onto
the core are still unclear. For low-mass stars it
has been postulated that magnetically focused
flows (Blandford \& Payne 1982) and/or
accretion columns (K\"onigl 1991, Edwards et al.
1993) and/or X-winds (Shu et al. 1994, 1995)
are involved. Because 50\,\% 
($\simeq$\,3/8\,${\rm G M_* \dot M_{S-acc}/R_*}$)
of the total gravitational energy
is converted into heat and radiated away within
1\,R$_*$ of the core, one can expect higher
temperatures and luminosities as either M$_*$ or
${\rm \dot M_{S-acc}}$ increases or as the core contracts. The
temperature of this hot shocked gas
\[
   {\rm T \simeq \frac{ G M_*/R_*}{k/\mu m_H}
   \simeq 10^6\,K\
   \left[ \frac{M_*}{1\,M_\odot} \right]
   \left[ \frac{R_*}{10\,R_\odot} \right]^{-1}}
\]
is sufficiently high to produce X-rays even for low-mass stars.

The above argument is only slightly
modified when the core and disk produce a
wind or outflow. We then speak of a net mass
accretion of ${\rm \dot M_{S-acc} - \dot M_{S-wind}}$.
A portion of the
core's rotational energy and angular momentum
can be converted into kinetic energy and
angular momentum of the wind.

A necessary condition to accrete sufficient
material to produce a massive star is thus,
\begin{equation}
  {\rm M_* = \int_0^t \left[ \dot M_{S-acc}(t') - \dot M_{S-wind}(t')
  \right]\, dt' \ga 8\,M_\odot} .
\end{equation}
That is, the accretion rate ${\rm \dot M_{S-acc}}$ onto an embryo
object must exceed the outflow rate ${\rm \dot M_{S-wind}}$ during a
significant proportion of the formation process. For
this to occur, the acceleration owing to gravity must
exceed the outward directed radiative acceleration of
the accreting core. Whereas gravity ${\rm G M_*/r^2}$ at each
radial point in an envelope increases linearly with
core mass, the radiative acceleration of dusty
material ${\rm \kappa L/4\pi r^2 c}$ is proportional to the core's
luminosity. This increases as a high power of stellar
mass. A lower limit to the core's luminosity is the
ZAMS luminosity (see {\bf Figure~4}).

Thus, to allow infall, we require as a necessary
condition
\begin{equation}
{\rm \kappa_{eff} L / 4 \pi r^2 c < G M_* / r^2}
\end{equation}
with L\,=\,L$_*$\,+\,L$_{\rm acc}$, which translates into
\begin{equation}
  {\rm \kappa_{eff} < 130\, cm^2 g^{-1}
  \left[ \frac {M_*}{10\,M_\odot} \right]
  \left[ \frac {L}{1000\,L_\odot} \right]^{-1}}  ,
\label{eq:critopacity}
\end{equation}
where the effective opacity for radiative acceleration
of accretable material is defined as
\begin{equation}
  {\rm \kappa_{eff} = {\int_0^\infty} \kappa_\nu^{\rm rad} F_\nu\,d\nu / F} ,
\end{equation}
here $\kappa_\nu^{\rm rad}$ is the frequency-dependent gram-opacity
(cgs-units: cm$^2$g$^{-1}$) of the material subject to
radiative acceleration and ${\rm F = \int_0^\infty F_\nu\,d\nu}$ is the
radiative flux. (For a more complete discussion of
the definition of $\kappa_\nu^{\rm rad}$, including effects of
nonisotropic scattering and photoejection of particles
from the dust, see Yorke 1988.) The (proto)star's
luminosity is given by the sum of its intrinsic
luminosity L$_*$  and the luminosity ${\rm L_{acc}}$ emitted by the
dissipation of kinetic energy of the material being
accreted.

For dusty gas, $\kappa_\nu^{\rm rad}$ is strongly frequency-dependent.
Depending on the hardness of
radiation emitted by the central source and the
accretion shocks, dusty gas could be repelled
from the star. The ISM extinction in the J-band,
for instance, corresponds to ${\rm \kappa}$\,=\,130\,cm$^2$\,g$^{-1}$.
In {\bf Figure~5} we indicate values of $\kappa_{\rm eff}$ for dusty
gas at solar abundances ($\kappa_\nu^{\rm rad}$ taken from the
Preibisch et al. 1993 dust model with and
without ice-coated grains) illuminated by 600\,K
and 6000\,K black bodies [i.e., ${\rm F_\nu \propto B_\nu (T_*)}$],
respectively, and for a fully ionized dustless
hydrogen plasma.

{\bf Figure~6} displays the mean effective opacity
of dusty gas as defined by {\bf Equation~4}, using the
Preibisch et al. (1993) dust model with and
without ice-coated grains ({\it blue} and {\it red} curves,
respectively) and assuming blackbody radiation
at the temperature T$_{\rm rad}$. Using the mass scale on
the right of {\bf Figure~6} we note that the net force
on dusty gas surrounding a deeply embedded
100\,M$_\odot$ main-sequence star is directed toward
the star as long as the star appears to the dust to
be a $\la$\,50-K source. By contrast, for an
unobscured 3\,M$_\odot$ main-sequence star with
T$_{\rm eff} \ga 10^4$\,K the net force on dusty gas would be
directed away from the star.

\subsection{Overcoming Radiative Acceleration}

Next we discuss how {\bf Equation~2} can be
satisfied, i.e., accretional growth of an already
existing stellar embryo can be enabled. At least
one of the following conditions must be met:
({\it a}) ${\rm \kappa_{eff}}$  must be sufficiently low, i.e.,
significantly lower than its ISM value for
optical/UV radiation; ({\it b}) the total luminosity L
must be reduced; or ({\it c}) gravity (i.e., the stellar
mass M$_*$ ) must be increased. Below, we
discuss each of these three possibilities.

\subsubsection{Reduce ${\rm \kappa_{eff}}$} Owing 
to the strong frequency
dependence of dust opacity, ${\rm \kappa_{eff}}$ 
can be significantly
lower than its ISM value if the radiation field seen
by the accreting material is that of a cold, embedded
object. When embedded the  protostellar radiation
field is shifted from the optical/UV -- where the dust
absorbs the photons -- into the far-IR where the dust
reemits the absorbed energy. Alternatively, ${\rm \kappa_{eff}}$ can
also be reduced if the average size of dust grains
increases (but remains compact rather than
becoming fractal) or if most of the dust is destroyed.
In their pioneering efforts, Kahn (1974) and Wolfire
\& Cassinelli (1987) studied the 1D, spherically
symmetric accretion problem for massive star
formation with an emphasis on the dust opacity.
Kahn concluded that a 40-M$_\odot$ star could be formed
by spherically symmetric accretion, but his assumed
dust destruction temperature T$_{\rm sub}$\,=\,3600\,K is too high
by a factor of about two (implying for the outer
regions greater protection from radiation pressure).
Furthermore, his assumed grain opacities, which can
be written in the form
${\rm \kappa \simeq 100 cm^2 g^{-1} (T_{dust}/9000 K)^2}$, 
were somewhat too low.

In a more careful treatment of the dust (but
still assuming steady-state spherically
symmetric infall), Wolfire \& Cassinelli (1987)
concluded that very massive stars can form
only if the dust has been significantly modified.
By contrast, Yorke \& Kr\"ugel (1977) showed in
a hydrodynamical simulation that spherically
symmetric accretion must be nonsteady for the
high-mass case. They were able to produce
stars of masses 17\,M$_\odot$ and 36\,M$_\odot$ from clouds
of masses 50\,M$_\odot$ and 150\,M$_\odot$, respectively, in
a highly variable accretion flow.

These early attempts to explain high-mass
star formation suffer significantly from the fact
that spherically symmetric infall was assumed.

The effective opacity of the accreting
material can also be reduced by density
inhomogeneities resulting from the photon
bubble instability. The radiation escapes readily
through the gaps between the shocks that are
driven by disturbances in the radiation flux
(Turner, Quataert \& Yorke 2007). Another
possibile explanation for reducing the effective
opacity is the accretion of optically thick blobs
or fingers. These can be expected to form via
Rayleigh-Taylor instabilities in the radiation-inflated
cavities produced by luminous
(proto)stars (Krumholz, McKee \& Klein 2005a;
Yorke 2002). In this case,
\[
  {\rm \kappa_{eff} = \pi R^2_{blob} / M_{blob}}.
\]
As a particular subset of this family of solutions,
Bonnell, Bate \& Zinnecker (1998) considered
building up massive stars by coalescence of
intermediate-mass stars within a very dense
protostellar cluster (see also Bonnell \& Bate 2002,
Zinnecker \& Bate 2002). We return to this issue in
more detail below.

Opacity modifications owing to coagulation
of dust and dust destruction processes during
the collapse phase were calculated by Suttner \&
Yorke (2001) for three different detailed dust
models: compact spherical particles, fractal
BPCA (ballistic particle-cluster agglomeration)
grains, and fractal BCCA (ballastic cluster-cluster
agglomeration) grains. (BPCA dust
consists of grains with widely different grains
sizes, whereas BCCA dust is formed by
coagulation of similar-sized grains). Assuming
axial symmetry, Suttner \& Yorke followed the
dynamics of gas and 30 individual dust
components. They found that even during the
early collapse and the first $\sim$\,10$^4$\,yr of
dynamical disk evolution, the initial dust size
distribution is strongly modified. Close to the
disk's midplane, coagulation produces dust
particles of several tens of  $\mu$m in size (for
compact spherical grains) up to several
millimeters in size (for fluffy BCCA grains). In
contrast, in the vicinity of the accretion shock
front located several density-scale heights
above the disk, large velocity differences
inhibit coagulation. Dust particles larger than
about 1\,$\mu$m segregate from smaller grains
behind the accretion shock. Owing to the
combined effects of coagulation and grain
segregation the IR dust emission is modified.
Within the accretion disk, an interstellar
medium dust size distribution (Mathis, Rumpl
\& Nordsieck 1977) provides a poor description
of the general dust properties. Nevertheless, the
radiative force acting on material infalling from
the envelope to the disk is hardly affected by
coagulation (see also Weidenschilling \&
Ruzmaikina 1994).

\subsubsection{Reduce the Effective Luminosity} Nakano,
Hasegawa \& Norman (1995) and Jijina \&
Adams (1996) pointed out that, because we expect
accretion to proceed through an accretion disk,
radiation pressure could blow away the tenuous
polar regions but not the massive disk. Yorke \&
Bodenheimer (1999) and Yorke \& Sonnhalter
(2002) studied this effect quantitatively and
substantiated this claim through numerical
simulations. They found that, whereas the central
object may emit radiation isotropically, the radiation
field quickly becomes anisotropic farther from the
center. For an outside observer, and in particular, for
a dust grain attempting to accrete onto an existing
protostellar disk, the radiative flux close to the
equatorial plane can be much smaller than the
component parallel to the rotation axis. This so-called
flashlight effect (beaming of radiation in the
polar direction) occurs whenever a circumstellar
disk forms.

Yorke \& Bodenheimer (1999) considered the
evolution of a 10-M$_\odot$ molecular fragment that
produced a 8.2-M$_\odot$ star (their case F). {\bf Figure~7}
depicts an intermediate stage of the evolution,
after 7\,M$_\odot$ of material have accreted onto the
central protostar, while 2.8\,M$_\odot$ have
accumulated in a circumstellar disk and 0.2\,M$_\odot$
still reside in an larger extended infalling
envelope. Inner and outer disk accretion shocks
(density and temperature discontinuities) are
visible above and below the disk. The dark red
regions with an opening angle of about 45$^\circ$
show preferential heating of polar regions by
the protostar. The light red regions within 45$^\circ$
of the equator beyond $\sim$\,1000\,AU indicate
shadowing by the disk.

Yorke \& Bodenheimer (1999) estimated that
the edge-on and pole-on bolometric fluxes can
differ by more than a factor of 30 after about
one-half of the mass of a 2-M$_\odot$ collapsing
protostellar clump has accreted onto the
protostar. The difference in radiative
acceleration is much greater than this factor of
$\ga$\,30, however, because the edge-on flux is
dominated by the far-IR, which is far less
effective at radiatively accelerating dusty gas
than mid- and near-IR light seen pole-on. Yorke
\& Sonnhalter (2002) showed by calculating
frequency-dependent radiation transfer that the
flashlight effect is further enhanced by the fact
that the central star's optical and UV radiation
blows out material in the polar direction,
reducing back-scattering of radiation toward the
disk. A similar effect has been reported by
Krumholz, McKee \& Klein (2005a) from polar
cavities blown free by outflows originating
close to the star. Thus, photons emitted or
scattered into these directions will not hinder
accretion of material within the disk or material
in the disk's shadow regions.

Although the flashlight effect allows dusty
material to come close to the central source via
a circumstellar disk, the material to be accreted
eventually encounters optical and UV radiation
from the central source. For this material to be
accreted rather than blown out by radiation, the
dust must be largely destroyed or it must have
coagulated into larger particles so that the
opacity is dominated by the gaseous
component.

Even though no massive disk has yet been
directly observed around a main-sequence O
star, there is much indirect evidence that such
disks exist (see the review by Zhang 2005 and
our more detailed discussion in {\bf Section~6}). A
compelling argument that disks exist during the
early phases of massive star formation is the
observation of massive bipolar outflows. Such
massive outflows are probably powered by disk
accretion, and, as in their low-mass
counterparts, the flow energetics appear to scale
with the luminosity of the source (see Shepherd
\& Churchwell 1996a,b; Richer et al. 2000;
Henning et al. 2000; Wu et al. 2005).

If the primary source of the massive star's
material is accretion from the surrounding
molecular core, then a circumstellar disk should
be the natural consequence of the star-formation
process even in the high-mass case.
However, it should be difficult to observe disks
around massive stars. The high far UV and
extreme UV fluxes associated with high-mass
stars will begin to photoevaporate the disks on
timescales of $\sim$\,10$^5$\,yr (Hollenbach, Yorke \&
Johnstone 2000). The results will be observable
as deeply embedded ultracompact HII-regions
with comparable lifetimes (Richling \& Yorke
1997). The fact that disks around O stars
photoevaporate so quickly provides negative
feedback for disk accretion. This limits the
build-up of more massive stellar objects and
may even imply an upper mass limit for star
formation.

\subsubsection{Increase Gravity} For completeness we
mention the fact that the gravitational acceleration is
enhanced with respect to radiative acceleration when
massive stars form within a dense cluster of
not-so-brightly-radiating objects. In this scenario, one
requires a density-peaked cluster of low-mass
objects embedded within a molecular cloud, with
${\rm \rho_{objects} \gg \rho_{gas}}$. 
The effective gravity near the
cluster's center is enhanced relative to an isolated
molecular cloud without the cluster and relative to
off-center regions of the molecular cloud (cf. Keto
2002). If this were the only way to form massive
stars, isolated massive stars would exist only in very
exceptional cases.

\subsection{Stellar and Protostellar Luminosity
Evolution}

Because the luminosity is so critical during
accretion up to high stellar masses, one must
also consider the luminosity evolution of the
accreting object. As discussed by Behrend \&
Maeder (2001) and Yorke (2002), {\bf Figure~8}
shows that protostars do not evolve along
premain-sequence tracks until they land on and
remain at a unique spot on the main sequence
where hydrogen burning starts. Rather, they
reach the main sequence -- that is, the
hydrogen-burning state -- well before they have
finished accreting mass. After that, they
continue growing in mass and evolve up the
main sequence until they run out of material to
accrete. This means that an initially low-mass
object that gains mass through accretion
evolves substantially differently in the HR
diagram than would a nonaccreting premain-sequence
star of the same final mass.

{\bf Figure~8} shows a number of protostellar
evolutionary tracks in the  HR diagram,
assuming a sequence of conceivable accretion
rates increasing by factors of 10, but each
constant in time. These accreting tracks were
calculated by Yorke (2002) and are
qualitatively similar to the more detailed stellar
evolution calculations by Norberg \& Maeder
(2000), Behrend \& Maeder (2001), and -- for
the lower accretion rates -- Palla \& Stahler
(1992). Differences to the two former
investigations can be attributed to Yorke's
(2002) starting mass, 0.1\,M$_\odot$ instead of 1\,M$_\odot$,
and the differing accretion rates. In all cases
published to date, not only do the tracks of
accreting objects consistently lie slightly below
the equilibrium deuterium-burning birthline,
but the qualitative effect of more rapid
accretion is to shift the tracks away from the
birthline. (In this context, the birthline is
defined as the point on the Hayashi track at
which equilibrium deuterium burning begins in
a nonaccreting star. Palla \& Stahler (1993),
however, use the concept of birthline as the
evolutionary track of accreting premain-sequence
stars in the HR diagram. Deuterium
burning of newly accreted material keeps these
tracks well above the main sequence until
accretion stops.) The tracks of accreting stars
eventually converge to the main sequence and
follow along the ZAMS as more material is
added. For example, at an accretion rate of 
10$^{-3}$\,M$_\odot$\,yr$^{-1}$,
hydrogen burning begins at
t\,$\simeq$\,1.3\,$\times$\,10$^4$\,yr, 
after $\sim$\,13\,M$_\odot$ have been
accreted.

The tracks discussed above do not display
the degree of bloating seen by Kippenhahn \&
Meyer-Hofmeister (1975) for accreting main-sequence
stars M$_*$\,$\sim$\,5\,--\,10\,M$_\odot$.
Yorke's (2002)
simplifying assumption of thermally adjusted
(pre)stellar objects is not strictly fulfilled,
especially for the 
${\rm \dot M_{S-acc}}$\,=\,10$^{-3}$\,M$_\odot$\,yr$^{-1}$ 
case.
Norberg \& Maeder (2000) did not consider
accretion rates 
${\rm \dot M_{S-acc}}$\,$>$\,10$^{-4}$\,M$_\odot$\,yr$^{-1}$ 
and
Behrend \& Maeder (2001) attained high
accretion rates 
${\rm \dot M_{S-acc}}$\,$\ga$\,10$^{-3}$\,M$_\odot$\,yr$^{-1}$
for high-mass
objects M$_*$\,$\ga$\,25\,M$_\odot$ only. Palla \& Stahler
(1992, 1993) found significant bloating for the
highest accretion rate $10^{-4}$\,M$_\odot$\,yr$^{-1}$ they
considered, attributable to shell deuterium
burning. We expect that (proto)stars accreting
at extremely high rates are not fully thermally
adjusted and thus bloat up. The degree of
bloating depends on the accretion rate and mass
of the accreting object 
(Kippenhahn \& Meyer-Hofmeister 1975).

We remind the reader that these theoretical
tracks in the HR diagram do not reflect the
actual observable bolometric luminosities of
accreting protostars nor do they reflect the
observable effective temperatures. Plotted in
the HR diagram are the intrinsic (proto)stellar
properties, from which the observed properties
must be derived by multidimensional radiative
transfer calculations (e.g. Indebetouw et al. 2006).
Note that much of the accretion luminosity,
${\rm L_{acc}}$, is indistinguishable from the intrinsic
luminosity, L$_*$, of the star. 
Also, owing to the
existence of shock fronts and their postshock
relaxation zones, strong deviations from a
stellar spectral energy distribution can be
expected. The importance of accretion
luminosity, ${\rm L_{acc} \sim G M_*\dot M/R_*}$, is seen in
{\bf Figure~5}. At high accretion rates the L/M ratio
can attain high values even for low-mass cores,
so that radiative acceleration can clear out part
of the infalling material in the polar regions.

What order of magnitude of mass accretion
rate can be expected? In order to produce a star
of mass M$_*$  within, say 200,000\,yr, an
average accretion rate of
${\rm 5 \times 10^{-6}\,M_\odot\,yr^{-1} [M_*/M_\odot]}$  
is necessary.
Assuming this average accretion rate, we note
that during the main accretion phase the
luminosity of low-mass stars is dominated by
accretion luminosity, whereas for high-mass
stars the luminosity is initially determined by
accretion but is eventually dominated by the
intrinsic stellar luminosity. Of course, the actual
accretion rate may vary strongly from this
average value. The maximum sub-Eddington
accretion rate possible onto a core hydrogen-burning
star, assuming electron scattering and
the effects of both the intrinsic stellar
luminosity and accretion luminosity, can be
inferred from {\bf Figure~5}. Whereas the dotted
curve corresponding to an accretion rate of
${\rm 10^{-3}\,M_\odot\,yr^{-1}}$
lies below the value permitted
by electron scattering, an accretion rate ten
times higher would clearly lie above this value
everywhere. Thus, accretion of ionized material
onto a stellar core at a rate
$\ga {\rm 10^{-2}\,M_\odot\,yr^{-1}}$
implies super-Eddington accretion.

\subsection{Stellar Evolution Beyond the Zero Age Main
Sequence}

As hydrogen burning proceeds, the stars begin
to evolve away from the ZAMS even as they
accrete material through a disk and lose
material through a stellar wind. The star's
evolution strongly depends on its mass loss rate
and on the internal mixing that is induced by
rotation. These effects must be taken into
account when modeling these stars and their
effect on their environment.

If the stars attain their final mass and become
optically visible on a timescale much shorter
than their main-sequence lifetime, they will not
have evolved far from the ZAMS. If, however,
the stars first become optically visible after
significant stellar evolution, the traditional
concept of the ZAMS as the starting point for
newly formed massive stars is flawed. To
illustrate this point we compare in {\bf Figure~9} the
current best estimate of the locations in the HR
diagram of spectral types O3 through O9.5 for
luminosity classes V, III, and I with theoretical
isochrones and evolutionary tracks. It is not
surprising that the O dwarfs do not fall on the
ZAMS, because we are dealing with average
properties over a range of ages. A quantitative
explanation of the magnitude and systematic
variation of the offset for early to late O dwarfs,
for example a shift of average ages from $\sim$\,1 to
5\,Myr, is still lacking (Martins, Schaerer \&
Hillier 2005). Such a shift could occur, because
stars of later spectral types have weaker winds
and lower UV fluxes and therefore remain
embedded in their parental cloud longer.

\begin{table}[h]
\begin{center}
\caption{O-stars spectral type versus mass for
different luminosity classes (Martins, Schaerer \&
Hillier 2005)}
\vspace{0.5cm}
\begin{tabular}{@{}ccccc@{}}
\hline\hline
{\bf SpT} & \hspace{1cm} & & {\bf Luminosity class} & \\
\hline
                   & & V    & III  & I\\
\hline
B0                 & & --   & --   & -- \\
O9.5               & & 16.5 & 21.0 & 30.4 \\
O9                 & & 18.0 & 23.1 & 32.0 \\
O8                 & & 22.0 & 26.9 & 36.8 \\
O7                 & & 26.5 & 31.2 & 40.9 \\
O6.5               & & 29.0 & 33.7 & 43.1 \\
O6                 & & 31.7 & 36.4 & 45.8 \\
O5                 & & 37.3 & 41.5 & 50.9 \\
O4                 & & 46.2 & 48.8 & 58.0 \\
O3                 & & 58.3 & 58.6 & 66.9 \\
O2.5               & & --   & --   & -- \\
\hline
\end{tabular}
\end{center}
\vspace{-1.0cm}
\begin{tabular}{@{}ll@{}}
\hspace{2.5cm} & \\
& Note: masses in solar units\\
\end{tabular}
\end{table}

We find it useful to end this stellar evolution
section with a compilation ({\bf Table~3}) of the
masses of massive stars as a function of spectral
type (from O9.5 through O3) and luminosity
class (V, III, and I).

\subsection{Ionization Evolution and Cloud Disruption}

Because massive stars are built by accretion,
whether from a monolithic collapse or
competitively, they will produce an increasing
number of hydrogen ionizing photons even
while they grow as they become hotter and
more luminous. Thus, in spite of accretion,
hydrogen-ionizing and helium-ionizing fluxes
similar to those expected from ZAMS stars are
likely (see {\bf Figure~10}). Indeed, the process of
accretion itself is likely to produce hard
ionizing radiation in a series of accretion
shocks close to the stellar surface. Moreover,
powerful winds interacting with surrounding
material will produce strong shocks and hard
radiation.

Thus, material in the immediate vicinity of
the accreting massive star, in particular material
in the circumstellar disk and in any nearby star-disk
systems, can be ionized. In the case of
circumstellar disks, a thin layer of ionized gas
on the surface of the disk results. Close to the
star, where the escape velocity is much greater
than the sound speed, ionized gas remains
bound. Ionized gas outside a radius r$_{\rm evap}$, that is,
the radius where the sound speed exceeds the
escape velocity, expands outward. The situation
depicted in {\bf Figure~11} is likely to result.

The photoionization of circumstellar disks
produces an outflowing disk wind of ionized
gas that interacts with the stellar wind.
Depending on its  strength (and the
photoionization rate), the stellar wind will be
collimated by the ionized disk wind (Richling
\& Yorke 1997). The stronger the stellar wind
(or the weaker the ionizing photon flux and
hence the ionised disk wind), the less
collimated the stellar wind will be.

This is different from the case of collimated
jets and outflows due to X-winds and disk
winds in young low-mass stars, where magnetic
fields and forces are involved (Shu et al. 1995).
In the high-mass case, although magneto-centrifugal
forces cannot be dismissed from
influencing the collimation process, we can
have rather collimated outflows without
magnetic fields. This being said, we suspect
that magnetic fields do play a role not only in
low-mass but also in high-mass star formation,
not least because there are by now at least two
cases where strong kilo-Gauss stellar magnetic
fields have been inferred from Zeeman spectro-polarimetry,
including the oblique magnetic
rotator $\theta^1$~Ori~C (Stahl et al. 1996, Donati et al.
2002) and the Of?p star HD 191612 (Walborn
et al. 2003, Donati et al. 2006). The mismatch
in $\theta^1$~Ori~C between the magnetic field
direction and the stellar spin axis could be an
indication that this well-known massive star
formed in a collisional process from a merger
of two lower mass stars. Finally, we mention
the detection of hard X-ray emission from the
W3 massive star-forming region, an indication
that embedded massive stars must be
magnetically active (Hofner et al. 2002).

Hoare et al. (2007) examined the suggestion
first made by Hollenbach et al. (1994) and later
calculated numerically by Yorke \& Welz
(1996) and Richling \& Yorke (1997) that
photoevaporating disks can explain the
existence of unresolved ultracompact HII
regions (sizes $\sim$\,0.1\,pc). Many of these have
since been resolved with long-baseline radio
interferometry and have been reclassifed as
hypercompact HII regions (sizes $\sim$\,0.01\,pc; see
Kurtz 2005). The latter most likely represent
the individual photoevaporating disks
associated with individual deeply embedded O
stars (Keto 2007). The phase of
photoevaporating disks is the first evolutionary
stage after the ionizing flux has turned on. It
remains to be seen whether the thermally
evaporating flow interacting with a stellar wind
can produce the high velocities seen in the
broad H92$\alpha$ recombination line seen toward
hypercompact HII regions (Sewilo et al. 2004).

Hoare et al. (2007) concluded that in a later
evolutionary phase most ultracompact HII
regions need to be interpreted as external
photoevaporation of a molecular clump. The
longevity of these HII regions is explained by a
relative motion of the ionizing source and its
stellar wind into the clump, producing a
combination of bow shock and unipolar flow (a
so-called champagne flow, see the review by
Yorke 1986). If this is the case, then the
transition from hypercompact to ultracompact
HII regions may mark the end of the existence
of the accretion disk. The ultracompact HII
region phase then marks the beginning of the
dissipation of the molecular cloud as the stellar
wind and ionizing radiation of the newly
formed massive star are able to interact with
lower density, more poorly gravitationally
bound molecular material.

An important point to make in the evolution
of an ultracompact HII region is that the
gravitational force of the star(s) responsible for
the HII region should be included (Keto 2002).
The gravitational attraction of the star(s) can
maintain the accretion flow within the ionized
gas and prevent the HII region from expanding
hydrodynamically. This is true as long as the
radius of the ionization equilibrium is smaller
than the radius where the sound speed of the
ionized gas, about 10\,kms$^{-1}$, approximates the
escape velocity (concept of the trapped HII
region). Indeed, observations of the H66$\alpha$
recombination line from the ultracompact HII
region G10.6-0.4 powered by a compact and
very luminous ($\sim$\,10$^6$\,L$_\odot$) cluster of newly
formed massive stars surprisingly show inward
motion (Keto 2002). This implies that, despite
the high luminosity and ionizing radiation of
several O stars, neither radiation pressure nor
thermal pressure has reversed the accretion
flow (Keto \& Wood 2006). The ram pressure of
the observed accretion flow, with a rate of
M$_*$\,=\,10$^{-3}$\,M$_\odot$\,yr$^{-1}$, 
can confine the bubbling
HII region and allow the massive stars to
continue growing in mass -- from ionized gas!
One may speculate whether the ionization
surrounding the moderately massive stars helps
rather than hinders their further growth by
accretion (Keto \& Wood 2006), the reason
being that ionized gas can couple strongly with
any magnetic fields. Magnetic fields can
transport angular momentum outward, thereby
allowing a high rate of infall to smaller radii to
be maintained. Eventually, of course, with the
stellar masses steadily increasing, the ionizing
flux will increase enough (roughly $\propto$\,M$_*^4$, see
{\bf Figure~10}) so that the radius of ionization
equilibrium grows beyond the critical radius  within
which the gas is gravitational bound. At this
point the HII region will burst free and start to
disrupt the dense cloud, preferentially
unidirectionally in a champagne flow, revealing
the initially deeply embedded cluster (both
high-mass and low-mass stars).

\noindent
\textcolor{blue}{
{\bf Interim Summary}}

\noindent
\textcolor{blue}{
Despite the obstructing effects of radiation
pressure on dust, which would prevent
massive stars with masses of more than
30\,--\,40\,M$_\odot$ to be formed in spherically
symmetric collapse and accretion models,
these and more massive stars can be formed
via accretion through a circumstellar disk
(i.e., in 2D and 3D models). An accreting
star quickly evolves to the main sequence
after about 10\,M$_\odot$ have been accreted, but
the star is not yet observable. Still obscured
by the material in its vicinity, its appearance
will be that of a hypercompact HII region.
Radiative acceleration, photoevaporation,
and stellar winds eventually destroy the
accretion disk, but prior to this, accretion
onto the star provides an additional source of
luminosity. (Realistically,  this accretion is
expected to be highly variable and episodic.)
A powerful radiation- or magnetically driven
outflow in the polar directions and a puffed-up
(thick) disk result from the high
luminosity of the central source. The details
of how disk material ultimately flows onto
the star are still unclear -- as is often the case
for accretion disks.}

\newpage

\section{MASSIVE STAR FORMATION: COMPETING CONCEPTS AND CALCULATIONS}

Three different concepts describing the origin
of massive stars have been discussed in the
recent literature, each of which may occur in
nature, depending on the initial and
environmental conditions for the parent
molecular clouds (e.g., the Mach number of
supersonic turbulence in the clouds and the
external pressure in the ISM). These are (1)
monolithic collapse and disk accretion, (2)
competitive accretion and runaway growth, and
(3) stellar collisions and mergers. Whether the
parent molecular cloud clumps are magnetically
subcritical or supercritical (Shu, Adams \&
Lizano 1987, Crutcher \& Troland 2007) may
play a crucial role in determining whether
massive stars form (1) in isolation or (2) are
strongly clustered. Furthermore, in extreme
star-forming environments (such as in massive
protoglobular cluster clouds), the initial gas
densities may have been so high that (3) stellar
collisions become an unescapable ingredient in
massive star formation. We discuss all these
routes toward massive star formation in this
section. In addition, we briefly discuss
competing, but not mutually exclusive
scenarios related to massive star formation
stimulated by the pressure of expanding HII
regions versus a supernova blast wave.
Similarly, we highlight the two competing, but
not mutually exclusive processes accounting for
runaway OB stars. Finally, the important
concept of mass segregation (the fact that
massive stars are often concentrated in the
centers of star clusters) is reviewed in terms of
nature (a birthmark) or nurture (subsequent
dynamical evolution).

\subsection{Monolithic Collapse and Disk Accretion}

Yorke \& Sonnhalter (2002) consider the
collapse of isolated, rotating, nonmagnetic,
massive molecular cores of masses 30\,M$_\odot$, 60\,M$_\odot$,
and 120\,M$_\odot$ using a  frequency-dependent
radiation hydrodynamics code. The
flashlight effect discussed in {\bf Section~3.5} allows
material to enter into the central regions
through a disk. For massive stars, it is important
to take into account the frequency-dependent
nature of the opacity and the flux within the
disk rather than assuming either Rosseland or
Planck gray opacities. For their 60\,M$_\odot$ case,
Yorke \& Sonnhalter find that 33.6\,M$_\odot$ are
accreted in the central regions as opposed to
20.7\,M$_\odot$ in a comparison gray calculation.
Because these simulations cannot spatially
resolve the innermost regions of the molecular
core, they cannot distinguish between the
formation of a dense central cluster, a multiple-star
system, or a single massive object. They
also cannot exclude significant mass loss from
the central object(s) that may interact with the
inflow into the central grid cell. With the basic
assumption that all material in the innermost
grid cell accretes onto a single object, they are
only able to provide an upper limit to the mass
of stars that could possibly be formed for the
cases considered.

Note that the $\sim$\,43\,M$_\odot$ star formed during the
collapse of the 120\,M$_\odot$ molecular clump
(Yorke \& Sonnhalter 2002) does not represent
an upper limit to the stellar mass that is enabled
by the flashlight effect. Larger initial masses or
a more focused flow along a filament could
conceivably lead to even more massive stars. In
an ongoing simulation of the collapse of 100-
and 200-M$_\odot$ clouds, Krumholz, Klein \&
McKee (2005) report that $\ga$\,27\,M$_\odot$ ($\ga$\,25\,M$_\odot$
for the turbulent case) accreted onto a stellar
core. As in the simulations of Yorke \&
Sonnhalter (2002), a disk formed around the
accreting (proto)stars.

One can speculate on the effect outflows
have on the accretion through an accretion disk.
The inner part of the accretion disk could well
look like the configuration shown in {\bf Figure~11}.
Radiation and the stellar wind from the central
star (presumably already hydrogen burning)
evacuate a cavity in the polar direction. At the
interface between the supersonic outflowing
stellar wind and the denser subsonic HII disk
atmosphere, some disk material will be
removed, but this cannot prevent inward flow
of disk material. Inward radial flow of dusty
molecular gas is allowed in the equatorial plane
of the disk as angular momentum is transfered
outward.

In cases of magnetized disks with high
radiation densities, photon bubbles can lead to
more efficient vertical transport of radiation in
the disk (Gammie 1998; Begelman 2001; Blaes
\& Socrates 2003; Turner et al. 2005; Turner,
Quataert \& Yorke 2007). This can explain
super-Eddington accretion in a variety of
luminous systems, including accreting compact
objects and very massive stars.

Efficient angular momentum transfer could
result from weak magnetic fields in the disk
(Balbus \& Hawley 1991, Hawley \& Balbus
1991, Balbus 2003), from turbulence and/or
spiral density waves (Bodenheimer 1995)
excited by gravitational instabilities in
nonmagnetized (Laughlin \& Bodenheimer
1994) or magnetized (Fromang et al. 2004)
disks, or from the tidal effects of nearby stars
(Terquem 2001). Indeed, rapid accretion
through a disk may be a direct consequence of
having nearby companions. This may explain
why massive stars are generally members of
multiple systems.

Once the disk material crosses 
r$_{\rm dust} \sim$\,25\,AU\,[M$_*$/30\,M$_\odot]^{1.6}$,
the radius of dust destruction,
its opacity decreases and it is not easily stopped
by radiation. It is, however, still unclear how
the disk material ultimately flows onto the star.
Surely, the disk puffs up close to the star, in
analogy to the accretion disks in active galactic
nuclei. Beyond ${\rm r_{evap} \sim 130 AU [M_*/30 M_\odot]}$,
where the escape velocity is less than 10\,km\,s$^{-1}$,
the disk loses material via photoevaporation on
a timescale of $\sim$\,10$^5$\,yr. This is of the same
order as the accretion timescale. These
competing effects (accretion and
photoevaporation) will determine the final mass
of the star and perhaps even the upper mass
limit.

\subsection{Competitive Accretion and Runaway Growth}

Bonnell et al. (1997, 2001a) present the first 3D
numerical simulations of the growth of stellar
masses by competitive accretion in small young
star clusters. The best way to visualize the idea
of competitive accretion is to compare it to an
economic model based on two complementary
concepts; the real estate concept -- "location,
location, location" -- and the capitalistic
concept -- "the rich get richer." The former
reminds us that environmental influences can
be very important, such that being in a fortunate
location can significantly promote growth,
whereas the latter simply means that the
gravitational attraction increases with success --
in this case, with the increasing mass of the
star. A protostar's ability to grow depends on
the size of its accretion domain, i.e., the region
from which gas can be gathered. A location in
the center of a protostellar cluster is beneficial,
as gas flowing down to the center of the cluster
increases the gas reservoir available to an
individual star. The early  birth of a protostar
may also give it an unfair advantage in the
competition to end up large and massive.

Imagine a large, dense molecular gas cloud
with a number of protostellar seeds distributed
inside the cloud that have initially condensed
from some denser portions of the cloud. These
condensations or cores subsequently have the
chance to grow in mass by accumulation
(accretion) of lower density cloud gas from
their individual accretion domains (Larson
1978, Zinnecker 1982). These accretion
domains are systematically larger for higher
mass seeds. With increasing mass, the
gravitational spheres of influence keep
growing.

Equally important, the amount of material
that enters an individual accretion domain
depends on the external environment. In
isolation, it would simply be proportional to the
mean gas density of the region. By contrast, in
a larger scale potential such as that provided by
a protocluster cloud, the gas density can
become significantly larger in the center of the
protocluster, as the gas settles into the deepest
part of the potential, there to be accreted by the
growing protomassive star.

The accretion domain of an off-center
protostar is tidally limited by the total mass in
the inner part of the cloud, whereas the
accretion domain of a protostar in the center of
the cloud is the whole cloud. Because the gas
reservoir is limited (the cloud has a finite
mass), the protostellar masses will eventually
compete for cloud gas, especially after the
accretion domains start to overlap. The action
of the cluster to gather matter from larger
distances and focus it toward the accreting
stars, combined with the increasing accretion
radii of these stars owing to their increasing
mass, is what makes competitive accretion such
a powerful mechanism.

This is also true in a molecular cloud with
hierarchical substructure, in particular for
subcluster clumps of gas in a bigger
protocluster cloud. The implication is that each
subcluster clump is likely to have one most
massive protostellar object in its center
surrounded by a hierarchy of lower mass
objects (Bonnell, Bate \& Vine 2003). When
and if those subcluster clumps merge to one big
protocluster (see {\bf Figure~12}), the complex
accretion history of each protostar is no longer
related to a single local cloud core (Schmeja \&
Klessen 2004). In this scenario, the massive
stars had accretion histories that were
priviledged at every stage -- from Jeans-instability
protostellar birth to location and
density in the gas cloud, all factors affecting
accretion were more favorable than the average.
This is why massive stars are rare. The rarest
and most massive of them probably formed in
the most favorable conditions by runaway
accretion until their gas reservoir was exhausted
or dissipated (e.g., by ionization feedback, see
von Hoerner 1968 and recently Clarke, Edgar \&
Dale 2005).

The above model of competitive accretion
has been critized by Krumholz, McKee \& Klein
(2005b) on the grounds that Bonnell's SPH
simulations start from very strongly
gravitationally bound protocluster clouds, while
observationally such clouds appear to be
supported by turbulent motions. In other words,
the simulations use a virial parameter
${\rm \alpha}$\,=\,E$_{\rm turb}$/E$_{\rm grav}$\,$\ll$\,1
while molecular observations
suggest ${\rm \alpha}$\,$\sim$\,1. 
Krumholz, McKee \& Klein
(2005b) argue from analytical considerations
that protostellar masses cannot grow in such a
turbulent medium, not even by a factor of two.
The Bondi-Hoyle accretion rate is far too low
owing to the high relative velocities between
the accreting stars and the turbulent gas. In
addition, radiative feedback from the incipient
massive star may prevent Bondi-Hoyle
accretion altogether (Edgar \& Clarke 2004).

In response to this criticism, Bonnell \& Bate
(2006) argue that global initial collapse versus
quasi-equilibrium support is not the issue, as
their more recent SPH calculations start with
initial conditions where the turbulent kinetic
energy is close to the gravitational energy. In
fact, the scaling laws of supersonic turbulence
(Larson 1981) imply small turbulent velocity
differences between protostars and their
neighboring gas. This then allows for
significant growth in stellar mass while more
distant high velocity turbulent gas cannot be
accreted. In other words, protostars are swept
along with their neigboring gas for some time
in a similar global motion before they hit more
distant and thus less-correlated gas. The above
Bondi-Hoyle accretion problem is also not as
serious as suspected, considering that the most
massive star forms in the center of the clump
with little motion relative to the surrounding
gas.

Future detailed observations of the motions
in massive ($\sim$\,1000\,M$_\odot$)  star-forming clumps
will discriminate between these different
dynamical scenarios (small versus large gas
motions relative to the protostars, turbulent
cloud support, or overall cloud collapse). A first
step in this direction was made by Peretto,
Andr\'{e} \& Belloche (2006) in their dust
continuum and molecular study of the NGC
2264 clumps, with the conclusion that the
observations are most consistent with "a picture
of massive star formation intermediate between
the scenario of stellar mergers of Bonnell et al.
(1998) and the massive turbulent core model of
McKee \& Tan (2003), whereby a turbulent,
massive ultra-dense core is formed by the
gravitational merger of two or more Class 0
protostellar cores at the center of a collapsing
protocluster".

\subsection{Stellar Collisions and Mergers}

Historically, the original reason for proposing
stellar collisions as a formation process for
massive stars was twofold. ({\it a}) At the time,
radiation pressure on dust was considered a
severe hindrance to gas accretion (then assumed
to occur spherically symmetrically). Today this
concern has gone away (see {\bf Section~4.1}). 
({\it b})
The packing of massive stars in dense clusters
was too tight, and so there was concern that a
sufficiently large gas reservoir for monolithic
collapse was not available. This is still a
concern, although not all massive stars form in
densely packed clusters -- many form in widely
spread OB associations.

In any case, it is possible that a collisional
build-up of high-mass stars can occur,
especially for the most massive stars in very
dense clusters. The problem is the very high
stellar density of already massive or at least
intermediate-mass stellar objects that is
required to get the process going. We can
estimate the threshold stellar number density
n$_{\rm star}$ or, equivalently, the average star-star
separation (s\,=\,n$_{\rm star}^{-1/3}$) 
in a dense cluster for stellar
collisions (encounters) to be important. Because
of the fundamental role of gravitational
focusing coupled with a distribution of
velocities (i.e., a finite probability of low
relative velocities at infinity v$_\infty$), the cross
section ${\rm \sigma_{grav}}$, where
\[
  {\rm \sigma_{grav} = \pi R^2_{min} \left(1 +
  \frac{2 G M_*}{v_\infty^2 R_{min}}\right)}
\]
for two stars of mass M$_*$ passing each other at
periastron within a minimum distance R$_{\rm min}$ is vastly
enhanced over the geometrical cross section.
Gravitational focusing can enhance the effective
cross section (by factors of $\sim$\,10$^4$), thus rendering
close stellar encounters in very young stellar clusters
realistic. Of course, the condition to meet for
massive star growth through mergers is that the
collision (close encounter) time t$_{\rm coll}$ for the stars to
collide must be shorter than the timescale for stellar
evolution of the most massive star in the cluster ($\sim$\,3\,Myr).

Using the formula for the stellar collision
time per star (Binney \& Tremaine 1987, Dale \&
Davies 2006), i.e.,
\[
  {\rm \tau_{coll} = \frac{1}{n_{star} \sigma_{grav} v_{rms}}
                  = \left[4 \sqrt{\pi} n_{star} v_{rms}
                    \left(R_{min}\right)^2 \left(1 + \frac{2 G M_*}
                    {R_{min} v^2_{rms}} \right)\right]^{-1}}
\]
we obtain, when gravitational focusing dominates
\begin{equation}
  {\rm \tau_{coll} = 7 \times 10^7
                    \left[\frac{n_{star}}{10^6\,pc^{-3}} \right]^{-1}
                    \left[\frac{M_*}{10\,M_\odot} \right]^{-1}
                    \left[\frac{R_{min}}{1\,R_\odot} \right]^{-1}
                    \times \left[\frac{v_{rms}}{10\,km\,s^{-1}} \right] yr}.
\end{equation}
Here we have normalized the expression with
reasonable numbers for the number density, mass,
size, and velocity dispersion of the stellar collision
partners, assumed to be of equal mass. We ignore
binary stars for the time being. Note that the velocity
dispersion is the 1D-velocity dispersion, which is 2.3\,km\,s$^{-1}$
in the Orion Nebula cluster (van Altena et al.
1988), but is $\sim$\,5\,km\,s$^{-1}$ in denser and more massive
cluster cores such as NGC 3603 and R136,
or even larger in young globular clusters such
as those in the Antennae (Mengel et al. 2002).

{\bf Equation~5} is based on equal mass collision
partners; however, recently the formula has
been extended to cover nonequal mass
encounters (Moeckel \& Bally 2006, 2007), as
well as larger cross sections due to
circumstellar disks and binary components
(Davies et al. 2006). This can decrease the
threshold stellar number density for collisions
from 10$^7$ or 10$^8$\,pc$^{-3}$ down to about 10$^6$\,pc$^{-3}$
(Bonnell \& Bate 2005).

In {\bf Figure~13} we show 3D SPH numerical
simulations of the collision process of two pairs
of stars, one pair of equal mass and one pair of
unequal mass.

\noindent
\textcolor{blue}{
{\bf Interim Summary}}

\noindent
\textcolor{blue}{
The primary difference between massive star
formation by monolithic collapse versus by
competitive accretion is that in monolithic
collapse the mass is assumed to be gathered
before the star-formation process begins,
whereas in competitive accretion the mass is
gathered during the star-formation process. If
the former is true, we should find massive prestellar
cores that live for long times. But
how did they form in the first place? Perhaps
turbulence could support a slow build-up, but
turbulence induces density fluctuations, so
why does the quasi-static core not fragment
into a cluster of low-mass stars? (This is the
Dobbs, Bonnell \& Clark 2005 versus
Krumholz,  Klein \& McKee 2007 debate, see
also Krumholz 2006). Magnetic fields may
help to stabilize massive cores against
subfragmentation.}

\textcolor{blue}{
The strength of competitive accretion is that it
provides a physical mechanism to gather the matter.
The gravitational potential of the protocluster clump
or cluster of stars funnels a significant fraction of
gaseous material to the cluster center, there to be
accreted by the protomassive stars. In addition, the
steep nonlinear mass dependence of the accretion
rate onto a gravitating point mass -- the Bondi-Hoyle
rate -- naturally gives a steep, power-law IMF, close
to the observed IMF in young clusters (see Bonnell,
Larson \& Zinnecker 2007). In this scenario, massive
star formation is explained in the context of
low-mass star formation (cospatial clusters).}

\textcolor{blue}{
Stellar mergers will be rare and only relevant for
the most massive stars in the richest young clusters
(such as young globular clusters). One reason to
invoke stellar mergers is that ongoing competitive
accretion (mass loading) increases the stellar density
in the cluster center (mass segregation at birth),
potentially to the point where grazing collisions
become unavoidable. Stellar collisions with small
impact parameters might be the process that forms
rapidly rotating massive stars and hence the
progenitors of (slow, long-duration) gamma-ray
bursts.}

\subsection{Triggered OB Star Formation}

The classical model of triggered OB star
formation goes back to Elmegreen \& Lada
(1977). The idea is that the ionization shock
front of one group of massive stars provides the
external pressure to compress adjacent
molecular cloud layers, thereby inducing the
formation of a new group of massive stars,
which in turn, by the same process, induces the
formation of another generation of massive
stars and so on. The Elmegreen \& Lada (1977)
model was developed to explain the sequence
of spatially distinct OB subgroups in nearby
OB associations such as Orion OB1 or
Sco-OB2 (Blaauw 1964, 1991). It is important to
note that the Elmegreen \& Lada model did not
predict the formation of the observed coeval,
low-mass T Tauri star population in the
subgroups of OB associations (Preibisch \&
Zinnecker 2007, Brice\~{n}o et al. 2007). The low-mass
population was assumed to form
independently and in many locations spread out
over the cloud. Hence the Elmegreen \& Lada
(1977)  process of sequential triggered star
formation, contrary to accepted wisdom, may
not be the main mechanism that accounts for
the existence of OB subgroups. Rather,
supernova triggering could be at work,
assuming that a supernova blast wave can
trigger both high-mass and low-mass stars at
the same time (which would then explain the
coevality of the high-mass and low-mass stars).
Numerical simulations along these lines are just
beginning (e.g., Melioli et al. 2006).

However, there are several clear cases known
in our Galaxy where expanding HII regions
have swept up molecular gas at their periphery
and in which new massive stars have formed or
are about to form. The latter is indicated by
luminous IR sources (e.g., Sh 104 and RCW 79,
Zavagno et al. 2005; RCW 108, Comer\'{o}n,
Schneider \& Russeil 2005). This is indeed
reminscent of the Elmegreen \& Lada (1977)
model of triggered sequential star formation
(the so-called collect and collapse scenario,
Elmegreen 1998; see also Whitworth et al. 1994
and Dale, Bonnell \& Whitworth 2007).
However, according to Elmegreen \& Lada
(1977), it may be expected that in a two-stage
starburst (cf. Walborn \& Parker 1992) the
second generation protocluster hosts only a
small group of high-mass OB stars without the
concomitant multitude of low-mass stars. Such
a small N-body group, much like the Orion
Trapezium system, would then be highly
dynamically unstable and could help explain
the occurrence of runaway OB field stars
(Clarke \& Pringle 1992, Allen, Poveda \&
Hern\'{a}ndez-Alc\'{a}ntara 2004). An important
implication of this scenario would be that the
HII region of a big star cluster (e.g., NGC
3603) is not able to trigger the formation of a
similarly massive second generation star
cluster, but only a smaller mass cluster (e.g.,
IRS9 in NGC 3603; see N\"urnberger 2003),
potentially with a top-heavy stellar IMF;
similarly, the R136 cluster in 30 Dor does not
seem to trigger the formation of a new massive
cluster but just a few small groups of embedded
protostars (Rubio et al. 1998, Walborn et al.
1999, Brandner et al. 2001, Walborn, Ma\'{i}z-Apell\'{a}niz
\& Barb\'{a}  2002).

A very interesting case is N81, one of the
most compact HII regions in the Small
Magellanic Cloud (SMC). N81 is isolated in an
area of low extinction in Shapley's wing, 1.2\,kpc
away from the main body of the SMC. In
contrast to other compact HII regions, typically
located within or at the edge of giant HII
regions, it appears that SMC N81 has been
formed in isolation. The study of its stellar
inventory shows that this 'high-excitation blob'
is ionized by at least eight near-ZAMS O stars
in an instantaneous burst (Heydari-Malayeri et
al. 2002, 2003). The question here is, Where is
the trigger? How did these massive stars form?

Back to the Galaxy, another region of note is
the Carina nebula powered by the most extreme
grouping of massive stars in the southern Milky
Way (Smith \& Brooks 2007).
Here $\sim$\,65 O stars (including many in the
clusters Tr14 and Tr16, but excluding $\eta$~Car)
provide a total of $\sim$\,10$^{52}$\,erg of kinetic energy
and a Lyman continuum luminosity of $\sim$\,10$^{51}$
photons\,s$^{-1}$, creating a giant superbubble. Most
of the bubble (seen as polycyclic aromatic
hydrocarbons in {\it Spitzer}/MSX images) resides
as atomic gas in the photodissociation regions
and not in dense molecular clouds. The
synchronized star formation around the
periphery of Carina strongly suggests that star
formation was triggered by stellar winds. The
second-generation population appears to
involve a cascade toward preferentially
intermediate-mass and low-mass stars (Smith \&
Brooks 2007), but the situation may change
soon when $\eta$~Car and its siblings explode as
supernovae, rejuvenating massive star
formation. The idea that it is supernovae rather
than HII regions and stellar winds that stimulate
wide-spread massive star formation is not new
(see Herbst \& Assousa 1977 for the Canis
Majoris star formation region, see also Gerola
\& Seiden 1978 for galactic spiral arms, and the
pioneering paper by \"Opik 1953). However,
what is new is the evidence from the Upper
Scorpius OB subgroup data that supernovae can
trigger both high-mass and low-mass star
formation in an OB subgroup at the same time
(Preibisch \& Zinnecker 2007). Supernovae
need a critical distance from a molecular cloud
-- not too close and not too distant -- to be an
effective trigger for star formation because of
momentum transfer of the blast wave onto
prestellar cores (Vanhala \& Cameron 1998,
Vanhala et al. 1998). Other triggering
mechanisms, like radiatively driven implosion
of globules, also operate, but seem to be
secondary processes, forming only small stellar
groups rather than whole OB subgroups with
thousands of stars (for a review of a whole
variety of triggering mechanisms, see
Elmegreen 1998).

The Carina star-forming region (diameter
150\,pc, age 3\,Myr, total IR-luminosity 
$\sim$\,10$^7$\,L$_\odot$)
may be the galactic analog of giant
extragalactic HII regions such as NGC 604 in
M33 (size 40\,arcsec or 140\,pc), a very large OB
association (see Ma\'{i}z-Apell\'{a}niz, P\'{e}rez \& Mas-Hesse
2004).

Closer to home, the Orion Nebula cluster is
likely an example of triggered massive star
formation. The appearance of the Orion A and
B molecular clouds suggest an interaction
caused by energy input from the Orion OB1ab
association subgroups (Blaauw 1991).
Furthermore, the formation of the nearest
massive star in the Orion BN/KL
region, which hosts a very bright IR source, has
probably been triggered by the Orion
Trapezium star's HII region pushing against the
Orion Molecular Cloud right behind the
Trapezium cluster (see Bally 2002).

Recent {\it Spitzer}/IRAC Galactic plane
observations (Churchwell et al. 2006) have
revealed dozens of parsec-sized bubbles,
formed by hot young stars in massive star
forming regions. Among the 80 or so ring-like
structures produced by young OB stars, several
show secondary bubbles on the rim of the
primary bubbles, suggestive of triggered star
formation. However, is this morphological
evidence conclusive, i.e., is triggering
necessary to explain such spatial correlations?
In an interesting paper, Dale, Clark \& Bonnell
(2007) attempt to address this and similar
issues. They conduct SPH simulations to
examine the difference between triggered and
revealed star formation. They study the impact
of irradiation by an external source of  ionizing
photons on a turbulent massive molecular cloud
and compare the results (the number and type
of stars formed) with a control simulation
where the turbulent cloud evolves without the
impact of the irradiation. They find that,
although the external ionization has a dramatic
effect on the morphology of the model cloud,
its impact in terms of extra star formation is
surprisingly minor; the feedback effects can be
both positive and negative, accelerating the
formation of some objects and delaying the
formation of others. Only a few objects form
that would otherwise not have formed, and the
effect of induced star formation on the overall
star formation efficiency is less than a factor of
two.

\subsection{Dynamical Evolution: 
Mass Segregation and Runaway OB Stars}

\subsubsection{Mass Segregation} Massive stars are often
found near the centers of star clusters but not
exclusively. Examples where massive stars are
preferentially located in and near the cluster center
include the Trapazium in the Orion Nebula cluster
(Hillenbrand 1997) or the WR and O stars in the
NGC 3603 cluster (Drissen et al. 1995). The issue is
whether this is a birthmark (we call this effect
prompt mass segregation) or whether this is an
N-body evolutionary effect after birth (in which case,
we call it dynamical mass segregation). The way to
decide between these two possibilities is to estimate
the dynamical time for massive stars to sink to the
center of the gravitational potential from the
half-mass radius of the cluster. If this timescale turns out
to be too long, i.e., longer than the age of the cluster,
dynamical mass segregation is ruled out and prompt
mass segregation is indicated (e.g., Bonnell \&
Davies 1998, who infer that mass segregation in the
Orion Nebula cluster must have been prompt and
hence an important constraint for understanding
massive star formation).

\subsubsection{Runaway OB Stars} Runaway OB stars
were defined by Blaauw (1961) as massive stars
with radial velocities in excess of 40\,km\,s$^{-1}$. Their
high space velocities prompt the question of how
they were accelerated to such kinetic energies. The
original proposal involved an asymmetric supernova
in a massive binary system that provided the kick to
eject the companion from the system (Blaauw 1961).
A strong prediction of this model is that the runaway
massive stars should themselves be single.

More recently, an alternative suggestion for the
runaway phenomenon was put forward (Poveda,
Ruiz \& Allen 1967, Gies 1987). Here the idea is that
massive stars in dense cluster cores (which is their
preferential location) undergo dynamical three-body
encounters, especially if a massive binary is
involved. In such gravitational interactions, potential
and kinetic energy are exchanged: The binary orbit
shrinks, and the intruder extracts the potential
energy and converts it into the corresponding kinetic
energy. It can also happen that an exchange reaction
occurs so that one of the binary components (usually
the lighter one, rather than the intruder) is ejected.
Conservation of momentum leads to a recoil for
each of the two system components, so they fly apart
in roughly opposite directions. In this scenario, the
runaway stars need not be single but can, in
principle, be binaries, at least in 50\,\% of the cases.

Hoogerwerf, de Bruijne \& de Zeeuw (2001) give
evidence that both processes for runaway OB stars
occur in nature: while  $\zeta$~Ophiuchi is a very high
proper motion massive O9.5 star without an
antipode in the opposite direction, and hence an
example for the supernova scenario, another famous
case, that of $\mu$~Columbae and AE Aurigae, argues
for the dynamical interaction scenario, where two
hard binary systems likely had a close encounter in
the 5\,Myr old NGC 1980 Orion cluster,  in which
two of the binary components were set off in
opposite directions with speeds of around 100\,km\,s$^{-1}$.
In their wake is $\iota$~Ori, a tight, eccentric, massive binary
(Gualandris, Portegies Zwart \&
Eggleton 2004), known to be classic, colliding-wind
X-ray binary. Other famous examples of massive runaway stars
include $\lambda$ Cep, $\xi$ Per, and $\alpha$ Cam (astronomy
picture of the day on 24 Nov 2006).

The dynamical ejection model for runaway stars
would predict not only very fast ejection speeds, but
also slower ones, as milder interaction events with
wider, i.e., softer, binary pairs can occur (Kroupa
2000). This must be kept in mind when discussing
the origin of massive field stars: with an escape
speed of only 5\,--\,10\,km\,s$^{-1}$, a massive O star with an
age of around 3\,--\,5\,Myr could still travel 15\,--\,50\,pc
from its cluster birth place. Some 40 nearby field O
stars are known and at least half of them can be
associated with a cluster origin (de Wit et al. 2004);
for the rest the situation is unclear. It is interesting to
recall that the spectroscopic and visual binary
frequency among these runaway O stars is very low
(Mason et al. 1998).

\newpage

\section{MASSIVE STAR FORMATION:\\BINARY AND MULTIPLE SYSTEMS}

Some of the most important clues toward
understanding the formation of massive stars,
that have been neglected in previous reviews
(e.g., Massey 2003), come from their high
frequency of binary and multiple systems,
together with an analysis of their properties
(period distributions, mass ratios, orbital
eccentricities); see {\bf Section~2.4} and the discussion
later in {\bf Section~6.2}. Massive stars often come in
hierarchical triples with an almost equal-mass
close massive binary and a third more distant
companion. One example includes $\theta^2$~Ori~A in
Orion's Bar (see Preibisch, Weigelt \&
Zinnecker 2001). Another is $\sigma$~Ori in the
center of the namesake cluster, whose
hierarchical configuration is even more
complex (see Sanz-Forcada, Franciosini \&
Pallavicini 2004). These close massive binaries
likely play an important role in the evolution
and age dating of starburst galaxies, as pointed
out by van Bever \& Vanbeveren (1998), and an
equally important role in estimating supernova
rates and the rates of high- and low-mass X-ray
binaries (see Verbunt 1993; see also H.A.
Kobulnicky, C.L. Fryer \& D.C. Kiminki,
submitted). Massive close binaries also bias and obfuscate
measurements of the velocity dispersion in
dense starburst clusters (see Bosch et al. 2001).

It seems worthwhile and appropriate to
summarize the various formation processes of
massive binary and multiple systems. The
following five processes could be relevant.

\subsection{Disk or Filament Fragmentation}

This is the most obvious mechanism; it is
similar to suggestions of how to form low-mass
T Tauri binaries from filament fragmentation
(Zinnecker 1991, Bonnell \& Bastien 1992,
Monin et al. 2007). A filamentary geometry
may play a key role in the fragmentation
process, because the isothermal case is a critical
one for the collapse of a cylinder: the collapse
and fragmentation of a cylinder can continue
freely as long as the temperature continues to
decrease, but not if it begins to increase (Larson
2005). As for disk fragmentation, gas thermal
physics controls the non-linear outcome of
gravitational instability in low-mass
circumstellar disks (Durisen 2001). The
fragmentation of massive disks has recently
been studied analytically by Kratter \& Matzner
(2006). They found that these disks are unstable
to fragment if they are cold enough, and
catastrophically so when 
${\rm \dot M}$\,$>$\,10$^{-3}$\,M$_\odot$\,yr$^{-1}$.
This mechanism accounts only for initially
wide binaries. Cold rotating Keplerian disks
produce circular orbits; sub-Keplerian disk
rotation or cold filaments tumbling end over
end produce highly eccentric orbits. $\theta^1$~Ori~C
may be an example of the latter (Kraus et al.
2007). Note that fragmentation is different from
fission, the splitting up of a hydrostatic, rapidly
rotating body. Fission does not work for a
compressible fluid (Tohline \& Durisen 2001); it
only leads to the ejection of spiral arms and
torques that slow down the rapid stellar
rotation.

\subsection{Accretion onto a Low-Mass, Wide Binary}

This is a less obvious but potentially key
mechanism to form tight massive binaries.
Speculated upon by Maeder \& Behrend (2002),
it was worked out by Bonnell \& Bate (2005).
These authors realized that the orbital
separation of the components of close binaries
is much smaller than the Jeans length at
reasonable gas densities and temperatures.
Therefore, they investigated the physical idea
that the orbital separation of an initially wide
binary would shrink by letting both components
accrete and grow in mass; the final separation
then depends on the orbital angular momentum
of the accreted material, as follows:

Let us consider the angular momentum
A\,=\,M\,v\,$\times$\,R of a binary system. Because of
v\,$\propto$\,(M/R)$^{1/2}$ (v is the Keplerian orbital speed),
\[
{\rm A \propto M^{3/2} R^{1/2}}
\]
where M is the instantaneous total mass of the
binary system and R is the separation of the binary
components. If the accreted material has zero net
angular momentum, as would be expected if the
infall were spherically symmetric, then A remains
constant (A\,=\,const), implying that R\,$\propto$\,M$^{-3}$,
 i.e., the
binary separation should be a strong function of the
binary mass. If, instead, the accreted material has
constant specific angular momentum (A/M\,=\,const),
then the total angular momentum will scale with the
mass of the binary (A\,$\propto$\,M), 
implying that R\,$\propto$\,M$^{-1}$.
Thus it is easy to see that accretion onto a binary
system can significantly decrease its separation at
the same time that it increases its mass (Bate 2000).
In the numerical SPH simulations, the early
evolution is well parameterized by the relation
R\,$\propto$\,M$^{-2}$, intermediate between the above two cases,
indicating that the binary is accreting some angular
momentum with the mass but that the net specific
angular momentum of the accreting gas is
decreasing with time. The reason for this is that
random velocities of gas at large radii tend to cancel
each other out, resulting in accretion of a lower net
angular momentum. With an R\,$\propto$\,M$^{-2}$ relation, we
infer that a low-mass (e.g., 3\,M$_\odot$) wide pair (e.g.,
separation of order 100\,AU) will become a binary
system with a semimajor axis of around 1\,AU by the
time the system has accreted up to 30\,M$_\odot$. That is,
accretional growth in mass by a factor of 10 will
make the separation decrease by a factor of 100, or
more if the growth in mass  is more than a factor of
ten. This may explain the observed very tight O-star
spectroscopic binaries (see {\bf Section~2.4}).

The component masses, even if unequal in
the beginning, would tend to become equal later
when higher angular momentum cloud material
falls preferentially on the secondary component
(Bate \& Bonnell 1997), as the lower mass
secondary is, by definition, further away from
the center of mass and would thus carry the
higher angular momentum in the system
initially.

Another very interesting new model to
explain the origin of close massive 'twins' is
proposed by Krumholz \& Thompson (2006).
They invoke mass transfer in close, rapidly
accreting protobinaries; this always pushes the
initial binaries toward mass ratio unity. Their
model is superficially similar to the model of
Bonnell \& Bate (2005), but in fact the physical
details are quite different, involving the
swelling of protostars undergoing deuterium
shell burning.

\subsection{Failed Mergers in Stellar Collisions}

If massive stars can form through stellar
mergers, then near misses might sometimes
form tight and eccentric massive binaries. A
necessary condition is that the kinetic energy of
the quasi-parabolic encounter can be dissipated
as tidal energy (Fabian, Pringle \& Rees 1975)
in a gravitationally focused grazing fly-by
(Zinnecker \& Bate 2002, Dale \& Davies 2006).
Apart from the very high stellar number
densities required for this process to be
significant and frequent, the main problem is
the extreme fine tuning of the collision impact
parameter. Indeed, the cross section is a very
small concentric annulus in impact parameter
space that is much smaller than the central area
inside the annulus relevant for massive mergers
(F.~Rasio, private communication).

One way out is disk formation during the
tidal disruption of a  lower density star by a
higher density star in a non-head-on collision;
see the example of a grazing collision between
a 3\,M$_\odot$ premain-sequence star and a 10\,M$_\odot$
main-sequence star in {\bf Figure~13}. This can lead
to subsequent disk-assisted capture of a
companion, to be discussed extensively in the
next subsection (the so-called 'shred and add'
scenario, Davies et al. 2006). A variant of the
failed merger scenario is the off-center collision
of two protostellar cores with extended
accretion envelopes (Stahler, Palla \& Ho 2000).
These protostellar envelopes offer a much
higher interaction cross section and can provide
the necessary orbital drag for collisional
massive binary formation (cf. Silk 1978).

\subsection{Disk-Assisted Capture}

If massive stars form via disk accretion, then
the large disk radii increase the interaction
cross section considerably. This suggests that
disk interactions with neighboring stars could
assist in capturing binary companions (Bally \&
Zinnecker 2005). Although this mechanism has
been found insufficient for solar mass stars with
disks (Heller 1995, Boffin et al. 1998), recent
SPH/N-body simulations by Moeckel \& Bally
(2006, 2007) convincingly showed that
disk-assisted capture is much more efficient in a
regime suited to massive stars (ca. 20\,M$_\odot$) with
large disks (ca. 500\,AU). We note that this
process works particularly well in providing
massive stars with  lower mass companions in
rather wide orbits (separation ca. 100\,AU). The
same authors also discuss the consequences of a
mass-dependent velocity dispersion and of an
initial mass segregation for the capture rates.
Furthermore, they considered the long-term
survival of these binaries in a dense cluster.

The point is that massive binaries with different
separations and mass ratios form by different
processes. Although accretion seems to be the only
way to form the observed tight equal-mass binaries
in young clusters, fragmentation and disk-assisted
capture can form the wider, unequal-mass binaries.

\subsection{N-Body Dynamical Evolution}

Small-N groups of massive and intermediate-mass
stars are seen forming in 3D SPH
simulations of the collapse and fragmentation
of gas-rich protoclusters (e.g., Bonnell, Bate \&
Vine 2003). Binary formation in these groups is
common and occurs through dynamical three-body
capture (different from tidal capture). To
begin with, a massive star typically has a lower
mass wide companion. With time, an exchange
interaction with a more massive third object
occurs, thus forming an almost equal mass but
still wide binary. The wide binary then shrinks
('hardens') as it takes up most of the binding
energy of the small group when the lower mass
group members get kicked and escape. The
typical final outcome is a tight equal-mass
massive binary, with a lower mass wide
companion (Bate, Bonnell \& Bromm 2002).
This scenario explains many aspects of massive
binaries, including the average companion star
fraction of 1.5 and the prevalence of
short-period spectroscopic systems. Also, the
timescales for the N-body dynamics are short
enough for the whole process to occur on or
before the ZAMS (van Albada 1968, Aarseth
2003).

However, it would seem that the frequency
of massive close binaries should  correlate with
the overall stellar number density in young
clusters, which is not what is observed. In fact,
at face value the opposite -- an anticorrelation --
is observed (Mermilliod \& Garc\'{i}a 2001). We
have no real explanation as to how this
environmental effect fits into the N-body
picture, unless most of the most massive
binaries in the densest young clusters have
managed to merge into a single object. Stellar
mergers are thought to occur in old globular
clusters (e.g., Dale \& Davies 2006), and this is
one possible explanation for the presence of
blue stragglers (younger stars with masses
higher than the cluster turnoff mass). Stellar
mergers may therefore be even more important
in dense, massive, young globular clusters, both
now and in the past.

\subsection{The Origin of Trapezium Systems}

Finally, we address the problem of how
Trapezium systems of massive stars, such as the
one in the center of the Orion Nebula Cluster
($\theta^1$~Ori), likely came into being. In brief, the
idea is as follows: numerical SPH simulations
of supersonic gravo-turbulent fragmentation of
a protocluster cloud (1000\,M$_\odot$)  suggest that a
collapsing cloud develops a few subclusters
(star+gas systems), which subsequently merge
into a single cluster entity. Each subcluster
carries one most massive star (likely already
part of a multiple). Hence the merging of
subclusters will result in a central Trapezium-type
system (see {\bf Figure~14}), as observed in the
core of the Orion Nebula Cluster (see {\bf Figure~15}).
Note that components A1 and B1 of the
Orion Trapezium are spatially unresolved
eclipsing spectroscopic binaries; for a summary
of the parameters of the Trapezium multiple
stars we refer to Preibisch et al. (1999) and
Schertl et al. (2003). Note also that component
B is itself a Trapezium-like system, indicating
the hierarchical nature of massive star
formation. The dynamical evolution of
Trapezium systems, including stellar ejections,
is discussed by Allen, Poveda \& Hern\'{a}ndez-Alc\'{a}ntara
(2004), taking into account their
multiplicity substructure. Future studies of
Trapezium-type systems will likely concentrate
on embedded systems, such as  W3-IRS5
(Megeath, Wilson \& Corbin 2005).

\newpage

\section{MASSIVE STAR FORMATION: DISCUSSION}

\subsection{Disks and Outflows}

Disks and outflows are a general phenomenon
in low-mass star formation that is explained by
the accretion-ejection connection (Camenzind
1990, Ferreira \& Pelletier 1995). That is, disk
accretion energy powers either a disk wind
(Pudritz \& Norman 1983, Pudritz et al. 2007) 
or an X-wind (Shu et al.
1994, 1995). These in turn are magnetically
collimated into a molecular jet. The jet then
runs into the parent molecular cloud and local
ISM, thereby accelerating entrained ambient
molecular gas into a wider momentum-driven
molecular flow.
Textbook examples include the famous
Hubble Space Telescope (HST) source HH30
(Burrows et al. 1996) and the most beautiful IR
jets HH211 and HH212 (McCaughrean, Rayner
\& Zinnecker 1994; Gueth \& Guilloteau 1999;
Zinnecker, McCaughrean \& Rayner 1998; Lee
et al. 2006).

Similar but usually less collimated outflows
have been observed around more luminous and,
hence, more massive young stars (Shepherd \&
Churchwell 1996a, b; see the compilation by
Beuther \& Shepherd 2005). This has led to the
notion that high-mass star formation is a scaled-up
version of low-mass star formation.
However, as noted by Arce et al. (2007), no
highly collimated outflow has been observed
for high-mass star-forming regions exceeding
10$^5$\,L$_\odot$ (corresponding to $\sim$\,25\,M$_\odot$).  
Likewise,
circumstellar disks have been detected around
young B-type massive stars (Zhang 2005), but
disks around the progenitors of O-type massive
stars in excess of 20\,M$_\odot$ have eluded detection
(Cesaroni et al. 2007). This suggests that
simple-minded claims of exact similarity
between high-mass star formation and low-mass
star formation are partly based on a casual
use of the term 'high-mass star' and partly on
wishful thinking. Yes, the rotating disks around
IRAS 20216+4104 (Cesaroni et al. 1997, 1999;
Cesaroni et al. 2005b), G192.16--3.82
(Shepherd, Claussen \& Kurtz 2002), Cep A
(Patel et al. 2005), or GL 490 (Schreyer et al.
2006) are wonderful examples of millimeter-interferometric
observations. But the central
sources are at best early B-type stars (below 20\,M$_\odot$)
rather than massive O stars where
radiation pressure and photoevaporation play a
role in the formation process. The same seems
to be true for the large silhouette disk in M17
(Chini et al. 2004), which is associated with a
molecular hydrogen jet (N\"urnberger et al.
2007).

Let there be no misunderstanding. It is likely
that the majority of massive stars forms by disk
accretion. But the issue is really whether stars
of 50\,--\,100\,M$_\odot$ can also form by some other,
more dramatic process like stellar or
protostellar collisions (Bonnell, Bate \&
Zinnecker 1998; Stahler, Palla \& Ho 2000).
One could imagine that the reason why no disks
and collimated outflows have been found for
the most massive stars in statu nascendi is
because their disks are being photoeroded at the
same time the stellar mass grows by disk
accretion. The combination of a powerful
stellar wind and radiative acceleration clear the
polar regions of material and magnetic fields,
thus leading to poorer collimation for their
molecular outflows. Maybe the outflow from
AFGL 2591 ({\bf Figure~16}) is such a case.

The most messy outflow from a massive star-forming
region is from the BN/KL region in
Orion. The OMC1 outflow (mass 10\,M$_\odot$,
velocity 30\,--\,100\,km\,s$^{-1}$) and the H$_2$ fingers
(Allen \& Burton 1993, McCaughrean \& Mac
Low 1997, Kaifu et al. 2000) have been
interpreted as the result of a powerful explosion
that occurred in the center of OMC1 within the
last 1000\,yr. It has been proposed that a
dynamical interaction 500\,yr ago, possibly
leading to a merger, may have produced the
OMC1 outflow and H$_2$ finger system (Bally \&
Zinnecker 2005). In this scenario, interactions
with surrounding gas have decelerated the
impulsive outflow powered by the stellar
collision by about a factor of two, thereby
reconciling the 1000-yr dynamical timescale
with the 500-yr timescale for the interaction.
The detection of oppositely directed motions in
radio sources IRc2-I and BN provides support
for models in which the OMC1 outflow was
powered by a dynamical interaction between
high-mass or intermediate-mass stars in an
ultradense environment (Rodr\'{i}guez et al.
2005a; see also Tan 2004).

Another issue that needs to be appreciated in
the context of massive outflows is the question:
to what extent do the observed massive
outflows trace the combination of stellar wind
and intrinsic disk wind? The latter is able to
entrain a lot of ambient gas, and care must be
taken not to mistake this gas mass for the
intrinsic wind mass. This is important because
the accretion rate is estimated from the outflow
rate, so if the outflow rate is overestimated, the
accretion rate is overestimated too.

Somewhat analogously, one must also be
careful when deriving the stellar mass from the
luminosity of the source (e.g., using IRAS and
submm data) when part of this luminosity is
nonstellar and comes from accretion onto the
disk and onto the (proto-)star (see {\bf Section~3.4.2}
and {\bf Figure~5}). Accretion occurs as long as
there is outflow activity, and the highest
accretion rate translates into the most violent
outflow. Spatially unresolved binary and multiple
systems compound the issue (see the next section).

A final issue is the inferred nonlinear mass
dependence of the disk accretion rate on the
instantaneous central stellar mass (Norberg \&
Maeder 2000, Behrend \& Maeder 2001). This
is a consequence of a misunderstanding
perpetuated in the literature. It dates back to the
observed relation that the rate of mass outflow
increases roughly linearly with the source
luminosity (for a recent plot see Wu et al.
2005). If the outflow rate scales with the disk
accretion rate and the luminosity varies with
some power of the stellar mass (e.g., square),
then we obtain ${\rm \dot{M}_* \sim M_*^2}$ (Zhang 2005).
However, it is a leap of faith to interpret this
relation as one of instantaneous quantities
rather than one regarding average quantities
(the latter simply means that the formation of
more massive stars requires higher average
accretion rates, whereas the former would
imply that the accretion rate keeps increasing
with increasing central stellar mass). By
completely ignoring the contribution of the
accretion luminosity in this scenario, one
implicitly assumes that a high luminosity
implies a high core mass, as opposed to a high
luminosity likely indicating that the accretion
rate onto a lower mass (proto)star is high!

A problem with an ever increasing accretion
rate is not only how to suddenly stop the
accretion, but also the observation that outflow
activity is strongest during the earliest phases of
star formation rather than during the final
phases. In fact, the same problem exists for the
turbulent core scenario of massive star
formation (McKee \& Tan 2003) where the
accretion rate is predicted to increase linearly
with time, implying that the most vigorous
outflows should occur toward the end of
accretion rather than at the beginning.

\subsection{Binary Statistics and the Most Massive Binary Systems}

The first spectroscopic binary survey among O
stars (V\,$\le$\,7) was conducted by Garmany, Conti
\& Massey (1980). They found that 24 out of
their 67 O stars are spectroscopic binaries
(36\,\%). Some 15 are double-lined and 4 are
single-lined, mostly with 1\,--\,10 day periods; the
remaining 5 O stars are binary candidates. A
later important work is the optical speckle
survey of O stars carried out by Mason et al.
(1998). This is a magnitude-limited sample
(V\,$<$\,8) of 227 bright O stars, observed at
0.1\,arcsec spatial resolution (15 new detections). In
addition, the previously known spectroscopic
binaries were included in a comprehensive
analysis. We summarize their results as follows.
Besides the 50 known spectroscopic binary
systems with periods less than 30 days and
typical mass ratios of 0.5\,--\,1.0, there is a similar
fraction of wide visual binaries (42 systems
with probable orbital motions, with periods
larger than 100\,--\,1000\,yr, and typical mass
ratios from 0.5 to 0.1). Clearly the distribution
of mass ratios is different for wide binaries and
for close binaries. However there is a huge gap
in orbital periods between close and wide
binaries. This is a selection effect: Such
binaries have been hard to detect with past
techniques. Interferometry will be able to find
such binaries in the near future. Mason et al.
(1998) estimate that the total O star
(spectroscopic + visual) binary frequency is
about 60\,\%, but they suggest that when the
above gap is filled, the frequency may well be
100\,\%. They also note that most binaries occur
in clusters and associations and that binaries are
less common among field stars and especially
among runaway stars.

10\,--\,25\,\% of all O stars are runaway stars,
whereas among the B stars this fraction is only
about 2\,\% (Gies \& Bolton 1986). The fact that
the O-star runaways are 10 times more common
than the B-star runaways implies that the
secondaries of O-star primaries (which are the
ones that get ejected, Leonhard \& Duncan
1990) should also preferentially be O stars.
That is, many O--O binaries must have mass
ratios skewed to unity (Clarke \& Pringle 1992),
in accordance with the observed high frequency
of double-lined spectroscopic binaries among
O-type stars in young clusters. However, the
binary frequency for O-type runaways is
generally lower than that in young clusters
(Mason et al. 1998). A possible explanation
could be that the runaway recoil process
produces a tight binary and a single massive
star ejected in almost opposite directions, thus
predicting a spectroscopic binary frequency for
runaway O stars that is half that of young
O-star clusters.

Wolf-Rayet (WR) stars are evolved massive
stars and, strictly speaking, do not belong to our
main-sequence statistics. On the other hand,
their progenitors were massive stars that lost
much of their mass because of strong stellar
winds. The mass loss rate and terminal wind
speed depend on metallicity; more metal-rich
stars lose more mass and lose it more quickly.
The initial masses of WR stars cannot easily be
determined from observations alone. Many new
WR star binaries as well as mixed WR-OB star
pairs have recently been discovered. The
important question is whether the binary
frequency of WR binaries and WR-OB star
pairs depends on metallicity. We leave a
discussion of WR-star binary statistics and
related topics to Crowther (2007, in this
volume).

The O3 stars HD 93129A and HD 93205 in
the Carina young star clusters Trumpler 14 and
16 were once thought to be the most massive
single stars in the Galaxy (Taresch et al. 1997,
Antokhina et al. 2000). In the case of HD
93129A this claim is based on the detailed non-local
thermodynamic equilibrium analysis of
the UV and hydrogen and helium optical
spectrum of this O-type supergiant 
(T$_{\rm eff}$\,=\,52,000\,K).
This analysis implies a very high
bolometric luminosity, 2.5\,$\times$\,10$^6$\,L$_\odot$. The
extreme dynamical stellar wind properties 
($\dot {\rm M}$\,=\,-20\,M$_\odot$\,Myr$^{-1}$,
v\,=\,3200\,km\,s$^{-1}$)
can be
used to infer a stellar mass of 130\,$\pm$\,20\,M$_\odot$
(Taresch et al. 1997). However, Walborn
(2003) and Nelan et al. (2004), using the Fine
Guidance Sensor on HST, recently resolved HD
93129A into two components with a separation
of 60\,mas or 165\,AU and a visual magnitude
difference of 0.5\,mag. This corresponds to
component masses of about 80 and 50\,M$_\odot$. It is
unclear how the new stellar parameters can be
consistent with the high luminosity and the
extreme wind properties, but this example
shows strikingly how we can be misled in our
conclusions if the binary nature of massive stars
is unrecognized.

As for HD 93205, this object consists of an
O3V and an O8V noneclipsing pair with an
orbital period of about 6 days. Because the
inclination angle of the orbit is unknown, we
only know that 
M$_1 \cdot \sin^3{\rm i}$\,=\,29\,M$_\odot$ and
M$_2\cdot\sin^3{\rm i}$\,=\,13\,M$_\odot$
(Antokhina et al. 2000). An
O8V star should correspond to about 
20\,--\,25\,M$_\odot$,
so one can tentatively infer a primary
mass M$_1$\,=\,50\,--\,60\,M$_\odot$. 
Morrell et al. (2001)
suggest a spectral type O3.5 V and a mass of 48\,M$_\odot$
for the primary of HD 93205, making it
one of the earliest main-sequence stars in the
Galaxy at this time. There is at least one other
massive double-lined spectroscopic binary in
the Trumpler 16 cluster (Tr 16-110 with P\,=\,3.5
days), as well as two massive eclipsing binaries
(Tr 16-1 and Tr 16-104) with P\,=\,2.2 days
(Rauw, personal communication). Also in the
same cluster, there is one of the most massive
WR stars in the Galaxy (WR 22, HD 92740),
which is the primary of a spectroscopic binary
with P\,=\,80 days. Its current mass is 45\,M$_\odot$
(Schweickhardt et al. 1999), corresponding to a
ZAMS mass $\sim$80\,M$_\odot$ (Rauw, personal
communication). We refer to Walborn et al.
(2002) for further discussion of the most
massive stars known (O2V stars). These
include Cyg OB2-22A and Pismis 24-1, with
masses in excess of 100\,M$_\odot$ and, up until
recently, believed to be single; however, the
latter has now been resolved by HST as a visual
triple system (see Ma\'{i}z-Apell\'{a}niz et al. 2007).
A similar case was LBV 1806--20: Eikenberry
et al. (2004) determined its mass to be close to
200\,M$_\odot$, but Figer, Najarro \& Kudritzki (2004)
dethroned it, finding that it is most likely a
spectroscopic binary. Therefore, at present, the
'Pistol Star' with an estimated initial mass of
200\,--\,250\,M$_\odot$ (Figer et al. 1998) in the
Quintuplet Cluster near the Galactic Center is
still considered to be the most massive star in
the Galaxy.

However, the most massive stars with
dynamically measured masses reside in the
double-lined spectroscopic, eclipsing binary
WR20a: the primary and secondary component
masses are 83\,M$_\odot$ and 82\,M$_\odot$, respectively,
with an error bar of 5\,M$_\odot$. Rauw et al. (2004)
originally discovered the spectroscopic binary
nature of the object with an orbital period of
about 3.7 days. Bonanos et al. (2004) obtained
an eclipse light curve in the I-band and so could
determine the inclination angle and hence the
masses. Both objects are slightly evolved (SpT
WN6ha), and still undergoing core hydrogen
burning, but on their way from early O-type to
WR stars -- with no previous phase of Roche
lobe overflow. The binary system  lies in the
Westerlund 2 cluster associated with the RCW
49 HII region, but interestingly not near the
cluster center. Speculation has it that this most
heavy close binary system was ejected from the
cluster core due to a dynamical interaction. A
spectacular {\it Spitzer}/IRAC image of the RCW 49
HII region and cluster can be found in the
Infrared Legacy Gallery at IPAC, courtesy of E.
Churchwell and NASA/JPL-Caltech. A glimpse
of some other spectacular {\it Spitzer}/IRAC images
of high-mass star forming regions is given in
Brandl et al. (2005), including the embedded
proto-OB association W49A.

Additional striking examples of multiplicity
among very massive stars can be found in the
30 Dor cluster in the LMC, where Massey,
Penny \& Vukovich (2002) have identified four
tight, double-lined spectroscopic binaries, three
of which are eclipsing systems (R136--38 is the
most massive system, with component masses
of 57 and 23\,M$_\odot$, while R136--42 is a close
second with component masses of 40 and 33\,M$_\odot$
(the latter is actually a physical pair of
O3V stars with a period of 2.89 days). Other
massive stars in 30 Dor, including Melnick 34,
are being monitored for radial velocity
variations.

Unlike the center of NGC 3603, which is dominated
by a massive WR binary system,
a recent, still unpublished near-IR integral
field spectroscopic survey of the R136 central
regions with the ESO-VLT has not identified
any 
WR spectroscopic binaries (T. Moffat \& O. Schnurr,
personal communication).
This result is contrary to the physical intuition
that would lead us to expect the heaviest
objects, i.e., massive binaries, to be located in
the very center (mass segregation due to
dynamical friction).

The center of the R136 cluster that was once
considered to host a supermassive star
(Feitzinger et al. 1980; Cassinelli, Mathis \&
Savage 1981) actually consists of a group of
eight massive stars within a projected radius of
0.4\,arcsec or equivalently 0.1\,pc (Weigelt \&
Baier 1985; Pehlemann, Hofmann \& Weigelt
1992), corresponding to a central mass density
in massive stars of 10$^5$\,\,M$_\odot$ per cubic parsec.
We do not know how much mass there is in
low-mass stars in the center, as two-body
dynamical mass segregation could have led to a
depletion of low-mass stars there. The density
of massive stars alone may not be high enough
at this time to enable a runaway stellar collision
process, leading to the formation of an
intermediate mass black hole, as envisaged by
Portegies Zwart \& McMillan (2002). It is
enough, though, for the occasional collision
between two massive stars, leading to the
formation of a rapidly rotating massive object,
possibly the progenitor for a gamma-ray burster
(Fryer \& Heger 2005, Zinnecker 2006b).
Although the R136 cluster is quite dense in its
center, more massive young globular clusters
such as those seen in the Antennae merging
galaxies may be denser still, and a runaway
collision process cannot be excluded.

\subsection{The Universality of the Upper Initial Mass Function}

How can we understand the universality of the
upper IMF? Wherever we look, the data seem
to be compatible with a Salpeter power-law
with a logarithmic slope of -1.35. Even
population studies at moderate to high redshift
are consistent with the assumption of a Salpeter
power law (e.g., Baldry \& Glazebrook 2003).

In essence, two completely different schools
of thought have attempted to explain the
robustness of the Salpeter IMF despite various
kinds of environmental factors that could be
expected to change this distribution (such as
metallicity, gas pressure, or the density of
stellar systems, to name but a few). These are
competitive accretion and random sampling of
fractal clouds.

As for competitive accretion, this has been
discussed extensively in the review of Bonnell,
Larson \& Zinnecker (2007) and will not be
repeated here. Suffice it to say that competitive
accretion of protostars for cluster gas can
explain the observed Salpeter stellar mass
distribution for massive stars, if most massive
stars form in dense clusters (Bonnell et al.
2001b, Klessen 2001a,b, Klessen \& Burkert
2001; for a discussion of the onset and richness of
clustering as a function of the most massive cluster star
see Testi et al. 1997 and Testi, Palla \& Natta
1999).

If, however, a substantial fraction of massive
stars forms in less dense OB associations
(Garmany 1994; Clark, Bonnell \&
Zinnecker 2005 ) or in isolated places in the
field (for example, where was Betelgeuse -- a
$\sim$\,20\,M$_\odot$ red supergiant -- born?), then competitive
accretion cannot be invoked, and another idea is
needed (e.g. Lamers et al. 2002;
Li, Klessen \& Mac Low 2003).

Elmegreen (1997) presented such an idea,
essentially a geometrical model, with a physical
icing on the (geometrical) cake. This model is
based on a random sampling of mass in
turbulent fractal interstellar clouds (see {\bf Figure~17}).
A star was assumed to get a fixed fraction
of the gas mass of the cloud piece in which it
formed. The mass distribution of the pieces in
any hierarchy has a logarithmic slope of -1
(i.e., equal mass in equal logarithmic intervals),
so the IMF would seem to end up with this
slope, but the sampling rate for pieces was
assumed to be proportional to the square root of
the local gas density to mimic the dynamical
processes that are involved. This local density
depends on the level in the hierarchy according
to the fractal scaling of density with size
(because mass scales with size to a power equal
to the fractal dimension, density scales with
size to a power equal to the fractal dimension
minus three). This density dependence means
that lower mass regions are sampled more
frequently, and the mass function slope
steepens from -1 to -1.35, which is the Salpeter
slope. The net result of this sampling is a mass
function for model stars that is
indistinguishable from the observed IMF for
young clusters and OB associations.

Another interesting attempt to connect star
formation to fractal cloud structure was made
by Henriksen (1991, see his figure 1), but the
details differ from Elmegreen's (1997) model.
An important variant of the fractal molecular
cloud IMF  model is the concept in which the
stellar mass function derives directly from the
mass distribution of cloud cores. Recent
observations show a surprising similarity
between the scaled cloud core mass distribution
and the stellar IMF (Alves, Lombardi \& Lada
2007).

Yet another alternative for explaining a
power-law IMF is due to Basu \& Jones (2004).
They note that the power-law tail in the mass
function of protostellar condensations and stars
arises from the accretion of ambient cloud
material on to the condensation, coupled with a
nonuniform (exponential) distribution of
accretion lifetimes (cf. radioactive decay).
Thus, this model assumes that not all
condensations accrete for the same time. If we
start with protostars with a log-normal mass
distribution (possibly a reasonable assumption
expected from the central limit theorem, see
Zinnecker 1985), this log-normal distribution
develops a power-law tail at high masses if the
accretion rate is directly proportional to the
instantaneous mass of the accreting object and
if the probability of stopping accretion is
constant in time. The latter implies an
exponential probability distribution of accretion
timescales with a constant death rate. How
general this random accretion model is and how
it relates to the competitive accretion model
above needs to be further explored (cf. Bate \&
Bonnell 2005).

Needless to say that the characteristic mass
scale of fragmentation and the IMF (a few
tenths of a solar mass) is beyond the scope of
this review, and we refer to Larson (1985,
2005) and also Whitworth, Boffin \& Francis
(1998) for an in-depth discussion of this issue.
Finally, we refer to Klessen, Spaans \& Jappsen
(2007) for the first numerical hydrodynamical
calculations of the characteristic stellar mass in
starburst regions, predicting a top-heavy IMF.
This is naturally explained as a consequence of
the elevated thermal Jeans mass in the warmer
and dustier starburst environment (including the
Galactic Center).

\subsection{The Number of Accreting Massive Protostars in the Galaxy}

In this subsection, we derive an estimate of
high-mass accreting stellar objects in the
Galaxy (note that we are trying to avoid the
term protostars). To this end, we first calculate
how much stellar mass is formed in the Galaxy
over a time-interval  t$_{\rm acc}$ corresponding to the
accretion phase (200,000\,yr). Using a total
Galactic star-formation rate (SFR)  of 5\,M$_\odot$\,yr$^{-1}$
(Smith, Mezger \& Biermann 1978, Diehl
et al. 2006) and multiplying by 200,000\,yr,
we obtain 10$^6$\,M$_\odot$.  Of these, we assume a
fraction f\,$\approx$\,0.10 ends up in massive stars (above
10\,M$_\odot$, say). The rest goes into intermediate-
and low-mass stars according to a Salpeter
(1955) field star IMF, with an effective lower
mass limit of 0.1\,M$_\odot$ and upper mass limit of
100\,M$_\odot$ (see Smith, Mezger \& Biermann 1978,
their table 2). A more realistic low-mass field
star IMF (Kroupa 2002, Chabrier 2003) would
increase the fraction f by about a factor of 2, yet
would decrease the SFR derived above by
about the same factor, thus leaving the product
f\,$\times$\,SFR approximately constant.

The number N$_*$\,($>$\,M$_*$) of stellar progenitors
in the Galaxy to be found in the accretion
phase, which eventually attain a final mass
above a given value M$_*$, can then be derived
from the two equations (the first solving for
N$_0$),
\newpage
\[
   {\rm N_0 \int^{100}_{0.1} m^{-x} dm = f \times SFR \times \Delta t_{acc}}
\]

\[
   {\rm N_* (> m_*) = N_0 \int^\infty_{m_*} m^{-x} d log m} 
\]
Here m$_*$\,=\,M$_*$/M$_\odot$ is the dimensionless
stellar mass and dN/dlog(m)\,=\,N$_0$\,m$^{\rm -x}$ is the
power-law upper stellar IMF with a slope x
(x\,=\,1.35 for a Salpeter IMF). Solving this
equation results in the numbers N$_*$\,($>$\,M$_*$)
given in {\bf Table~4}.

\begin{table}[h]
  \begin{center}
  \caption{Expected number N  of stellar progenitors
in the Galaxy to be found in the accretion phase$^a$,
which eventually attain a final mass above M$_*$, for
various logarithmic slopes x of the IMF$^b$}
  \vspace{0.5cm}
  \begin{tabular}{@{}cccc@{}}
     \hline\hline
       {\bf M$_*$}         &{\bf N$_*$}             &{\bf N$_*$}
	&{\bf N$_*$}          \\
       $\left[{\rm M}_\odot\right]$
                     &x\,=\,1.3        &x\,=\,1.35	&x\,=\,1.7      \\
     \hline
       $> 10$        &6300              &5400            &1600           \\
       $> 20$        &2400              &2000            &480            \\
       $> 30$        &1300              &1000            &220            \\
       $> 50$        &480               &390             &74             \\
     \hline
  \end{tabular}
  \end{center}
\vspace{-1.0cm}
\begin{tabular}{@{}ll@{}}
\hspace{0.8cm} & \\
& $^a$Assumed to last 200,000\,yr.\\
& $^b$Assuming a total Galactic star formation rate of 
   5\,M$_\odot$\,yr$^{-1}$.\\
\end{tabular}
\end{table}

The table shows that we can expect about
1000 accreting massive stars in the Galaxy with
masses in excess of 30\,M$_\odot$ (i.e., spectral types
earlier than O6.5V, see {\bf Table~3}). Assuming 10\,kpc
as the reference radius of the Milky Way,
we see that the average surface density of
accreting early O-type massive stars is about
3\,objects/kpc$^2$. This demonstrates the rareness of
massive protostars, but is in agreement with the
few luminous and totally embedded objects
known, such as BN/KL and Cep-A (see {\bf Table~5}
in the Appendix which provides a list of
massive star forming regions within 1\,kpc from
the Sun and some properties of the exciting
stars).

\subsection{Is There a Maximum Stellar Mass Set by Star Formation?}

In {\bf Section~2.5} we reviewed the observational
evidence for an upper stellar mass limit and
concluded there was such a limiting mass
somewhere in the range between 100 and 200\,M$_\odot$,
at least in Population I systems. Here we
present a few thoughts on the possible physics
of the stellar upper mass limit.

The first question is whether this limit is set by
stellar stability considerations or star-formation
theory. Ledoux (1941), and later Schwarzschild
\& H\"arm (1959), concluded from linear stability
analysis of radial stellar pulsations that there is
an upper mass limit of the order of 100\,M$_\odot$,
corresponding to the existence of a vibrational
instability owing to nuclear reactions. This
refers to massive stars of Population I chemical
composition, whereas for lower chemical
abundances the unstable mass is higher.
Appenzeller (1970) and Ziebarth (1970)
concluded, from nonlinear calculations, that
this vibrational instability 
(also known as $\epsilon$-instability)
does not have a global disrupting
effect, except perhaps in stars with masses
greatly exceeding the critical mass. A more
recent discussion can be found in Stothers \&
Chin (1993). These authors concluded that, for
a metallicity equal to or greater than 0.02, no
nuclear-induced pulsational instability
developed for masses up to at least 150\,M$_\odot$.
For a metallicity equal to 0.004, corresponding
to the value of the Small Magellanic Cloud, the
critical mass for such an instability was still
very high, around 140\,M$_\odot$.

Larson \& Starrfield (1971) were the first to
stress a possible upper mass limit resulting
from star-formation theory. They came up with
a value of 50\,--\,100\,M$_\odot$ as a larger and larger
fraction of the growing mass is thrown out by
radiation before the star reaches stellar
conditions.

Today, we realize that the opacity of dusty
gas does not determine the upper mass limit,
whereas the opacity of ionized gas (electron
scattering and UV line blanketing) certainly
may. The constraint imposed by dust opacity
(see {\bf Figure~6}) shows that a luminous object
with an L/M ratio consistent with a main-sequence
star M$_*$\,$>$\,100\,M$_\odot$ can still accrete
material as long as the object appears to be cool
($<$\,50\,K). This allows material to fall onto the
disk. Within the disk much of the radial
radiation flux is diverted into the polar direction
(flashlight effect). Provided angular momentum
transfer is adequate, dusty material can thus
flow radially inward until at some point the dust
is destroyed (sublimation radius). Dust-free
neutral gas has a lower opacity and it can
continue to flow inward. Eventually, as the disk
gas gets closer to the accreting star, the gas is
ionized and the opacity increases again sharply.
Disk models currently do not allow an accurate
estimate of the opacity of the ionized disk gas
close to the star; the lower limit to the opacity
provided by electron scattering limits the
maximum L/M to $\sim$\,6\,$\times$\,10$^4$, 
well in excess of that
expected from a main-sequence star of M$_*$\,=\,200\,M$_\odot$.

Because very massive stars also possess
strong radiation-driven winds, it is clear that the
opacity of ionized stellar gas is sufficiently high
that radiative acceleration exceeds surface
gravity. The upper mass limit could thus result
from the fact that mass loss from the star
matches or exceeds accretion. As the line-driven
wind mass loss from massive stars is
metallicity-dependent (Kudritzki 2002) the
upper mass limit would be metallicity-dependent,
too. However, if mass loss is by a
continuum-driven instability, as suggested for
$\eta$~Car and other Luminous Blue Variables (Smith
\& Owocki 2006), the upper mass limit would
be independent of metallicity.

It is unclear if stellar collisions and mergers
in dense stellar systems can beat the
opacity-limited accretion discussed above and form
stars with masses beyond 200\,M$_\odot$.  The
merging of a binary star with two 100-M$_\odot$
components is conceivable (Zinnecker 1986,
Bonnell \& Bate 2005), doubling the single star
upper mass limit. It is likely that a coalescence
process would deplete the number density of
stars to collide with near the center of a deep
cluster potential, thus imposing a density-dependent
limiting mass, especially for young
clusters undergoing core collapse (Portegies
Zwart et al. 2006). Interestingly, in this case,
too, the upper mass limit would be independent
of the heavy element abundance.  We refer to
Omukai \& Palla (2003) for a theoretical
prediction of the increased upper mass limit
($\sim$\,600\,M$_\odot$) in zero-metallicity, Population III
conditions, albeit based on 1D models.

\subsection{Evolutionary Sequence}

We can characterize the formation of massive
stars by the following crude four-stage
evolutionary sequence:
CDMC $\longrightarrow$ HDMC $\longrightarrow$ 
DAMS $\longrightarrow$ FIMS.
Here CDMC means cold
dense massive core, HDMC means hot dense
massive core, DAMS stands for disk-accreting
main-sequence star, and FIMS denotes the final
main-sequence star. The FIMS star is more
evolved than the theoretical concept of the
ZAMS star, which assumes no prior hydrogen
consumption. Of course, the sequence needs a
more detailed description. The CDMCs are
starless turbulent gravitationally bound
condensations; they have near-virial
equilibrium and are either on the verge of
collapsing or, indeed, already collapsing. In the
center an intermediate-mass protostar forms
that can heat up the dense massive core, turning
it into a HDMC. At this stage, collimated jets
and outflows first appear, traced by H$_2$O and
later by methanol maser emission. The central
star grows in mass primarily by disk accretion
and quickly becomes a DAMS object powered
more by hydrogen burning than by disk
accretion. The collimated outflows become less
collimated with widening opening angles. At
this stage, the accretion disk starts to get
photoionized and partly photoevaporated,
giving rise to a gravitationally confined
hypercompact HII region (HCHII) with broad
hydrogen recombination lines.

Eventually the star has accreted its final mass
and most of the disk has been dissipated. At this
point the ionizing radiation is no longer
quenched by the accretion flow and can expand
freely into the vicinity of the star, blending
together with ionized bubbles from other
similarly massive stars. This is when we speak
of an ultracompact HII region (UCHII), often
formed by a group of ionizing stars within a
volume of (0.1\,pc)$^3$ (e.g., W3-IRS5). UCHII
regions frequently exhibit OH maser 18-cm
radio emission [e.g., W3(OH)], which can be
used to obtain a magnetic field strength by
measuring the Zeeman splitting of the 1667\,MHz
line. Subsequently, the UCHII region
keeps expanding and evolves into a compact
radio HII region and finally into a normal
diffuse optical HII region like the Orion
Nebula.

How this simple-minded picture changes
when we deal with a massive starburst
protocluster, like the progenitor of R136 with
dozens of massive stars born in a small volume,
is anybody's guess. It is conceivable that
competitive accretion takes place in a
gravitationally bound dense HII region. It is
even conceivable that competitive accretion
turns into cooperative accretion (E. Keto,
private communication) which means the
ionized flow pulled in by the collective gravity
of the cluster is redirected onto the less massive
members of the cluster. Instead of being greedy
and competing for additional mass with their
siblings, the near-Eddington-limit massive stars
start to deflect the inflow onto neighboring
stars. By doing so, they aid the further growth
of their lower mass neighbors. Thus, accretion
becomes cooperative! Another possibility is the
hierarchical merging of subclusters which
naturally leads to prompt initial mass
segregation in the resulting final cluster
(McMillan, Vesperini \& Portegies Zwart 2007).
In this scenario, many but not all subclusters
(gas+stars) merge with each other and sink to
the center of gravity owing to the loss of kinetic
energy in these highly inelastic collisions.
There they form a dense stellar cluster of
massive stars, which will expel the residual
cluster gas. Some subclusters miss and won't
collide; these will form massive stars in the
cluster outskirts. Such a core-halo structure is
observed in many young clusters, including
R136 (e.g., Moffat, Drissen \& Shara 1994).
Super OB associations like NGC 604 in M33
have more widely spread subclusters that are
unable to merge, thus avoiding dense cluster
formation (cf. Hunter 1995).

\newpage

\section{WHY HIGH-MASS STAR FORMATION IS NOT A SCALED-UP VERSION OF
         LOW-MASS STAR FORMATION}

In this section, we try to convince the reader
that high-mass star formation is indeed different
from low-mass star formation, and not merely a
scaled-up version of star formation by disk
accretion for higher accretion rates. Many
complex new physical processes enter the scene
for high-mass stars.

To start with, radiative forces on gas and dust
play little or no role in the build-up of low-mass,
solar-type stars, whereas a substantial
fraction of the luminosity of high-mass stars is
emitted in ionizing radiation, which introduces
new effects such as the photoevaporation of the
star's accretion disk and protostellar envelope.
This dramatically limits late accretion and the
final stellar mass. In addition, the ionizing
photons can photoevaporate the disks of the
neighboring lower mass stars (cf. the proplyds
in the Orion Nebula). The nonionizing far-UV
radiation will influence the massive stars'
molecular cloud environment by dissociating
H$_2$ and CO molecules. This requires photons of
about 11.2\,eV; even early-type B stars can
produce these photons, but low-mass and
intermediate-mass stars cannot. The radiative
acceleration of dusty and gaseous matter also
leads to radiation-driven bipolar winds, and
ionizing radiation can escape through these
wind-blown cavities (flashlight effect);
however, the bipolar outflows from low-mass
stars are generated by magneto-centrifugal
forces.

The second big difference between the
formation of low-mass and high-mass stars is
the fact that massive stars are practically born
on the main sequence, whereas low-mass stars
spend a considerable part of their youth as
contracting premain-sequence objects (30\,Myr
for a solar-mass star). A massive star, forming
by accretional growth from an initially low-mass
star with an accretion rate of 
10$^{-4}$\,M$_\odot$\,yr$^{-1}$
begins central hydrogen burning after
about 9\,M$_\odot$ have accumulated -- 13\,M$_\odot$ for an
accretion rate of 10$^{-3}$\,M$_\odot$\,yr$^{-1}$. For low-mass
stars, circumstellar disk evolution proceeds
during the whole extended premain-sequence
phase, whereas for massive stars the disk
lifetime is very short (less than 1\,Myr). When
we see massive O stars close to the ZAMS
(Walborn 2007), their disks have been
dissipated (but see Kastner et al. 2006). It is
only in the embedded phase that circumstellar
CO-bandhead emission and hydrogen emission
lines are detected (Bik et al. 2005, Blum 2005),
indicating a dense neutral remnant disk with an
ionized upper layer. Bipolar outflows are
another strong indication of the existence of
disks.

A further significant difference between low-
and high-mass star formation is the role of
competitive accretion in protoclusters.
Competitive accretion is far more important for
high-mass stars than for low-mass stars. The
latter can form directly by Jeans-type
gravitational instability and turbulence-induced
cloud fragmentation. The former must accrete
large amounts of protocluster gas.

Gravitational dynamics (N-body interactions)
also have a much greater effect when massive
stars are involved. This can best be seen when
considering the dynamical ejection of members
of massive multiple systems producing
runaway OB stars. This phenomenon is largely
absent for low-mass stars. It appears that
massive stars, when they form in their own
local subclusters, are always accompanied by a
small group of lower mass stars, some of which
stay bound to their big parent star even after
subclusters merge. Thus the higher companion
star fraction observed for massive stars
compared to solar-type stars testifies to a more
dynamic scenario for the heavy objects thanks
to their higher than average gravitational
attraction.

Finally, massive stars have a much bigger
influence on triggering new star formation in
adjacent regions than low-mass stars. Massive
stars provide external pressure in the form of
expanding HII regions, stellar winds, and
supernovae explosions. They are capable of
sustaining sequential and self-propagating star
formation (Gerola \& Seiden 1978), a process
that low-mass stars are incapable of. Through
runaway OB stars, massive star formation can
trigger further massive star formation over large
(kpc) distances, an important feature in
sustaining large-scale nuclear starbursts. We
speculate that, if massive star formation is
massively triggered, the individual collapse of
massive cores is outside-in, instead of inside-out
(Banerjee, Pudritz \& Anderson 2006), and
always magnetically supercritical (Shu, Adams
\& Lizano 1987). In other words, magnetic
fields likely play a more passive role in massive
star formation, whereas in low-mass stars this is
opposite.

In defense of the notion of a scaled-up
formation picture, some arguments have been
raised. Foremost is the observation that
collimated outflows occur both in high-mass
and low-mass young stars (Beuther et al. 2002,
Davis et al. 2004). However, a closer look
reveals that collimated outflows and jets do not
occur in the most luminous sources, but only in
sources with total luminosities up to 10$^4$\,L$_\odot$
(Shepherd 2005), with one or two exceptions
(Garay et al. 2003, Rodr\'{i}guez et al. 2005b).
This implies that massive star formation can be
a scaled-up version of low-mass star formation,
but only up to early B stars. The outflow
morphology does not scale further for O stars,
which generate powerful, wide-angle, ionized
winds, calling into question the relationship
between outflow and accretion (Shepherd
2005).

Another point that has been made relates to
the IMF. It has been suggested (see the
discussion in Zinnecker 2004) that there should
be a feature (knee) in the IMF, at a critical mass
where the intermediate-mass and high-mass
star-formation processes diverge. However, this
feature has not been seen and the upper IMF is
a power-law with a constant slope (Kroupa
2002). It is unclear how to refute this argument,
but the constant slope of the upper IMF may
simply imply that stellar collisions and mergers
are not a dominant factor in massive star
formation, except for the most massive stars
(Bonnell \& Bate 2002) where statistical
fluctuations can hide a slope change.
Collisional growth of massive stars (Bonnell,
Bate \& Zinnecker 1998; Bally \& Zinnecker
2005) also seems to be ruled out by recent
observations of stellar rotation: There is a
continuous behavior of the specific stellar spin
angular momentum over the full range of stellar
masses, i.e., J/M\,$\sim$\,M$^{0.3}$ between 0.2\,M$_\odot$ and
50\,M$_\odot$, indicative of a single stellar formation
and angular momentum regulation mechanism
(Wolff et al. 2006). However, a collisional
process and stellar mergers are not ruled out for
the most massive early O-type stars (M\,$>$\,50\,M$_\odot$).

\newpage

\section{OUTLOOK: RELEVANT FUTURE OBSERVATIONS}

Here we suggest a number of future
observations that would help promote?advance
our understanding of massive star formation.
The list, however, is by no means complete.

\begin{enumerate}

\item IR-observations with an Extremely Large
      Telescope (ELT)\\ (diameter 30\,--\,42\,m)\\
      As described in Zinnecker (2006a), such a
      powerful telescope could penetrate the dust
      extinction of ultracompact HII regions
      (A$_{\rm V}$\,=\,100\,--\,200\,mag) 
      in the near-IR (A$_{\rm K}$\,=\,10\,--\,20)
      and see the stellar photospheres of
      massive stars, resolving very tight
      embedded clusters, such as W3-IRS5 (cf.
      Megeath, Wilson \& Corbin 2005). There is
      hope that an ELT might have sharp enough
      vision to test the prediction (Zinnecker
      2006b) that gamma-ray bursts occur in the
      centers of young massive protoglobular
      clusters.
\item Submillimeter observations with the
      {\it Atacama Large Millimeter Array}\\
      The spatial resolution of the {\it Atacama Large
      Millimeter Array} ({\it ALMA}) at 350 microns
      ($\sim$\,0.01\,arcsec, i.e., 50\,AU at 5\,kpc) would
      allow us to map dense molecular clumps,
      where massive stars are believed to form, in
      the dust continuum and see whether
      fragmentation is occurring (Dobbs, Bonnell
      \& Clark 2005) or not (Krumholz 2006).
      {\it ALMA} should also be able to measure
      rotation curves of circumstellar disk gas
      around massive stars and thus measure their
      enclosed stellar mass (M. Krumholz, R.I.
      Klein \& C.F. McKee, submitted).
\item Mid-infrared observations with the {\it James
      Webb Space Telescope}\\
      The {\it James Webb Space Telescope}  will
      have unparalleled background-limited
      sensitivity to reveal very embedded
      populations around young massive stars in
      the making in near- and far-galactic star
      forming regions as well as regions like 30
      Dor/R136 in the Magellanic Clouds.
      Progress may also be expected for
      understanding star formation in the Galactic
      Center Regions, including resolution of the
      Arches and Quintuplet clusters and the OB
      populations near Sgr~A$^*$  (Figer 2003, Kim
      et al. 2006, Paumard et al. 2006, Martins et
      al. 2006).
\item Proper motion observations with {\it Gaia}\\
      Precise proper motion data of young
      clusters and OB associations are eagerly
      awaited, surpassing the limited milli-arcsec
      precision of Hipparcos (de Zeeuw et al.
      1999). This will be crucial to understanding
      the dynamical nature of OB associations,
      such as the question of whether an OB
      association is just an expanding, dissolving
      massive star cluster or something of its own
      (Blaauw 1983, Brice\~{n}o et al. 2007). Hence
      {\it Gaia} will tell us if the stars of an
      association expand from a common center
      or from several centers or from no center at
      all. {\it Gaia} will also provide us with a
      complete sample of field O stars and their
      parallaxes within a few kiloparsecs.
\item Far-IR observations with {\it Herschel}\\
      The immediate future will see the launch of
      ESA's {\it Herschel} space observatory, whose
      Photodetector Array Camera \&
      Spectrometer (PACS) instrument will make
      it possible to follow up on {\it Spitzer}/MIPS
      observations and to detect highly embedded
      massive protostellar sources down to a
      spatial resolution of 3 arcsec at 60 microns.
      Surveys with the Spectral and Photometric
      Imaging Receiver (SPIRE) will allow us to
      identify and characterize pre-collapse cores
      at an early evolutionary stage (Boss \&
      Yorke 1995). PACS together with SPIRE
      can be used to better constrain the 'collect
      and collapse' model of triggered massive
      star formation (Zavagno et al. 2005),
      discriminating between true triggering
      versus the compression of pre-existing
      clumps.
\item Interferometric near-IR observations with
      VLTI and LBTI\\
      Very high angular resolution observations
      with closure phases will give us
      interferometric images (and not just
      visibilities) of ultracompact HII regions and
      the stellar content of hot molecular cores.
      This will allow us to obtain a clearer picture
      of the maternities of massive stars (Garay
      2005), including images of disks and jets as
      well as astrometric information, such as the
      stellar velocity dispersion, dynamics of the
      stars in protocluster centers, and the orbital
      motion of short-period embedded binary
      and multiple systems.
\item Long-term variability surveys
      The formation and evolution of massive stars
      is likely to involve a variety of cataclysmic
      or eruptive phenomena on short timescales
      (days to years), such as episodic FU Orionis-like
      accretion (Chini, private
      communication) and stellar mergers (Bally \&
      Zinnecker 2005), or $\eta$~Car-like LBV
      eruptions (Smith \& Owocki 2006) and even
      infrared supernovae in luminous starburst
      galaxies (Maiolino et al. 2002). Without
      continuous monitoring, these (embedded)
      events may go unnoticed. This calls for long-term
      variability compaigns, in the near-infrared,
      for example from a medium-sized
      survey telescope at Dome C in Antarctica.
\item "Observations" with a Numerical Telescope\\
      As numerical simulations of the star
      formation process become more refined in
      3D resolution and the microphysics involved
      (magnetic fields, chemistry, ionization,
      heating, cooling, dust physics, etc.) the
      ability to accurately "observe" the evolution
      at various stages with respect to dust
      continuum features, line intensities and line
      profiles increases in importance. In other
      words, theoretical modeling will have to
      keep pace with the rapid development of
      observing capabilities as described above.
   
\end{enumerate}

\newpage

\section{FINAL SUMMARY}

Massive star formation is not a simple scaled-up
version of low-mass star formation, particularly
when it comes to ZAMS O stars ($>$\,20\,M$_\odot$). The
formation of early-type B stars, however, may still
be considered a continuation of low-mass star
formation. In terms of formation processes, the
monolithic collapse and disk accretion model and
the competitive accretion scenario are the two
opposite ends of a continuum of cases in the
accretion theory of O-type massive star formation.
The coalescence process, introduced to circumvent
the obstacle of radiation pressure, is no longer
generally necessary but may still occur
in exceptional circumstances, especially for very
massive stars in the centers of very dense
protoclusters or subclusters with strong initial mass
segregation. The observed multiplicity and
clustering of massive stars suggests complex N-body
and gas dynamical interactions (tidal, drag, and
capture effects) among the youngest stars or
protostars that would contribute to the argument
against a monolithic collapse of isolated massive
protostellar cores.

Massive stars are rare and correspond to the tail of
a power-law stellar mass distribution. For a Salpeter
IMF, the number of all newborn ZAMS OB stars
($>$\,8\,M$_\odot$) is a mere 10\,\% of all stars in the mass range
of 1\,--\,2\,M$_\odot$. At present, competitive accretion (i.e.,
runaway growth of a few objects from cluster gas)
provides the best physical explanation of the high-mass
IMF and its surprisingly universal Salpeter
slope. The maximum stellar mass (around 150\,M$_\odot$),
long believed to be a result of stellar instability
(Eddington-Ledoux limit), might be due to the
negative feedback caused by the increasingly
destructive radiative erosion of massive accretion
disks at increasingly higher stellar masses. This self-limiting
star formation can, however, be beaten by
stellar collisions under very special circumstances.

Rapid external shock compression (i.e.,
supersonic gas motions) generating high column
densities in less than a local free-fall time rather than
slow quasi-static build-up of massive cores may be
the recipe to set up the initial conditions for local
and global bursts of massive star formation. Such
bursts can both enhance or quench further massive
star formation (positive and negative feedback),
depending on environment (gas density and
pressure), but such a discussion is beyond the scope
of this review. The question of how feedback from
massive stars can influence proto galaxy evolution
and morphological types (cf. Sandage 1986,
Elmegreen 1999, Silk 1997, 2005) should be the
topic of a future Annual Reviews article.

\newpage

\section{APPENDIX -- NEARBY ($<$\,1\,kpc)\\ MASSIVE STAR-FORMATION REGIONS}

In this Appendix, we attempt to give a complete
table of star formation regions within 1\,kpc that
contain at least one massive star (see {\bf Table~5}).
Some of these nearby regions of massive star
formation are described in more detail by Bally
et al. (2005). Moreover, the massive stellar
content of nearby (distance $<$\,650\,pc) OB
associations are disscussed in the classic
Hipparcos paper by de Zeeuw et al. (1999),
including Sco OB2, Vel OB2, Per OB2, and
Cep OB2 (see their figures 1 and 29). We also
should mention the Galactic O-star catalog of
Ma\'{i}z-Apell\'{a}niz, P\'{e}rez \& Mas-Hesse (2004),
which includes spectral classification, optical-NIR
photometry, multiplicity, and astrometric
information, as well as cluster?association
membership. (See also the living O-star Web-catalogue
http://www-int.stsci.edu/~jmaiz/GOS/GOSmain.html).
O stars in open cluster regions can also be
selected from the combined photometric and
astrometric membership analysis of 520
previously known (Kharchenko et al. 2004) and
130 newly detected (Kharchenko et al. 2005)
open clusters. As a final highlight, we show
here Westerlund 1, the most massive young star
cluster in the Galaxy ({\bf Figure~18}).

\begin{table}
\begin{center}
\caption{Star formation regions (within 1\,kpc)
         with massive stars (B2 and earlier)}
\vspace{0.5cm}
\begin{tabular}{@{}cccccc@{}}
\hline\hline

{\bf Region} & {\bf Sources} & {\bf Distance} & {\bf Age} & {\bf L/SpT} & {\bf References}\\

\hline

OMC-1 & BN/KL & 450\,pc & emb$^a$ & 10$^5$\,L$_\odot$ & Menten \& Reid 1995\\

 & & & & & Greenhill et al. 2004\\

OMC-1S & FIR4 & 450\,pc & emb & 10$^4$\,L$_\odot$ & Schmid-Burgk et al. 1990\\

 & & & & & Smith et al. 2004\\

NGC1977 & HD 37018 & 450\,pc & 5\,Myr & B1V & Makinen et al. 1985\\

 & & & & & Howe et al. 1991\\

NGC1980 & $\iota$ Ori & 550\,pc & 5\,Myr & O9III & Johnstone \& Bally 2006\\

 & & & & & Piskunov et al. 2006\\

NGC2023 & HD 37903 & 475\,pc & $\sim$\,5\,Myr & B1.5V & Howe et al. 1991\\

 & & & & & Wyrowski et al. 2000\\

NGC2024 & IRS2b & 360\,pc & emb & O8V & Bik et al. 2003\\

 & & & & & Lenorzer et al. 2004\\

$\sigma$ Ori & $\sigma$ Ori AB & 350\,pc & $\sim$\,3\,Myr & O9.5V & Sanz-Forcada et al. 2004\\

 & & & & & Hernandez et al. 2007 \\

Coll69 & $\lambda$ Ori & 400\,pc & $\sim$\,6\,Myr & O8III & Dolan \& Mathieu 2002\\

 & & & & & Barrado y Navascu\'{e}s 2005\\

S106 & IRS4 & 600\,pc & emb & O9V & Hodapp \& Rayner 1991\\

 & & & & & Furuya et al. 1999\\

W40 & OS1a, 2a, 3a & 600\,pc & obsc$^b$ & O9V & Smith et al. 1985\\

 & & & & & Vall{\'e}e \& MacLeod 1994\\

\hline
\end{tabular}
\label{tab:regions}
\end{center}
\end{table}

\begin{table}
\begin{center}
\begin{tabular}{@{}cccccc@{}}
\hline\hline

{\bf Region} & {\bf Sources} & {\bf Distance} & {\bf Age} & {\bf L/SpT} & {\bf References}\\

\hline

NGC1579 & LkH$\alpha$101 & 700\,pc & obsc & B0.5V & Barsony et al. 1991\\

 & & & & & Herbig et al. 2004\\

NGC2264 & IRS1 & 800\,pc & emb & B2V & Schwartz et al. 1985\\

 & & & & & Schreyer et al. 2003\\

Mon R2 & IRS3 & 830\,pc & emb & B1V & Carpenter et al. 1997\\

 & & & & & Preibisch et al. 2002\\

GGD 12-15 & IRAS06084 & 830\,pc & emb & B0.5V & Gomez et al. 1998\\

 & & & & & Gutermuth et al. 2005\\

S140 & IRS1 & 900\,pc & emb & B2V & Lester et al. 1986\\

 & & & & & Preibisch et al. 2001\\

IC 1396 & HD 206267 & 800\,pc & $\sim$\,3\,Myr & O6.5V & Schulz et al. 1997\\

 & & & & & de Zeeuw et al. 1999\\

Cep-OB3 & Cep-A/HW2 & 725\,pc & emb & 2\,$\times$\,10$^4$\,L$_\odot$ & Garay et al. 1996\\

 & & & & & Hartigan et al. 2000\\

Cyg-X & AFGL 2591 & 1000\,pc & emb & 2\,$\times$\,10$^4$\,L$_\odot$ & Tamura \& Yamashita 1992\\

 & & & & & van der Tak et al. 2006\\

\hline
\end{tabular}
\end{center}
$^a$emb stands for embedded, i.e., a very young phase ($<$\,1\,Myr);
sources optically invisible\\
$^b$obsc stands for obscured, i.e., a young phase ($\sim$\,1\,Myr);
sources optically visible but heavily dust extincted\\
Note: AFGL 2591 may be at a distance of 1700\,pc (Schneider
et al. 2006) and part of the Cyg OB2 association, one of the
richest regions of massive stars in the Galaxy with $\sim$\,100 O stars,
including a hidden very massive OB cluster, shown in
{\bf Figure~18}. Note that this massive star forming region is 
a source of significant diffuse $\gamma$ ray emission, i.e. the
1.8\,MeV line from the radioactive decay of $^{26}$Al (Pl\"uschke et al. 2002)

\end{table}

\newpage

\section{GLOSSARY}

\begin{itemize}

\item Accretion\\
gas accumulation of a star or protostar,
increasing the mass of the object

\item Competitive accretion\\
gas initially unbound to a star or protostar and
moving with relative speed, v$_{\rm {rel}}$, w.r.t. to the object
is added to it, i.e., to its gravitational sphere or column 
of influence (defined by an impact parameter 2\,GM/v$^2_{\rm {rel}}$) 

\item Coalescence\\
growth of the mass of a stellar or protostellar object
by a physical
collision and merger with another object

\item Merger\\
inelastic collision between two stars or protostars
leading to the amalgamation of the two bodies

\item Massive star\\
star more massive than about 8\,M$_{\odot}$
that ends its life with a type II supernova
(unless it is in a close binary system with mass transfer);
a massive star on the main sequence has spectral type B3 or earlier.

\item OB star\\
star of spectral type O ($>$\,16\,M$_{\odot}$) or
type B (B\,$>$\,4\,M$_{\odot}$)

\item Ultracompact HII region\\
small blob (typical density $10^4$\,cm$^{-3}$,
size 0.1\,pc) of ionized gas
emitting radio continuum radiation
of high emission measure (L\,n$_{\rm e}^{2}$\,=\,10$^7$\,pc\,cm$^{-6}$)

\item Hypercompact HII region\\
very small blob (density 10$^6$\,cm$^{-3}$,
typical size 0.01\,pc) of ionized gas
emitting radio continuum radiation
of very high emission measure 
(L\,n$_{\rm e}^{2}$\,=\,10$^{10}$\,pc\,cm$^{-6}$)

\item Protostar\\
object on its way to become a star,
with more than half of its final mass
still to be accumulated

\item Infrared Dark Cloud (IRDC)\\
dense interstellar cloud (often filamentary) seen in absorption
against background thermal IR emission;
detected with {\it ISO}, {\it MSX}, and {\it Spitzer}

\item Core\\
basic molecular cloud unit (small dense gas fragment, 
mass 10\,--\,100\,M$_\odot$, size $\sim$\,0.1\,pc)
to form one or a few stars

\item Clump\\
basic molecular cloud unit (large dense gas fragment, mass 
$\sim$\,1000\,--\,5000\,M$_\odot$, size $\sim$\,0.5\,pc)
to form a young cluster of stars

\item Hot Molecular Core (HMC)\\
dense warm (T\,$>$\,100\,K) molecular gas
in emission (CO, NH$_3$, CH$_3$OH)
heated from a protostar inside;
a compact region (size $<$\,0.1\,pc)
in a star forming molecular clump such as the BN/KL region in Orion

\item Orion Nebula Cluster (ONC)\\
elongated 1\,$\times$\,2\,pc cluster of young stars,
centered on the high-mass Trapezium stars
in Orion; about 4000 cluster members with
a total stellar mass of $\sim$1800\,M$_{\odot}$

\item Trapezium Cluster\\
The 4\,--\,5 stars in the center of the Orion nebula,
$\theta^1$ C, A, B, D, E in order of decreasing luminosity,
the main exciting/ionising source being $\theta^1$ C (O5.5V, 
ca. 35\,M$_{\odot}$)

\item 30 Doradus (30 Dor)\\
prime example of a giant (200\,pc) extragalactic HII region,
powered by $\sim$\,100 massive O stars stars in the NGC 2070 cluster
(cluster mass some 10$^4$\,M$_{\odot}$, age $<$\,3\,Myr, 
cluster radius about 1\,arcmin or 15\,pc)

\item R136\\
nominally the 1\,arcsec core of the 30 Dor nebula and HII region
(Radcliffe object No. 136), but R136 also often designates
the central dense core of 1\,pc (4\,arcsec) radius of the exciting
star cluster NGC 2070

\item Jeans mass\\
critical mass that must be exceeded for a gas cloud to collapse
dynamically, with the self-gravity of the gas cloud being
opposed only by thermal pressure; if squeezed by an external
pressure, the critical mass for the onset of collapse is called
the Bonnor-Ebert mass. Other anisotropic forces such as rotation
or magnetic fields can oppose the cloud's self-gravity and thus
impede star formation.

\item Free-fall timescale\\
timescale it takes for a pressure-free cloud to collapse dynamically
to a very small size (nominally a point) under its own self-gravity;
the timescale depends on the average cloud
gas density only (for number density n\,=\,10$^5$\,cm$^{-3}$, 
t$_{\rm {ff}}$\,=\,10$^5$\,yr).

\item Kelvin-Helmholtz (KH) timescale\\
time it takes for a young stellar object or stellar core to radiate
its thermal energy content, typically 10$^4$\,yr for young high-mass
stars but 10$^7$\,yr for solar-type stars (t$_{\rm {KH}}$\,=\,GM$^2$/RL).
Nuclear energy generation eventually offsets the thermal energy loss.

\item Premain-Sequence evolution\\
While high-mass stars are born on the Main-Sequence (i.e. they
reach hydrogen burning while still accreting matter onto their
stellar surface during their deeply embedded phase),
the collapse of a low-mass star does not immediately lead to a
hydrogen-burning star, rather to a quasi-static object that keeps
contracting in radius by factors of a few before hydrogen burning
ignites in the center. During this contraction, the young stellar
object has a time-dependent radius, effective temperature, and
luminosity that can be calculated by pre-MS evolutionary models.

\item ZAMS\\
zero-age Main Sequence: a star that has contracted enough to start
hydrogen burning in its central region

\item Initial Mass Function (IMF)\\
mass distribution of (single) stars at birth
introduced by Salpeter in 1955

\end{itemize}

\newpage

\begin{itemize}

\item {\it Spitzer}\\
NASA mid-infrared space telescope of 85\,cm diameter launched in 2003

\item {\it MSX}\\
mid-infrared satellite explorer with an aperture of 20\,cm diameter

\item {\it Chandra}\\
0.5\,--\,10\,keV NASA X-ray telescope in space with particularly good
imaging quality comparable to optical seeing 
(angular resolution $\sim$\,1\,arcsec)

\item {\it HST}\\
{\it Hubble Space Telescope}, 2.4\,m aperture, 
workhorse for high angular resolution
space astronomy in UV, optical, and near-IR; in orbit since 1990 

\end{itemize}

\newpage

{\bf Acknowledgments}

The present review benefited greatly from
insightful discussions with John Bally, Henrik
Beuther, Adriaan Blaauw, Ian Bonnell,
Bernhard Brandl, Wolfgang Brandner, Anthony
Brown, Ed Churchwell, Paul Crowther, Melvyn
Davies, Lise Deharveng, Bruce Elmegreen,
Neal Evans, Simon Glover, Wolf-Rainer
Hamann, Thomas Henning, George Herbig,
David Hollenbach, Jim Jackson, Eric Keto,
Spyros Kitsionas, Ralf Klessen, Pavel Kroupa,
Mark Krumholz, Richard Larson, Hendrik Linz,
Andre Maeder, Eric Mamajek, Jorge Melnick,
Karl Menten, Vincent Minier, Tony Moffat,
Dieter N\"urnberger, Tolya Piskunov, Simon
Portegies Zwart, Thomas Preibisch, Gregor
Rauw, Hugues Sana, Fernando Selman, Joe
Silk, Nathan Smith, Bringfried Stecklum,
Jonathan Tan, Neal Turner, Virpi Niemela (deceased),
Nolan Walborn, and Gerd Weigelt. We thank
Geoffrey Burbidge and Allan Sandage for their
vote of confidence, and Ewine van Dishoeck
and particularly John Kormendy for their
critical reading of the manuscript.

Portions of this research was conducted at the Jet
Propulsion Laboratory, California Institute of
Technology, which is supported by the National
Aeronautics and Space Administration (NASA). The
assistance of U. Hanschur at AIP Potsdam was
indispensible in completing this review and is
gratefully acknowledged. H.Z. would like to thank
Andrea Lagarini and the Magalhaes family in Rio de
Janeiro and ESO in Santiago de Chile for their
hospitality during the very final stages of writing
this review. This collaboration has a long history,
dating back to 1990\,--\,1995 when the authors were
colocated at the University of W\"urzburg.

\newpage

\begin{figure}[ht]
\begin{center}
\includegraphics[width=12.0cm]{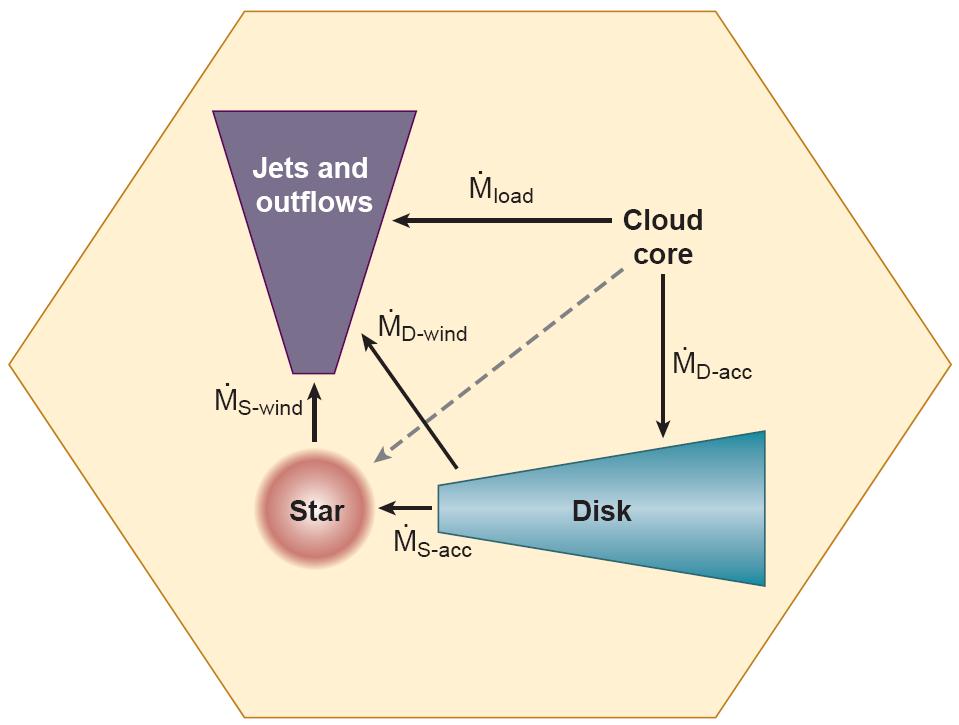}
\caption{
Accretion and mass loss as exchange
between components: the accretion disk as
reservoir and interface between the molecular
cloud core and the forming star.}
\end{center}
\end{figure}

\newpage

\begin{figure}[ht]
\begin{center}
\includegraphics[width=12.0cm]{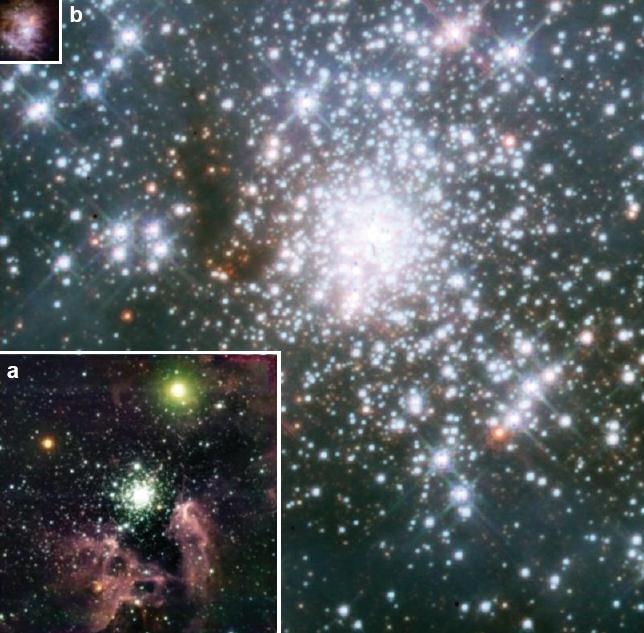}
\caption{
Hubble Space Telescope optical/IR
image of the dense massive young cluster
R136/30 Dor (courtesy of M.J. McCaughrean;
FOV  $\sim$\,30\,arcsec\,$\times$\,30\,arcsec or
7.5\,pc\,$\times$\,7.5\,pc).
Dozens of massive O stars are found crowded
within the half-light radius of 2\,pc (Brandl et al.
1996). ({\it a}) A VLT image of NGC 3603 (Brandl et
al. 1999) and ({\it b}) a VLT image of the Trapezium
Cluster in Orion (McCaughrean 2001) are shown,
as these two galactic clusters would be seen if
they were located at the distance of R136 in the
Large Magellanic Cloud (50\,kpc) and imaged
with similar angular resolution (see Zinnecker
2002).}
\end{center}
\end{figure}

\newpage

\begin{figure}[ht]
\begin{center}
\includegraphics[width=12.0cm]{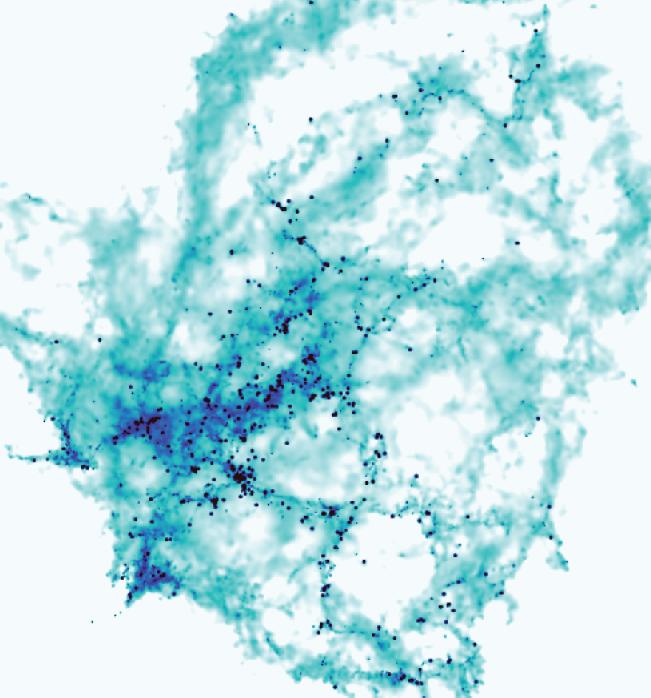}
\caption{Collapse and fragmentation of a giant
molecular cloud, simulated with
smoothed particle hydrodynamics (SPH) using sink
particles (I.A. Bonnell, P.C. Clark \& H. Zinnecker, in
preparation). For a description of SPH, see the
box. Plotted is the spatial  distribution of gas
column density, color-coded such that deep blue
refers to the highest values. The initial conditions
for this three-dimensional simulation included a
molecular gas cloud of mass  10$^6$\,M$_\odot$ and
diameter 100\,pc, somewhat centrally condensed
(factor of 20, with a Gaussian radial profile). The
cloud's turbulent kinetic energy was equal to its
gravitational energy. Note the filamentary
structure and the associated dense cores (blue
dots). These cores technically are represented by
sink particles ($\sim$2400 at this stage of evolution). With
typical masses of 10 to 100\,M$_\odot$ and typical sizes
of 0.1\,pc they could be the initial fragments for
high-mass star formation. A total of 2.5 million
SPH particles was used. Some similarity to
cosmological simulations of structure formation
is noted.}
\end{center}
\end{figure}

\newpage

\begin{figure}[ht]
\begin{center}
\includegraphics[width=12.0cm]{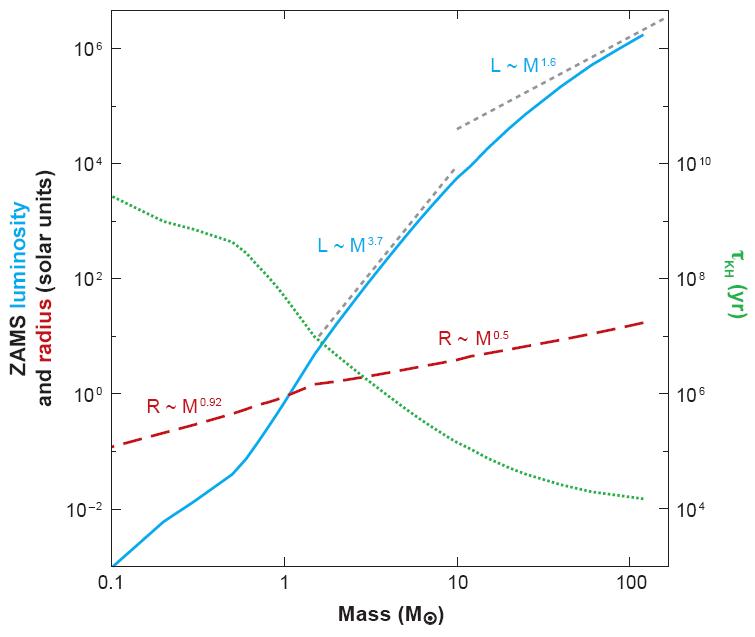}
\caption{
ZAMS luminosity (blue), ZAMS radius
(red dashed), and Kelvin-Helmholtz
quasi-static gravitational contraction
timescale toward the ZAMS
(green dotted) as a function of stellar mass
(ZAMS means zero-age main sequence). These
values are extracted from t\,=\,0\,yr models given
by Meynet \&
Maeder (2005) for rotating stars M$_*$/M$_\odot$\,$\ge$\,12
with mass loss, by Pietrinferni et
al. (2004) for stars 0.5\,$\le$\,M$_*$/M$_\odot$\,$\le$\,10, and for
completeness from unpublished
tracks by Yorke for masses M$_*$\,=\,0.1, 0.2, and 0.3\,M$_\odot$,
using the computer code provided in the
book by Bodenheimer et al. (2007). ZAMS radii
are well represented by two power laws with a
break at M$_*$/M$_\odot$\,=\,1.5. The power law slope of
the luminosity-mass relation for massive stars
varies from about 3.7 to 1.6 in the range
8\,$<$\,M$_*$/M$_\odot$\,$<$\,120.}
\end{center}
\end{figure}

\newpage

\begin{figure}[ht]
\begin{center}
\includegraphics[width=12.0cm]{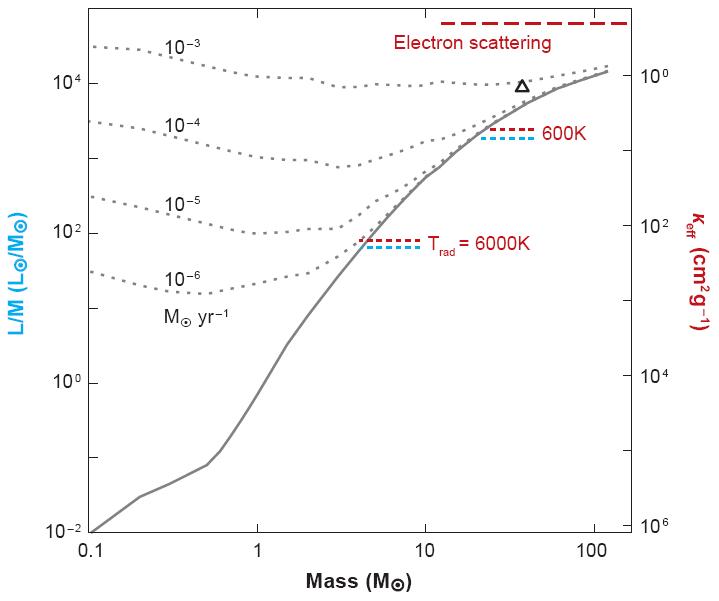}
\caption{
ZAMS luminosity to mass ratio (left
scale) and corresponding critical
effective opacity (right scale) as defined by
{\bf Equation~3} as a function of stellar
mass. The solid grey line depicts the ZAMS
models shown in {\bf Figure~4}. Dotted grey lines
indicate the total luminosity (including accretion
luminosity) of stars accreting at the indicated
constant rate as discussed in the text (cf. {\bf Figure~8}).
The triangle denotes the position of an O5V
star (see {\bf Figure~9}). Dashed lines denote the
opacities of dusty gas in the light of black body
sources at the temperature
indicated (for two different grain types, see
{\bf Figure~6}) and the contribution from electron
scattering in a fully ionized plasma. Because of
the combined effect of UV lines (UV line
blanketing), the actual opacity of a hot plasma
can be greater than that from electron scattering
alone. This may be relevant for the existence of
an upper stellar mass limit.}
\end{center}
\end{figure}

\newpage

\begin{figure}[ht]
\begin{center}
\includegraphics[width=12.0cm]{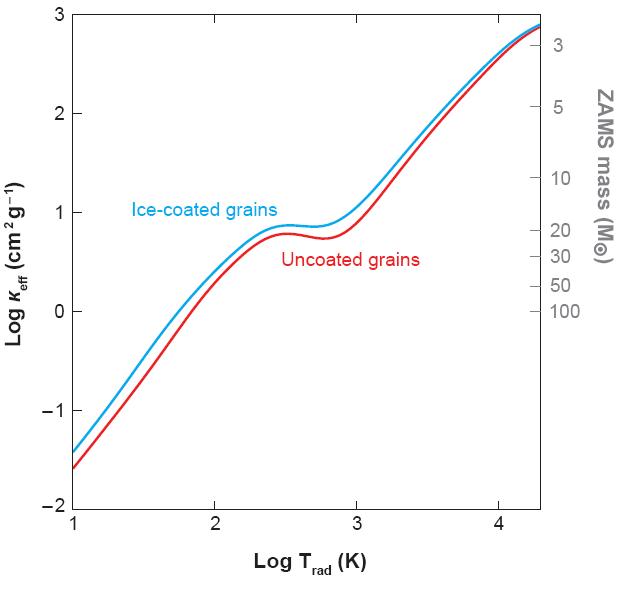}
\caption{
Planck-weighted mean effective opacity
of dusty gas, using the
(Preibisch et al. 1993) dust model with ice-coated
grains (blue curve) and grains
without ice mantels (red curve), assuming black-body
radiation at the temperature T$_{\rm rad}$ and solar
abundances. The right-hand scale (grey) is based
on the ZAMS models shown in {\bf Figure~5} (solid
grey line).}
\end{center}
\end{figure}

\newpage

\begin{figure}[ht]
\begin{center}
\includegraphics[width=12.0cm]{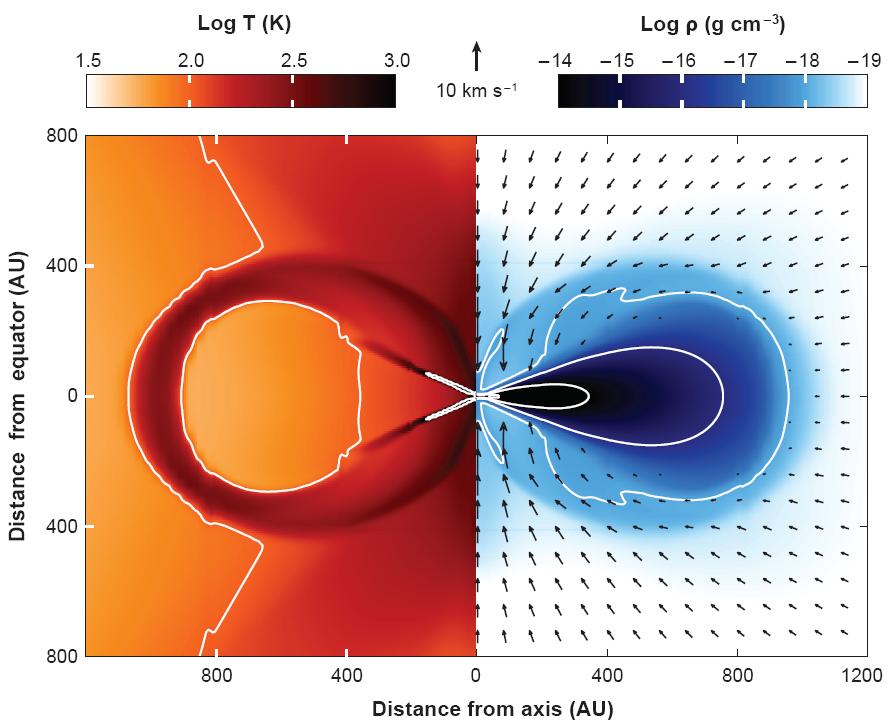}
\caption{Edge-on cut through an accreting
protostar and its circumstellar disk
and envelope. The protostar is located at (0,0)
and is too tiny to see. Temperature
(red scale, upper-left), density (blue scale, upper-right),
and velocity
(length and direction of arrows) distributions of
the accreting material are displayed (case F of
Yorke \& Bodenheimer 1999, recalculated using
an improved
version of their code). White contour lines are
plotted for $\log$\,T\,=\,2.0, 2.5, and
3.0 and for $\log \rho$\,=\,-18, -16, -14, and -12. At this
evolutionary time t = 65,000 yr
after formation of the protostellar core, 7.0\,M$_\odot$
of material have accreted onto
the protostar, 2.8\,M$_\odot$ are in the disk, and 0.2\,M$_\odot$
are in the infalling envelope.}
\end{center}
\end{figure}

\newpage

\begin{figure}[ht]
\begin{center}
\includegraphics[width=12.0cm]{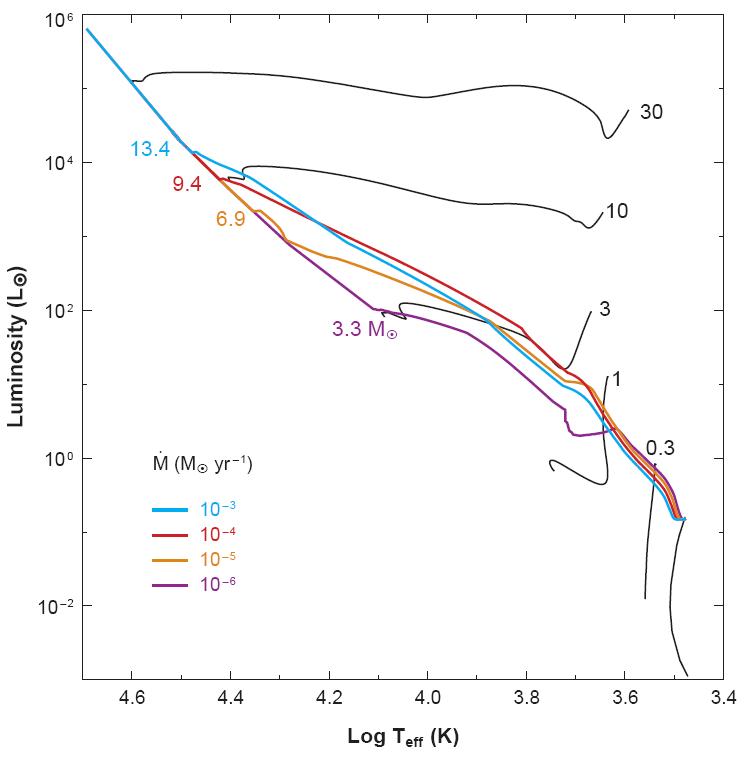}
\caption{
Evolutionary tracks in the HR diagram
for (proto-)stars accreting at a
constant rate (colored lines) are contrasted to the
tracks of non-accreting stars
(black lines). All accreting tracks are assumed to
begin at the birthline of an
equilibrium deuterium burning 0.1\,M$_\odot$ pre-main-sequence
star. Non-accreting
tracks up to H burning were calculated by Yorke
for M$_*$/M$_\odot$\,=\,0.1, 0.3, 1, 3, 10,
and 30, using the computer code supplied with
the book by Bodenheimer et al.
(2007). [Adapted from Yorke (2002)]}
\end{center}
\end{figure}

\newpage

\begin{figure}[ht]
\begin{center}
\includegraphics[width=12.0cm]{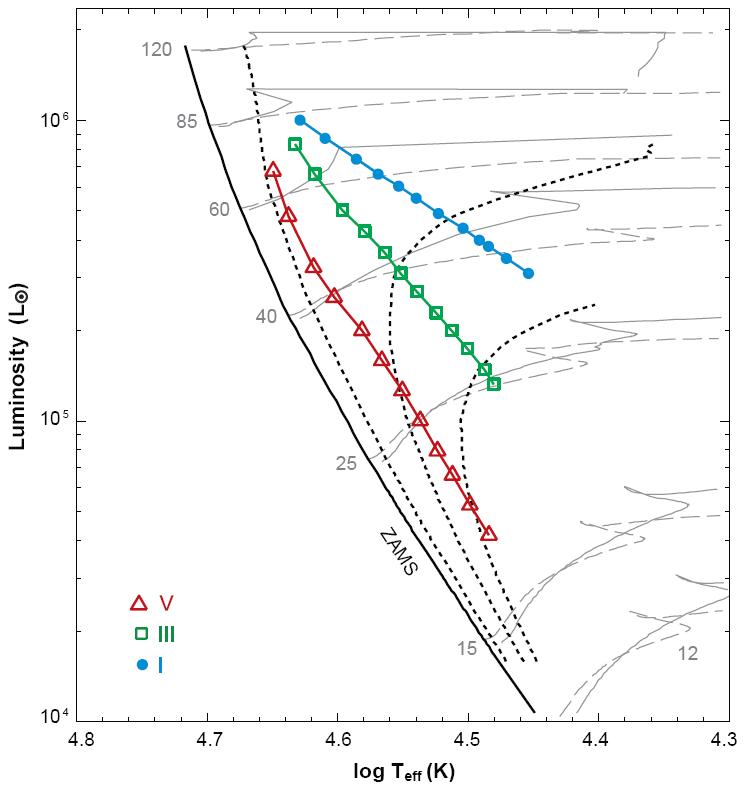}
\caption{
Luminosity and effective temperature
calibration of luminosity classes
I (blue-filled circles), III (green squares), and V
(red triangles) for stars of spectral classes O3,
O4, O5, O5.5, O6, O6.5, O7, O7.5, O8, O8.5, O9,
and O9.5. For comparison, theoretical
evolutionary tracks from Meynet \& Maeder
(2003) for non-rotating (dashed grey) and for
rotating stars (continuous grey) with mass loss
are labeled by their starting ZAMS mass. Meynet
et al. (1994) isochrones for non-rotating stars are
also plotted for 0 (solid black line labeled ZAMS),
1, 3, and 5\,Myr (dotted black lines). [Adapted
from Figure 14 of Martins, Schaerer \& Hillier
(2005)]}
\end{center}
\end{figure}

\newpage

\begin{figure}[ht]
\begin{center}
\includegraphics[width=12.0cm]{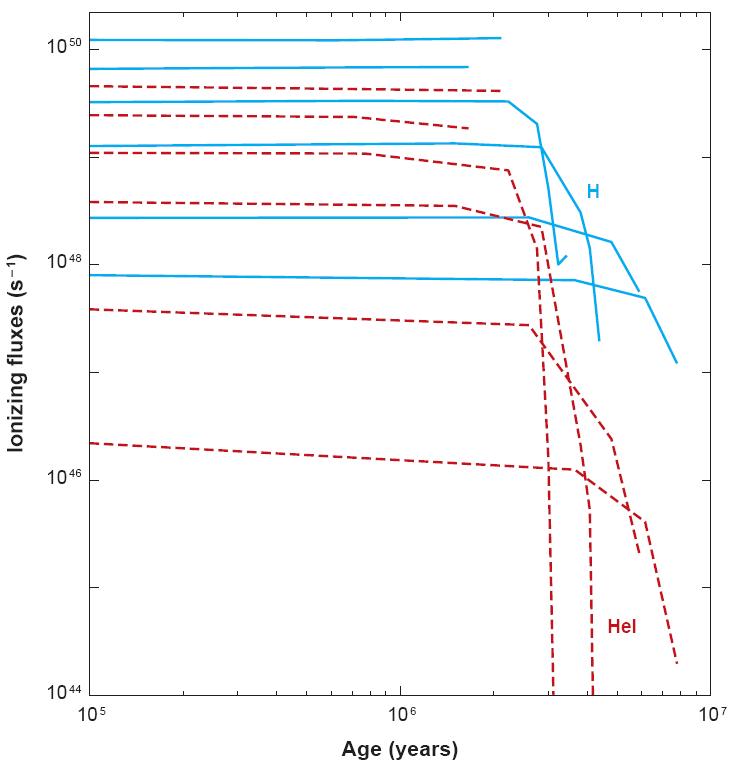}
\caption{
Flux of radiation that can ionize
hydrogen (blue lines) and neutral
helium (red dashed lines) as a function of age
(counted from the moment of
arrival on the ZAMS), for stars of ZAMS masses
(top to bottom) of M$_*$/M$_\odot$\,=\,120, 85, 60, 40, 25,
and 20. Mass loss is included in the evolution.
[Based on models from Schaerer \& de Koter
(1997)]}
\end{center}
\end{figure}

\newpage

\begin{figure}[ht]
\begin{center}
\includegraphics[width=12.0cm]{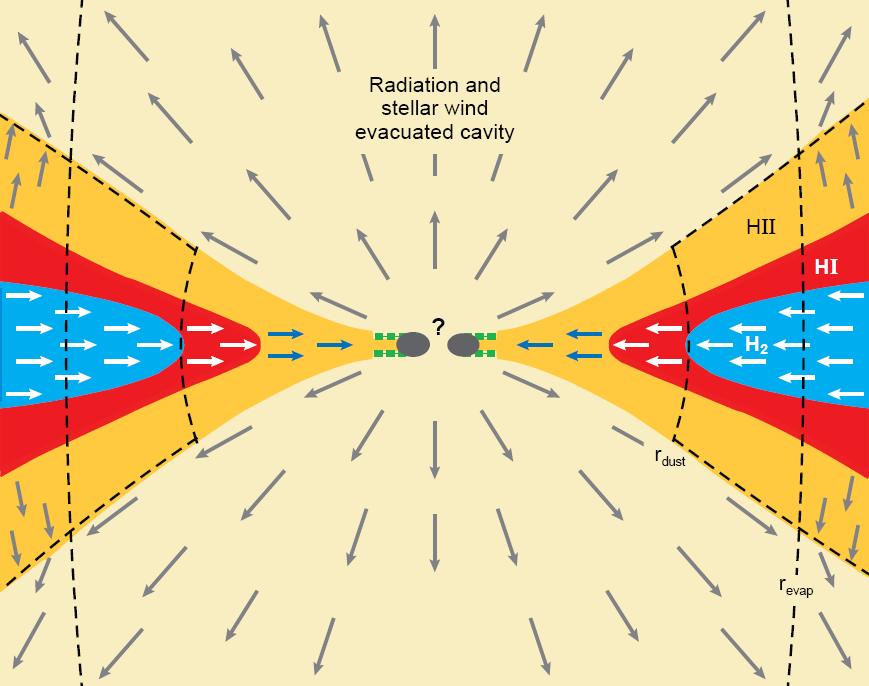}
\caption{The inner accretion disk around a
close massive accreting binary pair: inward
radial flow is allowed in the equatorial plane. A
polar cavity is evacuated by a combination of
radiation and the stellar wind. The disk is self-shielded
from the intense EUV  field by an
ionization front separating HII and HI gas; and
from the FUV  field interior to the HI/H$_2$
interface by dust, molecular hydrogen, and CO.
The dust is destroyed at r$_{\rm dust}$.
Interior to r$_{\rm evap}$, the
radius where the sound speed exceeds the escape
velocity, even the ionized gas is gravitationally
bound. Sizes are not to scale.}
\end{center}
\end{figure}

\newpage

\begin{figure}[ht]
\begin{center}
\includegraphics[width=12.0cm]{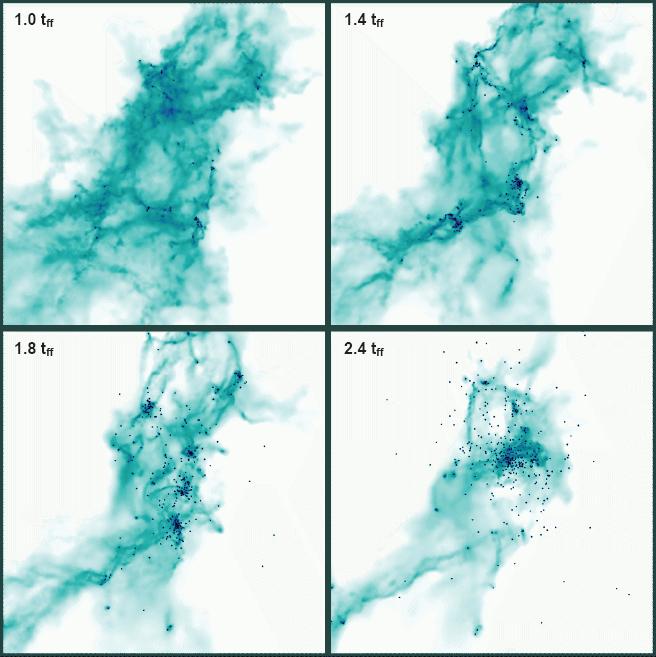}
\caption{
Time-dependent evolution of a
turbulent, self-gravitating 1000-M$_\odot$
cloud showing the formation of sheets, filaments,
and cores, the latter of which
become gravitationally unstable. Newly formed
stars are shown in dark blue,
the gas is shown in blue-green. Stars tend to
cluster, they continue to accrete
material in competition with other stars. Times
shown are 1.0, 1.4, 1.8, and
2.4 initial free-fall times (t$_{\rm ff}$), from left to right
and top to bottom. [Adapted from
Bonnell, Bate \& Vine (2003)].}
\end{center}
\end{figure}

\newpage

\begin{figure}[ht]
\begin{center}
\includegraphics[width=12.0cm]{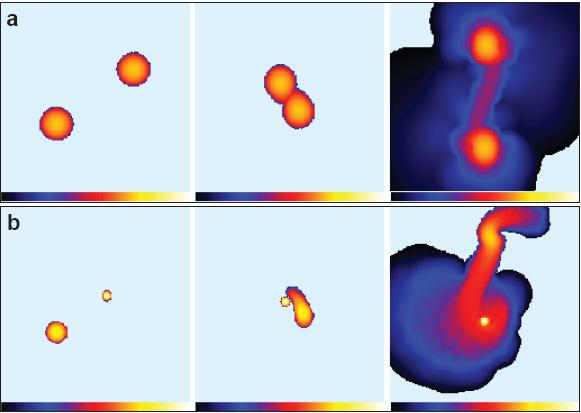}
\caption{({\it a}) A grazing encounter between two 3\,M$_\odot$
premain-sequence stars
that results in the formation of a binary
(r$_{\rm min}$\,=\,25.8 solar radii).
({\it b}) A
detached encounter between a 3\,M$_\odot$ premain-sequence
star and a 10\,M$_\odot$ zero age main-sequence
star with the same minimum periastron
distance as in the (a)
encounter. The stars have radii of 12.9 and 3.92
solar radii, respectively. The
greater density of the 10\,M$_\odot$ star results in tidal
disruption of the low-density
3\,M$_\odot$ star to form a disk around the massive
star. The encounters have zero
relative velocity at infinity (i.e., they are
parabolic encounters). [ Adapted from
Zinnecker \& Bate (2002), Davies et al. (2006)].
The cross section for a subsequent
collision with (say) another 10\,M$_\odot$ star is
significantly increased, and a runaway
collisional growth in mass is possible. This is the
so-called 'shred and add' process.}
\end{center}
\end{figure}

\newpage

\begin{figure}[ht]
\begin{center}
\includegraphics[width=12.0cm]{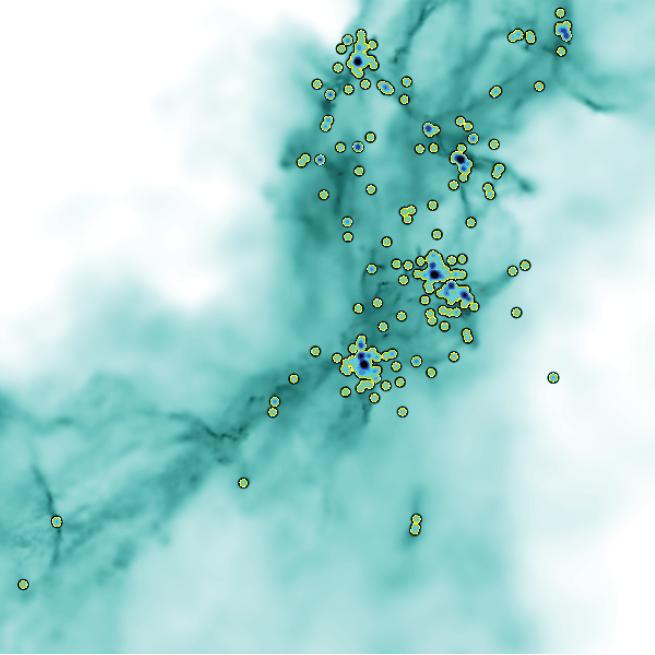}
\caption{
Massive stars (dark blue circles) are
formed in the center of individual subclusters of
low-mass stars (light circles) because of
competitive accretion. These subclusters evolve
by merging with the other subclusters. The  final
state of the simulation is a single, centrally
condensed cluster with little substructure but
with 4 massive stars, one from each subcluster.
This then is a model for the origin of Trapezium-type
systems. [Adapted from Bonnell, Vine \&
Bate (2004)]}
\end{center}
\end{figure}

\newpage

\begin{figure}[ht]
\begin{center}
\includegraphics[width=12.0cm]{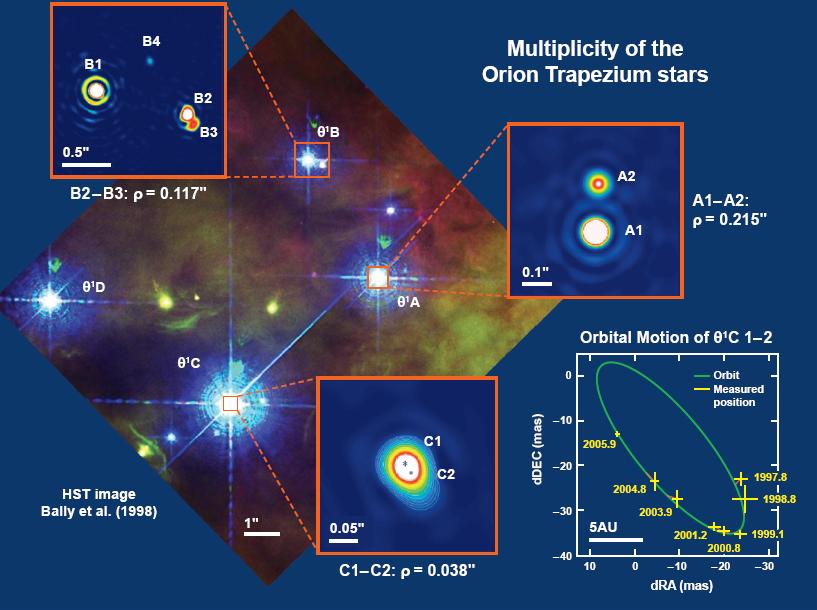}
\caption{
The multiple star systems of the Orion
Trapezium as revealed by bispectrum speckle
interferometry. [Courtesy of G. Weigelt \& Th.
Preibisch; inserts adapted from Schertl et al.
(2003) and Kraus et al. (2007)]}
\end{center}
\end{figure}

\newpage

\begin{figure}[ht]
\begin{center}
\includegraphics[width=12.0cm]{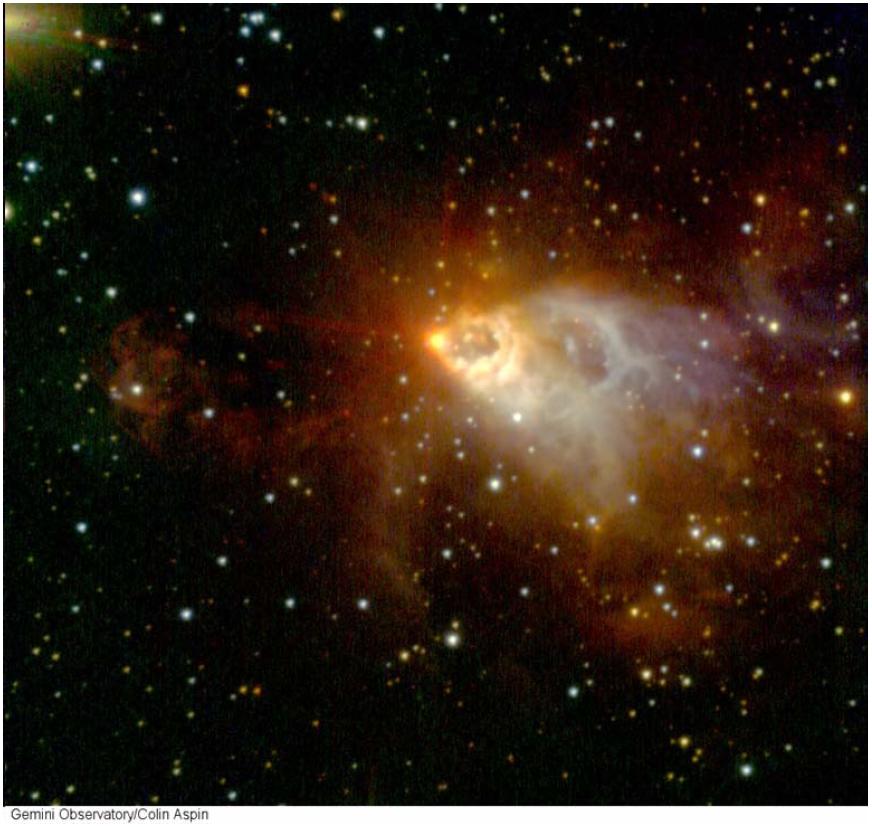}
\caption{
A wide-field JHKs composite
image of the AFGL 2591 massive outflow
taken with the NIRI camera at the Gemini
North telescope in excellent seeing (courtesy
of C. Aspin and Gemini Observatory; FOV $\sim$\,2\,arcmin,
seeing 0.35\,arcsec).The broad-band
Ks filter includes the ro-vibrational
molecular hydrogen v\,=\,1\,--\,0 S(1) emission line
at 2.12\,$\mu$m, which is indicative of shock-excited
gas. However, much of the nebulosity is
likely seen in reflection because it is also
detected in J and H. Note the multiple
poorly-collimated flows, loops, and cavities
emanating to the west from this massive (ca.
20\,M$_\odot$) star (see also the speckle image in
Preibisch et al. 2003 from the 6\,m SAO
telescope). Nothing of this kind is seen to the
east of the source, except for a few faint bow-shocks,
as the (redshifted) counterflow is
deeply embedded and probably partly
obscured by a dense circumstellar disk
(Trinidad et al. 2003) around this very young
star. The fact that the flow and counterflow
both reach the same distance of $\sim$\,40\,arcsec
(0.2\,pc) from the central source indicates
that this is intrinsically a bipolar outflow,
confirmed by high-velocity CO radio
observations (Lada et al. 1984; see also the
H$_2$ and optical spectroscopic observations of
the associated Herbig-Haro objects by
Tamura \& Yamashita 1992 and Poetzel,
Mundt \& Ray 1992, respectively). It seems
that the bipolar outflow from this late O-type
massive star has both a well-collimated
and a wide-angle component!}
\end{center}
\end{figure}

\newpage

\begin{figure}[ht]
\begin{center}
\includegraphics[width=12.0cm]{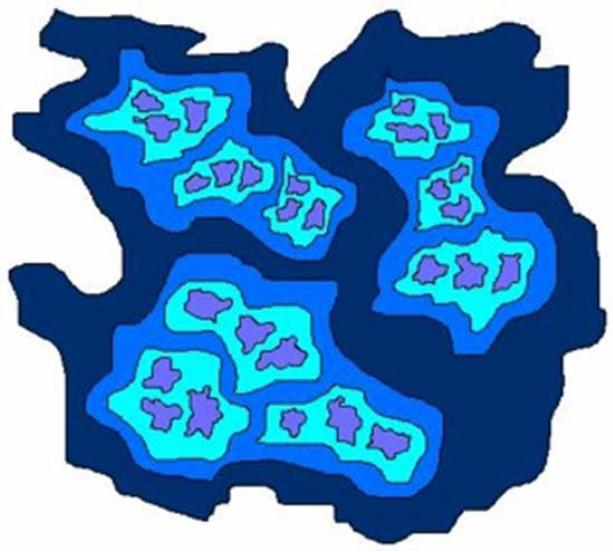}
\caption{
Schematic representation of a
self-similar fractal hierarchy of dense gas
cores in a turbulent molecular cloud with
three branches per level (courtesy J.
Melnick). The chance of randomly picking a
core of mass M in a cloud with such
hierarchical structure is proportional to 1/M.
Because the cloud branches correspond to
log intervals, random sampling yields a mass
spectrum of (prestellar) cores with equal
mass in equal logarithmic intervals, i.e., a
power-law dN/dlogM with index -1. If there
is competition for mass, i.e., if some smaller
(and presumably denser) cores are turned
into stars before the bigger cores, of which
they are part, have time to collapse, then
there will no longer be enough mass
available to form a star with the mass of the
undiminished larger core. Thus, high-mass
cores (stars) get depleted from the mass
function if low-mass cores (stars) form first.
The faster formation of low-mass stars,
combined with the mass depletion of
subparts inside clouds, converts the power-law
mass spectrum with a power index of -1
into one with an index of -1.35 (or so), i.e.,
into a Salpeter initial mass function. Note
that in this reasoning (following Elmegreen
1997) it was assumed that a fixed fraction of
each and every core mass is turned into
stellar mass (single, binary, or multiple).}
\end{center}
\end{figure}

\newpage

\begin{figure}[ht]
\begin{center}
\includegraphics[width=12.0cm]{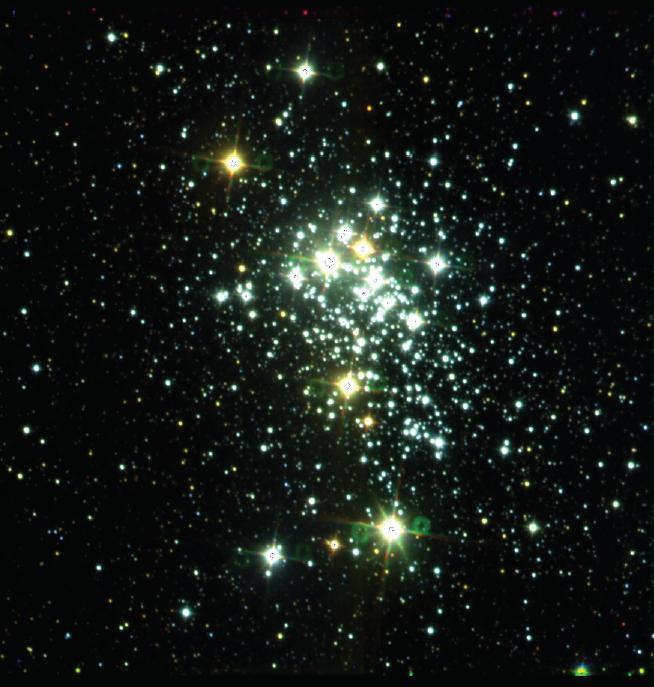}
\caption{
Composite colour JHKs image of the
$\sim$\,4\,Myr old Galactic starburst
cluster Westerlund 1 (Wd 1) in Cyg OB2
obtained with the ESO NTT telescope (courtesy
W. Brandner). This obscured object is held to be
the most massive young cluster in the Galaxy,
with ca. 100 O-type stars and a mass of at least
20,000\,M$_\odot$ (see Kn\"odelseder 2000, Clark et al.
2005, Mengel \& Tacconi-Garman 2007, Brandner
et al. 2007). A number of massive eclipsing binary
systems have been identified (Bonanos 2007) as
well as a very massive runaway star, an Of-type
supergiant, whose origin can be traced back to
the cluster (Comer\'{o}n \& Pasquali 2007). Like the
Orion Nebula Cluster, or the NGC 3603 and 30 Dor
central clusters, Wd 1 is strongly mass segregated
with the high mass stars being
more centrally concentrated than the low mass
stars. The field of view is 4'\,$\times$\,4',
corresponding to 4.2\,pc$\times$\,4.2\,pc at the distance of
Wd 1.}
\end{center}
\end{figure}


\newpage

\begin{thebibliography}{}

\bibitem{}
	Aarseth SJ. 2003. {\it Gravitational N-Body 
	Simulations.} Cambridge, UK: Cambridge Univ. 
	Press

\bibitem{}
	Abel T, Bryan GL, Norman ML. 2000. {\it Ap. J.} 
	540:39--44

\bibitem{}
	Adams FC, Fatuzzo M. 1996.  {\it Ap. J.} 464:256--71

\bibitem{}
	Allen DA, Burton MG. 1993. {\it Nature.} 363:54--56

\bibitem{}
	Allen C, Poveda A, Hern\'{a}ndez-Alc\'{a}ntara A. 2004. 
	{\it Rev. Mex. Astron. Astrofis. Ser. Conf.} 21:195--99

\bibitem{}
	Alves J, Lombardi M, Lada CJ. 2007. {\it Astron. 
	Astrophys.Lett.} 462:L17--21

\bibitem{}
	Antokhina EA, Moffat AFJ, Antokhin II, Bertrand J-F,
	Lamontagne R. 2000. {\it Ap. J.} 529:463--76

\bibitem{}
	Apai D, Bik A, Kaper L, Henning T, Zinnecker H. 
	2007. {\it Ap. J.} 655:484--91

\bibitem{}
	Appenzeller I. 1970. {\it Astron. Astrophys.} 9:216--20

\bibitem{}
	Arce HG, Shepherd D, Gueth F, Lee C-F, Bachiller 
	R, et al. 2007. See Reipurth et al. 2007, pp. 245--60

\bibitem{}
	Balbus SA. 2003. {\it Annu. Rev. Astron. Astrophys.} 
	41:555--97

\bibitem{}
	Balbus SA, Hawley JF. 1991. {\it Ap. J.} 376:214--22

\bibitem{}
	Baldry IK, Glazebrook K. 2003. {\it Ap. J.} 593:258--71

\bibitem{}
	Ballesteros-Paredes J, Hartmann L, V\'{a}zquez-Semadeni
	E. 1999. {\it Ap. J.} 527:285--97

\bibitem{}
	Bally J. 2002. See Crowther 2002, pp. 219--33

\bibitem{}
	Bally J, Moeckel N, Throop H. 2005. In {\it Chondrites 
	and the Protoplanetary Disk,} ed. AN Krot, ERD 
	Scott, B Reipurth. {\it ASP Conf. Ser.} 341:81--106

\bibitem{}
	Bally J, Cunningham N, Moeckel N, Smith N. 2005. 
	See Cesaroni et al. 2005a, pp. 12--22

\bibitem{}
	Bally J, Zinnecker H. 2005. {\it Astron. J.} 129:2281--93

\bibitem{}
	Banerjee R, Pudritz RE, Anderson DW. 2006. 
	{\it MNRAS} 373:1091--06

\bibitem{}
	Banerjee R, Pudritz RE. 2007. {\it Ap. J.} 660:479--88

\bibitem{}
	Baraffe I, Chabrier G, Allard F, Hauschildt PH. 
	2002. {\it Astron. Astrophys.} 382:563--72

\bibitem{}
	Barrado y Navascu\'{e}s D. 2005. {\it Rev. Mex. Astron. 
	Astrofis. Ser. Conf.} 24:217--18

\bibitem{}
	Barsony M, Schombert J, Kis-Halas K. 1991. {\it Ap. J.} 
	379:221--31

\bibitem{}
	Bastian N, Goodwin SP. 2006. {\it MNRAS Lett.} 
	369:L9--13

\bibitem{}
	Basu S, Jones CE. 2004. {\it MNRAS Lett.} 347:L47--51

\bibitem{}
	Bate MR. 2000. {\it MNRAS} 314:33--53

\bibitem{}
	Bate MR, Bonnell IA, Price NM. 1995. {\it MNRAS} 
	277:362--76

\bibitem{}
	Bate MR, Burkert A. 1997. {\it MNRAS} 288:1060--72

\bibitem{}
	Bate MR, Bonnell IA. 1997. {\it MNRAS} 285:33--48

\bibitem{}
	Bate MR, Bonnell IA, Bromm V. 2002. {\it MNRAS} 
	336:705--13

\bibitem{}
	Bate MR, Bonnell IA. 2005. {\it MNRAS} 356:1201--21

\bibitem{}
	Begelman MC. 2001. {\it Ap. J.} 551:897--906

\bibitem{}
	Behrend R, Maeder A. 2001. {\it Astron. Astrophys.} 
	373:190--98

\bibitem{}
	Belkus H, Van Bever J, Vanbeveren D. 2007. {\it Ap. J.} 
	659:1576--81

\bibitem{}
	Benjamin RA, Churchwell E, Babler BL, Bania TM, 
	Clemens DP, et al. 2003. {\it Publ. Astron. Soc. Pac.} 
	115:953--64

\bibitem{}
	Bertoldi F, McKee CF. 1992. {\it Ap. J.} 395:140--57

\bibitem{}
	Beuther H, Schilke P, Sridharan TK, Menten KM, 
	Walmsley CM, Wyrowski F.  2002. {\it Astron. 
	Astrophys.} 383:892--904

\bibitem{}
	Beuther H, Shepherd DS. 2005. In {\it Cores to 
	Clusters: Star Formation with Next Generation 
	Telescopes,} ed. MS Nanda Kumar, M Tafalla, P 
	Caselli. {\it Astrophys. Space Sci. Libr.} 324:105--19

\bibitem{}
	Beuther H, Sridharan TK, Saito M. 2005. {\it Ap. J. Lett.} 
	634:L185--88

\bibitem{}
	Beuther H, Churchwell E, McKee CF, Tan JC. 2007. 
	See Reipurth et al. 2007, pp. 165--80

\bibitem{}
	Bik A, Lenorzer A, Kaper L, Comer\'{o}n F, Waters 
	LBFM. 2003. {\it Astron. Astrophys.} 404:249--54

\bibitem{}
	Bik A, Kaper L, Hanson MM, Smits M. 2005. 
	{\it Astron. Astrophys.} 440:121--37

\bibitem{}
	Binney J, Tremaine S. 1987. {\it Galactic Dynamics.} 
	Princeton, NJ: Princeton Univ. Press. 755 pp.

\bibitem{}
	Blaauw A. 1961. {\it Bull. Astron. Inst. Neth.} 15:265--90

\bibitem{}
	Blaauw A. 1964. {\it Annu. Rev. Astron. Astrophys.} 
	2:213--46

\bibitem{}
	Blaauw A. 1983. {\it Irish Astron. J.} 16:141--47

\bibitem{}
	Blaauw A. 1991. In {\it The Physics of Star Formation 
	and Early Stellar Evolution,} ed. CJ Lada, ND 
	Kylafis. {\it NATO ASI Ser. C,} 342:125--54 
	Dordrecht: Kluwer

\bibitem{}
	Black DC, Bodenheimer P. 1975. {\it Ap. J.} 199:619--32

\bibitem{}
	Blaes O, Socrates A. 2003. {\it Ap. J.} 595:509--37

\bibitem{}
	Blandford RD, Payne DG. 1982. {\it MNRAS} 199:883--903

\bibitem{}
	Blum RD. 2005. See Cesaroni et al. 2005a, pp. 216--24

\bibitem{}
	Bodenheimer P. 1995. {\it Annu. Rev. Astron. Astrophys.} 
	33:199--238

\bibitem{}
	Bodenheimer P, Laughlin GP, R\'{o}\.{z}yczka M, Yorke 
	HW. 2007. {\it Numerical Methods in Astrophysics. 
	An Introduction.} New York: Taylor \& Francis 

\bibitem{}
	Boffin HMJ, Watkins SJ, Bhattal AS, Francis N, 
	Whitworth AP. 1998. {\it MNRAS} 300:1189--204

\bibitem{}
	Bonanos AZ. 2007. {\it Astron. J.} 133:2696--708

\bibitem{}
	Bonanos AZ, Stanek KZ, Udalski A, 
	Wyrzykowski L, \.{Z}ebru\'{n} K, et al. 2004. {\it Ap. J. 
	Lett.} 611:L33--36 

\bibitem{}
	Bonnell I, Bastien P. 1992. {\it Ap. J.} 401:654--66

\bibitem{}
	Bonnell IA, Bate MR, Clarke CJ, Pringle JE. 1997. 
	{\it MNRAS} 285:201--08

\bibitem{}
	Bonnell IA, Bate MR, Zinnecker H. 1998. {\it MNRAS} 
	298:93--102

\bibitem{}
	Bonnell IA, Davies MB. 1998. {\it MNRAS} 295:691--98

\bibitem{}
	Bonnell IA, Bate MR, Clarke CJ, Pringle JE. 2001a. 
	{\it MNRAS} 323:785--94

\bibitem{}
	Bonnell IA, Clarke CJ, Bate MR, Pringle JE. 2001b. 
	{\it MNRAS} 324:573--79

\bibitem{}
	Bonnell IA, Bate MR. 2002. {\it MNRAS} 336:659--69

\bibitem{}
	Bonnell IA, Bate MR, Vine SG. 2003. {\it MNRAS} 
	343:413--18

\bibitem{}
	Bonnell IA, Vine SG, Bate MR. 2004. {\it MNRAS} 
	349:735--41

\bibitem{}
	Bonnell IA, Bate MR. 2005. {\it MNRAS} 362:915--20

\bibitem{}
	Bonnell IA, Bate MR. 2006. {\it MNRAS} 370:488--94

\bibitem{}
	Bonnell IA, Larson RB, Zinnecker H. 2007. See 
	Reipurth et al. 2007, pp. 149--64

\bibitem{}
	Bosch G, Selman F, Melnick J, Terlevich R. 2001. 
	{\it Astron. Astrophys.} 380:137--41

\bibitem{}
	Boss AP, Yorke HW. 1995. {\it Ap. J. Lett.} 439:L55--58

\bibitem{}
	Brandl B, Sams BJ, Bertoldi F, Eckart A, Genzel R, 
	et al. 1996. {\it Ap. J.} 466:254--73

\bibitem{}
	Brandl B, Brandner W, Eisenhauer F, Moffat AFJ, 
	Palla F, Zinnecker H.  1999. {\it Astron. Astrophys. 
	Lett.} 352:L69--72

\bibitem{}
	Brandl BR, Townsley LK, Churchwell E, Carey S, 
	Zinnecker H, et al. 2005. See Cesaroni et al. 
	2005a, pp. 311--17

\bibitem{}
	Brandner W, Grebel EK, Barb\'{a} RH, Walborn NR, 
	Moneti A. 2001. {\it Astron. J.} 122:858--65

\bibitem{}
	Brandner W, Clark JS, Stolte A, Waters R, 
	Neguerela I, Goodwin SP. 2007. {\it Astron. 
	Astrophys.} In press

\bibitem{}
	Brice\~{n}o C, Preibisch T, Sherry WH, Mamajek EA, 
	Mathieu RD, et al. 2007. See Reipurth et al. 
	2007, pp. 345--60

\bibitem{}
	Bromm V, Larson RB. 2004. {\it Annu. Rev. Astron. 
	Astrophys.} 42:79--118

\bibitem{}
	Bronfman L, Casassus S, May J, Nyman L-\r{A}. 2000. 
	{\it Astron. Astrophys.} 358:521--34

\bibitem{}
	Brown A. 2001. {\it Astron. Nachr.} 322:43--46 

\bibitem{}
	Burrows CJ, Stapelfeldt KR, Watson AM, Krist JE, 
	Ballester GE, et al. 1996. {\it Ap. J.} 473:437--51

\bibitem{}
	Burton MG, Hill T, Longmore SN, Purcell CR, 
	Walsh AJ. 2005. See Cesaroni et al. 2005a, pp. 
	157--62

\bibitem{}
	Camenzind M. 1990. {\it Rev. Mod. Astron.} 3:234--65

\bibitem{}
	Carpenter JM, Snell RL, Schloerb FP, Skrutskie MF. 
	1993. {\it Ap. J.} 407:657--79

\bibitem{}
	Carpenter JM, Meyer MR, Dougados C, Strom SE, 
	Hillenbrand LA. 1997.  {\it Astron. J.} 114:198--221

\bibitem{}
	Cassinelli JP, Mathis JS, Savage, BD. 1981. {\it Science} 
	212:1497--501

\bibitem{}
	Cassinelli JP, Churchwell EB, ed. 1993. {\it Massive 
	Stars: Their Lives in the Interstellar Medium, 
	ASP Conf. Ser.,} Vol. 35. San Francisco: ASP 

\bibitem{}
	Cesaroni R. 2005. {\it Ap. Space Sci.} 295:5--17

\bibitem{}
	Cesaroni R, Felli M, Testi L, Walmsley CM, Olmi 
	L. 1997. {\it Astron. Astrophys.} 325:725--44

\bibitem{}
	Cesaroni R, Felli M, Jenness T, Neri R, Olmi L, et 
	al. 1999. {\it Astron. Astrophys.} 345:949--64

\bibitem{}
	Cesaroni R, Felli M, Churchwell E, Walmsley M, 
	ed. 2005a. {\it Massive Star Birth: A Crossroads of 
	Astrophysics, IAU Symp. 227.} Cambridge: 
	Cambridge University Press 

\bibitem{}
	Cesaroni R, Neri R, Olmi L, Testi L, Walmsley CM, 
	Hofner P. 2005b. {\it Astron. Astrophys.} 434:1039--54

\bibitem{}
	Cesaroni R, Galli D, Lodato G, Walmsley CM, 
	Zhang Q. 2007. See Reipurth et al. 2007, pp. 
	197--212

\bibitem{}
	Chabrier G. 2003. {\it Publ. Astron. Soc. Pac.} 115:763--95

\bibitem{}
	Chini R, Hoffmeister V. Kimeswenger S, 
	Nielbock M, N\"urnberger D, et al. 2004. {\it Nature} 
	429:155--57

\bibitem{}
	Churchwell E. 2002. {\it Annu. Rev. Astron. Astrophys.} 
	40:27--62

\bibitem{}
	Churchwell E, Povich MS, Allen D, Taylor MG, 
	Meade MR, et al. 2006. {\it Ap. J.} 649:759--78

\bibitem{}
	Clark JS, Negueruela I, Crowther PA, Goodwin SP. 
	2005. {\it Astron. Astrophys.} 434:949--69

\bibitem{}
	Clark PC, Bonnell IA. 2005. {\it MNRAS} 361:2--16
	
\bibitem{}
	Clark PC, Bonnell IA, Zinnecker H. 2005. {\it MNRAS} 
	359:809--18

\bibitem{}
	Clarke CJ, Pringle JE. 1992. {\it MNRAS} 255:423--30

\bibitem{}
	Clarke CJ, Edgar RG, Dale JE. 2005. See Corbelli et 
	al. 2005, pp. 449--54

\bibitem{}
	Comer\'{o}n F, Schneider N, Russeil D. 2005. {\it Astron. 
	Astrophys.} 433:955--77

\bibitem{}
	Comer\'{o}n F, Pasquali A. 2007. {\it Astron. Astrophys. 
	Lett.} 467:L23--27

\bibitem{}
	Corbelli E, Palla F, Zinnecker H. ed. 2005. {\it The 
	Initial Mass Function 50 Years Later, Astrophys. 
	Space Sci. Libr. 327.} Dordrecht: Springer

\bibitem{}
	Costero R, Echevarria J, Richer MG, Poveda A, 
	Li W. 2006. {\it IAU Circ.} 8669

\bibitem{}
	Crowther PA, ed. 2002. {\it Hot Star Workshop III: The 
	Earliest Stages of Massive Star Birth, ASP Conf. 
	Proc.,} Vol. 267. San Francisco: ASP 

\bibitem{}
	Crowther PA. 2007. {\it Annu. Rev. Astron. Astrophys.} 
	45:In press 

\bibitem{}
	Crutcher RM. 1999. {\it Ap. J.} 520:706--13
	
\bibitem{}
	Crutcher RM. 2005. See Cesaroni et al. 2005a, pp. 
	98--107

\bibitem{}
	Crutcher RM, Troland TH. 2007. See Elmegreen, 
	Palous. 2007, pp. 141--47

\bibitem{}
	Dale JE, Davies MB. 2006. {\it MNRAS} 366:1424--36

\bibitem{}
	Dale JE, Bonnell IA, Whitworth AP. 2007. {\it MNRAS} 
	375:1291--98

\bibitem{}
	Dale JE, Clark PC, Bonnell IA. 2007. {\it MNRAS} 
	377:535--44

\bibitem{}
	Davies MB, Bate MR, Bonnell IA, Bailey VC, Tout 
	CA. 2006. {\it MNRAS} 370:2038--46

\bibitem{}
	Davis CJ, Varricatt WP, Todd SP, 
	Ramsay Howat SK. 2004. {\it Astron. Astrophys.} 
	425:981--95

\bibitem{}
	De Becker M, Rauw G, Manfroid J, Eenens P. 2006. 
	{\it Astron. Astrophys.} 456:1121--30

\bibitem{}
	De Buizer JM. 2003. {\it MNRAS} 341:277--98

\bibitem{}
	De Buizer JM, Minier V. 2005. {\it Ap. J. Lett.} 
	628:L151--54

\bibitem{}
	de Geus EJ. 1992. {\it Astron. Astrophys.} 262:258--70

\bibitem{}
	de Wit WJ, Testi L, Palla F, Vanzi L, Zinnecker H. 
	2004. {\it Astron. Astrophys.} 425:937--48

\bibitem{}
	de Wit WJ, Testi L, Palla F, Zinnecker H. 2005. 
	{\it Astron. Astrophys.} 437:247--55

\bibitem{}
	de Zeeuw PT, Hoogerwerf R, de Bruijne JHJ, Brown 
	AGA, Blaauw A. 1999. {\it Astron. J.} 117:354--99

\bibitem{}
	Diehl R, Halloin H, Kretschmer K, Lichti GG, 
	Sch\"onfelder V, et al. 2006. {\it Nature} 439:45--47

\bibitem{}
	Dobbs CL, Bonnell IA, Clark PC. 2005. {\it MNRAS} 
	360:2--8

\bibitem{}
	Dolan CJ, Mathieu RD. 2002. {\it Astron. J.} 123:387--403

\bibitem{}
	Donati J-F, Babel J, Harries TJ, Howarth ID, Petit P, 
	Semel M.  2002. {\it MNRAS} 333:55--70

\bibitem{}
	Donati J-F, Howarth ID, Bouret J-C, Petit P, Catala 
	C, Landstreet J.  2006. {\it MNRAS Lett.} 365:L6--10

\bibitem{}
	Drissen L, Moffat AFJ, Walborn NR, Shara MM. 
	1995. {\it Astron. J.} 110:2235--41

\bibitem{}
	Duch\^{e}ne G, Simon T, Eisloffel J, Bouvier J. 2001. 
	{\it Astron. Astrophys.} 379:147--61

\bibitem{}
	Durisen RH. 2001. See Zinnecker, Mathieu. 2001, 
	pp. 381--90

\bibitem{}
	Edgar R, Clarke C. 2004. {\it MNRAS} 349:678--86

\bibitem{}
	Edwards S, Strom SE, Hartigan P, Strom KM, 
	Hillenbrand LA, et al. 1993. {\it Astron. J.} 106:372--82

\bibitem{}
	Egan MP, Shipman RF, Price SD, Carey SJ, Clark 
	FO, Cohen M.  1998. {\it Ap. J. Lett.} 494:L199--202

\bibitem{}
	Eikenberry SS, Matthews K, LaVine JL, Garske 
	MA, Hu D, et al. 2004. {\it Ap. J.} 616:506--18

\bibitem{}
	Eisenhauer F, Quirrenbach A, Zinnecker H, Genzel 
	R. 1998. {\it Ap. J.} 498:278--92

\bibitem{}
	Elmegreen BG. 1997. {\it Ap. J.} 486:944--54

\bibitem{}
	Elmegreen BG. 1998. In {\it The Origin of Stars and 
	Planetary Systems,} ed. CE Woodward, JM Shull, 
	HA Thronson Jr. {\it ASP Conf. Ser.,} 148:150--83

\bibitem{}
	Elmegreen BG. 1999. {\it Ap. J.} 517:103--07

\bibitem{}
	Elmegreen BG. 2000. {\it Ap. J.} 539:342--51

\bibitem{}
	Elmegreen BG, Lada CJ. 1977. {\it Ap. J.} 214:725--41

\bibitem{}
	Elmegreen BG, Palous J. ed. 2007. {\it Triggered Star 
	Formation in a Turbulent ISM, IAU Symp. 237.} 
	Cambridge: Cambridge University Press

\bibitem{}
	Evans NJ. 1999. {\it Annu. Rev. Astron. Astrophys.} 
	37:311--62

\bibitem{}
	Evans NJ. 2005. See Cesaroni et al. 2005a, pp. 443--48

\bibitem{}
	Fabian AC, Pringle JE, Rees MJ. 1975. {\it MNRAS} 
	172:15--18

\bibitem{}
	Feitzinger JV, Schlosser W, Schmidt-Kaler T, 
	Winkler C. 1980. {\it Astron. Astrophys.} 84:50--59

\bibitem{}
	Ferreira J, Pelletier G. 1995. {\it Astron. Astrophys.} 
	295:807--32

\bibitem{}
	Figer DF. 2003. In {\it A Massive Star Odyssey: From 
	Main Sequence to Supernova,} ed. K van der 
	Hucht, A Herrero, E C\'{e}sar. {\it IAU Symp.} 212:487--96.
	San Francisco: ASP

\bibitem{}
	Figer DF. 2005. {\it Nature} 434:192--94

\bibitem{}
	Figer DF, Najarro F, Morris M, McLean IS, Geballe 
	TR, et al. 1998. {\it Ap. J.} 506:384--404

\bibitem{}
	Figer DF, Najarro F, Kudritzki RP. 2004. {\it Ap. J. Lett.} 
	610:L109--12

\bibitem{}
	Fromang S, Balbus S, Terquem C, de Villiers JP. 
	2004. {\it Ap. J.} 616:364--75

\bibitem{}
	Fryer CL, Heger A. 2005. {\it Ap. J.} 623:302--13

\bibitem{}
	Furuya RS, Kitamura Y, Saito M, Kawabe R, 
	Wootten HA. 1999. {\it Ap. J.} 525:821--31

\bibitem{}
	Gammie CF. 1998. {\it MNRAS} 297:929--35

\bibitem{}
	Garay G. 2005. See Cesaroni et al. 2005a, pp. 86--91

\bibitem{}
	Garay G, Ramirez S, Rodr\'{i}guez LF, Curiel S, 
	Torrelles JM. 1996. {\it Ap. J.} 459:193--208

\bibitem{}
	Garay G, Lizano S. 1999. {\it Publ. Astron. Soc. Pac.} 
	111:1049--87

\bibitem{}
	Garay G, Brooks KJ, Mardones D, Norris RP. 2003. 
	{\it Ap. J.} 587:739--47

\bibitem{}
	Garay G, Fa\'{u}ndez S, Mardones D, Bronfman L, 
	Chini R, Nyman L-\r{A}. 2004. {\it Ap. J.} 610:313--19

\bibitem{}
	Garc\'{i}a B, Mermilliod JC. 2001. {\it Astron. Astrophys.} 
	368:122--36

\bibitem{}
	Garmany CD. 1994. {\it Publ. Astron. Soc. Pac.} 106:25--37

\bibitem{}
	Garmany CD, Conti PS, Massey P. 1980. {\it Ap. J.} 
	242:1063--76

\bibitem{}
	Garmany CD, Conti PS, Chiosi C. 1982. {\it Ap. J.} 
	263:777--90

\bibitem{}
	Genzel R, Stutzki J. 1989. {\it Annu. Rev. Astron. 
	Astrophys.} 27:41--85

\bibitem{}
	Gerola H, Seiden PE. 1978. {\it Ap. J.} 223:129--35

\bibitem{}
	Gies DR. 1987. {\it Ap. J. Suppl.} 64:545--63

\bibitem{}
	Gies DR, Bolton CT. 1986. {\it Ap. J. Suppl.} 61:419--54

\bibitem{}
	Glover SCO. 2005. {\it Space Sci. Rev.} 117:445--508

\bibitem{}
	Glover SCO, Mac Low M-M. 2007. {\it Ap. J.} 
	659:1317--37

\bibitem{}
	Gomez Y, Lebron M, Rodr\'{i}guez LF, Garay G, 
	Lizano S, et al. 1998.  {\it Ap. J.} 503:297--306

\bibitem{}
	Greenhill LJ, Gezari DY, Danchi WC, Najita J, 
	Monnier JD, Tuthill PG.  2004. {\it Ap. J. Lett.} 
	605:L57--60

\bibitem{}
	Gualandris A, Portegies Zwart S, Eggleton PP. 2004. 
	{\it MNRAS} 350:615--26

\bibitem{}
	G\"usten R, Mezger PG. 1982. {\it Vistas Astron.} 26:159--224

\bibitem{}
	Gueth F, Guilloteau S. 1999. {\it Astron. Astrophys.} 
	343:571--84

\bibitem{}
	Gutermuth RA, Megeath ST, Pipher JL, 
	Williams JP, Allen LE, et al. 2005. {\it Ap. J.} 
	632:397--420

\bibitem{}
	Habing HJ, Israel FP. 1979. {\it Annu. Rev. Astron. 
	Astrophys.} 17:345--85

\bibitem{}
	Hartigan P, Morse J, Bally J. 2000. {\it Astron. J.} 
	120:1436--48

\bibitem{}
	Hawley JF, Balbus SA. 1991. {\it Ap. J.} 376:223--33

\bibitem{}
	Heger A, Fryer CL, Woosley SE, Langer N, 
	Hartmann DH. 2003. {\it Ap. J.} 591:288--300

\bibitem{}
	Heitsch F, Mac Low M-M, Klessen RS. 2001. {\it Ap. J.} 
	547:280--91

\bibitem{}
	Heller CH. 1995. {\it Ap. J.} 455:252--59

\bibitem{}
	Henning T, Schreyer K, Launhardt R, Burkert A. 
	2000. {\it Astron. Astrophys.} 353:211--26

\bibitem{}
	Henning T, Stecklum B. 2002. In {\it Modes of Star 
	Formation and the Origin of Field Populations,} 
	ed. EK Grebel, W Brandner. {\it ASP Conf. Proc.} 
	285:40--48

\bibitem{}
	Henriksen RN. 1991. {\it Ap. J.} 377:500--09

\bibitem{}
	Herbig GH, Andrews SM, Dahm SE. 2004. {\it Astron. 
	J.} 128:1233--53

\bibitem{}
	Herbig GH, Griffin RF. 2006. {\it Astron. J.} 132:1763--67
	
\bibitem{}
	Herbst W, Assousa GE. 1977. {\it Ap. J.} 217:473--75

\bibitem{}
	Hernandez J, Hartmann L, Megeath T, Gutermuth R, 
	Muzerolle J, et al. 2007. {\it Ap. J.} In press 
	(astro-ph/0701476)

\bibitem{}
	Heydari-Malayeri M, Rosa MR, Schaerer D, Martins 
	F, Charmandaris V. 2002. {\it Astron. Astrophys.} 
	381:951--58

\bibitem{}
	Heydari-Malayeri M, Meynadier F, Charmandaris 
	V, Deharveng L, Le Bertre T, et al. 2003. {\it Astron. 
	Astrophys.} 411:427--35

\bibitem{}
	Hill T, Burton MG, Minier V, Thompson MA, 
	Walsh AJ, et al. 2005. {\it MNRAS} 363:405--51

\bibitem{}
	Hillenbrand LA. 1997. {\it Astron. J.} 113:1733--68

\bibitem{}
	Hillenbrand LA, Hartmann LW. 1998. {\it Ap. J.} 
	492:540--53

\bibitem{}
	Hoare MG, Kurtz SE, Lizano S, Keto E, Hofner P. 
	2007. See Reipurth et al. 2007, pp. 181--96

\bibitem{}
	Hodapp K-W, Rayner J. 1991. {\it Astron. J.} 102:1108--17

\bibitem{}
	Hofmann K-H, Seggewiss W, Weigelt G. 1995. 
	{\it Astron. Astrophys.} 300:403--14

\bibitem{}
	Hofner P, Delgado H, Whitney B, Churchwell E, 
	Linz H. 2002. {\it Ap. J. Lett.} 579:L95--98

\bibitem{}
	Hollenbach D, Johnstone D, Lizano S, Shu F. 1994. 
	{\it Ap. J.} 428:654--69

\bibitem{}
	Hollenbach D, Yorke HW, Johnstone D. 2000. See 
	Mannings et al. 2000, pp. 401--28 

\bibitem{}
	Hoogerwerf R, de Bruijne JHJ, de Zeeuw PT. 2001. 
	{\it Astron. Astrophys.} 365:49--77

\bibitem{}
	Howe JE, Jaffe DT, Genzel R, Stacey GJ. 1991. {\it Ap. 
	J.} 373:158--68

\bibitem{}
	Hunter DA. 1995. In {\it Gaseous Nebulae and Star 
	Formation. Rev. Mex. Astron. Astrofis. Ser. 
	Conf.} 3:1--7

\bibitem{}
	Indebetouw R, Whitney BA, Johnson KE, Wood K. 
	2006. {\it Ap. J.} 636:362--80

\bibitem{}
	Irvine NJ. 1989. {\it Ap. J. Lett.} 337:L33--35

\bibitem{}
	Jappsen A-K, Klessen RS, Larson RB, Li Y, Mac 
	Low M-M. 2005. {\it Astron. Astrophys.} 435:611--23

\bibitem{}
	Jijina J, Adams FC. 1996. {\it Ap. J.} 462:874--87

\bibitem{}
	Johnstone, Doug; Bally, John. 2006. {\it Ap. J.} 653:383--97

\bibitem{}
	Kahn FD. 1974. {\it Astron. Astrophys.} 37:149--62

\bibitem{}
	Kaifu N, Usuda T, Hayashi SS, Itoh Y, Akiyama M, 
	et al. 2000. {\it Publ. Astron. Soc. Jpn.} 52:1--8 

\bibitem{}
	Kastner JH, Buchanan CL, Sargent B, Forrest WJ. 
	2006. {\it Ap. J. Lett.} 638:L29--32

\bibitem{}
	Kennicutt RC. 1998. {\it Annu. Rev. Astron. Astrophys.} 
	36:189--232

\bibitem{}
	Kennicutt RC. 2005. See Cesaroni et al. 2005a, pp. 
	3--11

\bibitem{}
	Kessel-Deynet O, Burkert A. 2003. {\it MNRAS} 
	338:545--54

\bibitem{}
	Keto E. 2002. {\it Ap. J.} 580:980--86

\bibitem{}
	Keto E. 2007. {\it Ap. J.} In press (astro-ph/0603856)

\bibitem{}
	Keto E, Wood K. 2006. {\it Ap. J.} 637:850--59

\bibitem{}
	Kharchenko NV, Piskunov AE, R\"oser S, Schilbach 
	E, Scholz R-D. 2004. {\it Astron. Nachr.} 325:740--48

\bibitem{}
	Kharchenko NV, Piskunov AE, R\"oser S, Schilbach 
	E, Scholz R-D. 2005. {\it Astron. Astrophys.} 
	440:403--08

\bibitem{}
	Kim SS, Figer DF, Kudritzki RP, Najarro F. 2006. 
	{\it Ap. J. Lett.} 653:L113--16

\bibitem{}
	Kippenhahn R, Meyer-Hofmeister E. 1975. {\it Astron. 
	Astrophys.} 54:539--42

\bibitem{}
	Kitsionas S, Whitworth AP. 2002. {\it MNRAS} 330:129--36

\bibitem{}
	Klein RI, Fisher R, McKee CF. 2004. In 
	{\it Gravitational Collapse: From Massive Stars to 
	Planets,} ed. G Garcia-Segura, G Tenorio-Tagle, 
	J Franco, HW Yorke. {\it Rev. Mex. Astron. 
	Astrophys. Ser. Conf.,} 22:3--7

\bibitem{}
	Klessen RS. 2001a. {\it Ap. J. Lett.} 550:L77--80

\bibitem{}
	Klessen RS. 2001b. {\it Ap. J.} 556:837--46 

\bibitem{}
	Klessen RS, Burkert A. 2001. {\it Ap. J.} 549:386--401

\bibitem{}
	Klessen RS, Ballesteros-Paredes J, V\'{a}zquez-Semadeni E,
	Dur\'{a}n-Rojas C. 2005. {\it Ap. J.} 
	620:786--94

\bibitem{}
	Klessen RS, Spaans M, Jappsen A-K. 2007. {\it MNRAS
	Lett.} 374:L29--33

\bibitem{}
	Knodlseder J. 2000. {\it Astron. Astrophys.} 360:539--48

\bibitem{}
	Koen C. 2006. {\it MNRAS} 365:590--94

\bibitem{}
	Konigl A. 1991. {\it Ap. J. Lett.} 370:L39--43

\bibitem{}
	Kratter KM, Matzner CD. 2006. {\it MNRAS.}  373:1563--76

\bibitem{}
	Kraus S, Balega YY, Berger J-P, Hofmann K-H, 
	Millan-Gabet R, et al. 2007. {\it Astron. Astrophys.} 
	466:649--59 

\bibitem{}
	Kritsuk AG, Norman ML, Padoan P. 2006. {\it Ap. J.} 
	638:L25--28

\bibitem{}
	Kroupa P. 2000. In {\it Massive Stellar Clusters,} ed. A 
	Lancon, C Boily. {\it ASP Conf. Ser.,} 211:233--40 

\bibitem{}
	Kroupa P. 2002. {\it Science} 295:82--91

\bibitem{}
	Kroupa P, Aarseth S, Hurley J. 2001, {\it MNRAS} 
	321:699--712

\bibitem{}
	Krumholz MR. 2006. {\it Ap. J. Lett.} 641:L45--48

\bibitem{}
	Krumholz MR. 2007.  In {\it Massive Stars: From Pop 
	III and GRBs to the Milky Way,} ed. M Livio, E 
	Villaver. In press (astro-ph/0607429) 

\bibitem{}
	Krumholz MR, McKee CF, Klein RI. 2004. {\it Ap. J.} 
	611:399--412
	
\bibitem{}
	Krumholz MR, Klein RI, McKee CF. 2005. See 
	Cesaroni et al. 2005a, pp. 231--36 

\bibitem{}
	Krumholz MR, McKee CF, Klein RI. 2005a. {\it Ap. J.} 
	618:L33--36

\bibitem{}
	Krumholz MR, McKee CF, Klein RI. 2005b. {\it Nature} 
	438:332--34

\bibitem{}
	Krumholz MR, Klein RI, McKee CF. 2007. {\it Ap. J.} 
	656:959--79

\bibitem{}
	Krumholz MR, Thompson TA. 2007. {\it Ap. J.}  In press 
	(astro-ph/0611822) 

\bibitem{}
	Kudritzki R. 2002. {\it Ap. J.} 577:389--408

\bibitem{}
	Kurtz S. 2005. See Cesaroni et al. 2005a, pp. 111--19

\bibitem{}
	Kurtz S, Cesaroni R, Churchwell E, Hofner P, 
	Walmsley CM. 2000. See Mannings et al. 2000, 
	pp. 299--326

\bibitem{}
	Lada CJ, Thronson HA Jr., Smith HA, Schwartz PR, 
	Glaccum W. 1984. {\it Ap. J.} 286:302--09

\bibitem{}
	Lada CJ, Lada EA. 2003. {\it Annu. Rev. Astron. 
	Astrophys.} 41:57--115

\bibitem{}
	Lamers HJGLM, Panagia N, Scuderi S, Romaniello M,
	Spaans M, et al. 2002. {\it Ap. J.} 566:818--32

\bibitem{}
	Larson RB. 1969. {\it MNRAS} 145:271--95
	
\bibitem{}
	Larson RB. 1978. {\it MNRAS} 184:69--85

\bibitem{}
	Larson RB. 1981. {\it MNRAS} 194:809--26

\bibitem{}
	Larson RB. 1982. {\it MNRAS} 200:159--74

\bibitem{}
	Larson RB. 1985. {\it MNRAS} 214:379--98

\bibitem{}
	Larson RB. 2005. {\it MNRAS} 359:211--22

\bibitem{}
	Larson RB, Starrfield S. 1971. {\it Astron. Astrophys.} 
	13:190--97

\bibitem{}
	Laughlin G, Bodenheimer P. 1994. {\it Ap. J.} 436:335--54

\bibitem{}
	Ledoux P. 1941. {\it Ap. J.} 94:537--38

\bibitem{}
	Lee C-F, Ho PTP, Beuther H, Bourke TL, Zhang Q, 
	et al. 2006. {\it Ap. J.} 639:292--302

\bibitem{}
	Lee H-T, Chen WP. 2007. {\it Ap. J.} 657:884--96

\bibitem{}
	Lenorzer A, Bik A, de Koter A, Kurtz SE, Waters 
	LBFM, et al. 2004. {\it Astron. Astrophys.} 414:245--59

\bibitem{}
	Leonard PJT, Duncan MJ. 1990. {\it Astron. J.} 99:608--16

\bibitem{}
	Lester DF, Harvey PM, Joy M, Ellis HB Jr. 1986. 
	{\it Ap. J.} 309:80--89

\bibitem{}
	Li Y. Klessen RS, Mac Low M-M. 2003. {\it Ap. J.} 
	592:975--85

\bibitem{}
	Li Z-Y, Nakamura F. 2006. {\it Ap. J. Lett.} 640:L187--90

\bibitem{}
	Linz H, Stecklum B, Henning T, Hofner P, Brandl 
	B. 2005. {\it Astron. Astrophys.} 429:903--21

\bibitem{}
	Mac Low M-M, Klessen RS. 2004. {\it Rev. Mod. Phys.} 
	76:125--94

\bibitem{}
	Maeder A, Behrend R. 2002. See Crowther 2002,  
	pp. 179--92

\bibitem{}
	Maiolino R, Vanzi L, Mannucci F, Cresci G, 
	Ghinassi F, Della Valle M. 2002. {\it Astron. 
	Astrophys.} 389:84--92

\bibitem{}
	Ma\'{i}z-Apell\'{a}niz J, P\'{e}rez E, Mas-Hesse JM. 2004. 
	{\it Astron. J.} 128:1196--218

\bibitem{}
	Ma\'{i}z-Apell\'{a}niz J, Walborn NR, Morrell NI, 
	Niemela VS, Nelan EP. 2007. {\it Ap. J.} In press 
	(astro-ph/0612012) 

\bibitem{}
	Makinen P, Harvey PM, Wilking BA, Evans NJ II. 
	1985. {\it Ap. J.} 299:341--50

\bibitem{}
	Mannings I, Boss AP, Russell SS, ed. 2000. 
	{\it Protostars and Planets IV.} Tucson: Univ. Ariz. 
	Press

\bibitem{}
	Martel H, Evans NJ II, Shapiro PR. 2006 {\it Ap. J. 
	Suppl.} 163:122--44

\bibitem{}
	Martins F, Schaerer D, Hillier DJ. 2005. {\it Astron. 
	Astrophys.} 436:1049--65

\bibitem{}
	Martins F, Trippe S, Paumard T, Ott T, Genzel R, et 
	al. 2006. {\it Ap. J. Lett.} 649 :L103--06

\bibitem{}
	Mason BD, Gies DR, Hartkopf WI, Bagnuolo WG 
	Jr., ten Brummelaar T, McAlister HA.  1998. 
	{\it Astron. J.} 115:821--47

\bibitem{}
	Massey P. 1998. In {\it The Stellar Initial Mass 
	Function,} ed. G Gilmore, D Howell. {\it ASP Conf. 
	Ser.,} 142:17--44

\bibitem{}
	Massey P. 2002. {\it Ap. J. Suppl.} 141:81--122

\bibitem{}
	Massey P. 2003. {\it Annu. Rev. Astron. Astrophys.} 
	41:15--56

\bibitem{}
	Massey P, Johnson KE, Degioia-Eastwood K. 1995. 
	{\it Ap. J.} 454:151--71

\bibitem{}
	Massey P, Hunter DA. 1998. {\it Ap. J.} 493:180--94

\bibitem{}
	Massey P, Penny LR, Vukovich J. 2002. {\it Ap. J.} 
	565:982--93

\bibitem{}
	Masunaga H, Miyama SM, Inutsuka S-I. 1998. {\it Ap. 
	J.} 495:346--69

\bibitem{}
	Masunaga H, Inutsuka S-I. 2000. {\it Ap. J.} 531:350--65

\bibitem{}
	Mathis JS, Rumpl W, Nordsieck KH. 1977. {\it Ap. J.} 
	217:425--33

\bibitem{}
	McCaughrean MJ. 2001. In {\it From Darkness to Light: 
	Origin and Evolution of Young Stellar Clusters,} 
	ed. T Montmerle, P Andr\'{e}. {\it ASP Conf. Ser.,} 
	243:449--60

\bibitem{}
	McCaughrean MJ, Rayner JT, Zinnecker H. 1994. 
	{\it Ap. J. Lett.} 436:L189--92

\bibitem{}
	McCaughrean MJ, Mac Low M-M. 1997. {\it Astron. J.} 
	113:391--400

\bibitem{}
	McKee CF, Tan JC. 2003. {\it Ap. J.} 585:850--71

\bibitem{}
	McKee CF, Ostriker EC. 2007. {\it Annu. Rev. Astron. 
	Astrophys.} 45:In press

\bibitem{}
	McMillan SLW, Vesperini E, Portegies Zwart SF. 
	2007. {\it Ap. J. Lett.} 655:L45--49

\bibitem{}
	Megeath ST, Wilson TL, Corbin MR. 2005. {\it Ap. J. 
	Lett.} 622:L141--44

\bibitem{}
	Meixner M, Gordon KD, Indebetouw R, Hora JL, 
	Whitney B, et al. 2006. {\it Astron. J.} 132:2268--88
	
\bibitem{}
	Melioli C, de Gouveia Dal Pino EM, de La Reza R, 
	Raga A. 2006. {\it MNRAS} 373:811--18

\bibitem{}
	Mengel S. Lehnert MD, Thatte N, Genzel R. 2002. 
	{\it Astron. Astrophys.} 383:137--52 

\bibitem{}
	Mengel S, Tacconi-Garman LE. 2007. {\it Astron. 
	Astrophys.} In press (astro-ph/0701415)

\bibitem{}
	Menten KM. 1991. {\it Ap. J. Lett.} 380:L75--78

\bibitem{}
	Menten KM, Reid MJ. 1995. {\it Ap. J. Lett.} 445:L157--60

\bibitem{}
	Menten KM, Pillai T, Wyrowski F. 2005. See 
	Cesaroni et al. 2005a, pp. 23--34

\bibitem{}
	Mermilliod J-C, Garc\'{i}a B. 2001. See Zinnecker, 
	Mathieu. 2001, pp. 191--98

\bibitem{}
	Meynet G, Maeder A. 2003. {\it Astron. Astrophys.} 
	404:975--90

\bibitem{}
	Meynet G, Maeder A. 2005. {\it Astron. Astrophys.} 
	429:581--98

\bibitem{}
	Meynet G, Maeder A, Schaller G, Schaerer D, 
	Charbonnel C. 1994. {\it Astron. Astrophys. Suppl.} 
	103:97--105

\bibitem{}
	Mezger PG, Altenhoff W, Schraml J, Burke BF, 
	Reifenstein EC III, Wilson, TL.1967. {\it Ap. J.} 
	150:L157--66

\bibitem{}
	Minier V, Burton MG, Hill T, Pestalozzi MR, 
	Purcell CR, et al. 2005. {\it Astron. Astrophys.} 
	429:945--60

\bibitem{}
	Moeckel N, Bally J. 2006. {\it Ap. J.} 653:437--46

\bibitem{}
	Moeckel N, Bally J. 2007. {\it Ap. J.} 656:275--86

\bibitem{}
	Moffat AFJ, Drissen L, Shara MM. 1994. {\it Ap. J.} 
	436:183--93

\bibitem{}
	Moffat AFJ, Poitras V, Marchenko SV, Shara MM, 
	Zurek DR, et al. 2004. {\it Astron. J.} 128:2854--61

\bibitem{}
	Monaghan JJ. 1992. {\it Annu. Rev. Astron. Astrophys.} 
	30:543--74

\bibitem{}
	Monaghan JJ. 2005. {\it Rep. Prog. Phys.} 68:1703--59

\bibitem{}
	Monin J-L, Clarke CJ, Prato L, McCabe C. 2007. 
	See Reipurth et al. 2007, pp. 395--409

\bibitem{}
	Morrell NI, Levato H. 1991. {\it Ap. J. Suppl.} 75:965--85

\bibitem{}
	Morrell NI, Barb\'{a} RH, Niemela VS, Corti MA, 
	Albacete Colombo JF, et al. 2001. {\it MNRAS} 
	326:85--94

\bibitem{}
	Motte F, Bontemps S, Schilke P, Lis DC, Schneider N, Menten KM.
	2005. See Cesaroni et al. 2005a, pp. 151--56

\bibitem{}
	Muench AA, Lada EA, Lada CJ, Alves J. 2002. {\it Ap. 
	J.} 573:366--93

\bibitem{}
	Nakano T, Hasegawa T, Norman C. 1995. {\it Ap. J.} 
	450:183--95

\bibitem{}
	Nelan EP, Walborn NR, Wallace DJ, Moffat AFJ, 
	Makidon RB, et al. 2004. {\it Astron. J.} 128:323--29

\bibitem{}
	Nielbock M, Chini R, Hoffmeister VH, Scheyda 
	CM, Steinacker J, et al. 2007. {\it Ap. J. Lett.} 
	656:L81--84

\bibitem{}
	Norberg P, Maeder A. 2000. {\it Astron. Astrophys.} 
	359:1025--34

\bibitem{}
	N\"urnberger DEA. 2003. {\it Astron. Astrophys.} 404:255--65

\bibitem{}
	N\"urnberger DEA, Chini R, Eisenhauer F, Kissler-Patig M,
	Modigliani A, et al. 2007. {\it Astron. 
	Astrophys.} 465:931--36 

\bibitem{}
	Oey MS, Clarke CJ. 2005. {\it Ap. J.} Lett. 620:L43--46

\bibitem{}
	Omukai K, Palla F. 2003. {\it Ap. J.} 589:677--87

\bibitem{}
	\"Opik EJ. 1953. {\it Irish Astron. J.} 2:219--33

\bibitem{}
	Padgett DL, Strom SE, Ghez A. 1997. {\it Ap. J.} 
	477:705--10

\bibitem{}
	Padoan P, Nordlund \r{A}. 2002. {\it Ap. J.} 576:870--79

\bibitem{}
	Palla F, Stahler SW. 1992. {\it Ap. J.} 392:667--77

\bibitem{}
	Palla F, Stahler SW. 1993. {\it Ap. J.} 418:414--25

\bibitem{}
	Parker JW, Garmany CD. 1993. {\it Astron. J.} 
	106:1471--83

\bibitem{}
	Patel NA, Curiel S, Sridharan TK, Zhang Q, 
	Hunter TR, et al. 2005. {\it Nature} 437:109--11

\bibitem{}
	Paumard T, Genzel R, Martins F, Nayakshin S, 
	Beloborodov AM, et al. 2006. {\it Ap. J.} 643:1011--35

\bibitem{}
	Pehlemann E, Hofmann K-H, Weigelt G. 1992. 
	{\it Astron. Astrophys.} 256:701--14

\bibitem{}
	Perault M, Omont A, Simon G, Seguin P, Ojha D, et 
	al. 1996. {\it Astron. Astrophys. Lett.} 315:L165--68

\bibitem{}
	Peretto N, Andr\'{e} P, Belloche A. 2006. {\it Astron. 
	Astrophys.} 445:979--98

\bibitem{}
	Pflamm-Altenburg J, Kroupa P. 2006. {\it MNRAS} 
	373:295--304

\bibitem{}
	Pietrinferni A, Cassisi S, Salaris M, Castelli F. 2004. 
	{\it Ap. J.} 612:168--90

\bibitem{}
	Piskunov AE, Kharchenko NV, R\"oser S, 
	Schilbach E, Scholz RD. 2006.  {\it Astron. 
	Astrophys.} 445:545--65

\bibitem{}
	Plume R, Jaffe DT, Evans NJ II, Martin-Pintado J, 
	Gomez-Gonzalez J. 1997. {\it Ap. J.} 476:730--49

\bibitem{}
	Pl\"uschke S, Cervi{\~n}o M, Diehl R, Kretschmer K, 
	Hartmann DH, Kn\"odlseder J. 2002. {\it New Astron. Rev.}
	46:535--39

\bibitem{}
	Poetzel R, Mundt R, Ray TP. 1992. {\it Astron. 
	Astrophys.} 262:229--47

\bibitem{}
	Portegies Zwart SF, McMillan SLW. 2002. {\it Ap. J.} 
	576:899--907

\bibitem{}
	Portegies Zwart SF, Baumgardt H, McMillan SLW, 
	Makino J, Hut P, Ebisuzaki T.  2006. {\it Ap. J.} 
	641:319--26

\bibitem{}
	Poveda A, Ruiz J, Allen C. 1967. {\it Bol. Obs. 
	Tonantzintla Tacubaya} 4:86--90

\bibitem{}
	Preibisch T, Ossenkopf V, Yorke HW, Henning T. 
	1993. {\it Astron. Astrophys.} 279:577--88
	
\bibitem{}
	Preibisch T, Balega Y, Hofmann K-H, Weigelt G, 
	Zinnecker H. 1999. {\it New Astron.} 4:531--42

\bibitem{}
	Preibisch T, Zinnecker H. 1999. {\it Astron. J.} 
	117:2381--97

\bibitem{}
	Preibisch T, Weigelt G, Zinnecker H. 2001. See 
	Zinnecker \& Mathieu 2001, p. 69--78 

\bibitem{}
	Preibisch T, Balega YY, Schertl D, Smith MD, 
	Weigelt G. 2001. {\it Astron. Astrophys.} 378:539--45

\bibitem{}
	Preibisch T, Balega YY, Schertl D, Weigelt G. 2002. 
	{\it Astron. Astrophys.} 392:945--54

\bibitem{}
	Preibisch T, Balega YY, Schertl D, Weigelt G. 2003. 
	{\it Astron. Astrophys.} 412:735--43

\bibitem{}
	Preibisch T, Zinnecker H. 2007. See Elmegreen, 
	Palous. 2007, pp. 270--77

\bibitem{}
	Pudritz RE, Norman CA. 1983. {\it Ap. J.} 274:677--97

\bibitem{}
	Pudritz RE, Ouyed R, Fendt C, Brandenburg A. 
	2007. See Reipurth et al. 2007, pp. 277--94

\bibitem{}
	Rathborne JM, Jackson JM, Simon R. 2006. {\it Ap. J.} 
	641:389--405

\bibitem{}
	Reipurth B, Zinnecker H. 1993. {\it Astron. Astrophys.} 
	278:81--108

\bibitem{}
	Rauw G, De Becker M, Naz\'{e} Y, Crowther PA, 
	Gosset E, et al. 2004. {\it Astron. Astrophys. Lett.} 
	420:L9--13

\bibitem{}
	Reipurth B, Jewitt D, Keil K, ed. 2007. {\it Protostars 
	and Planets V.} Tucson: Univ. Ariz. Press
	
\bibitem{}
	Richer JS, Shepherd DS, Cabrit S, Bachiller R, 
	Churchwell E. 2000. See Mannings et al. 2000, 
	pp. 867--96

\bibitem{}
	Richling S, Yorke HW. 1997. {\it Astron. Astrophys.} 
	327:317--24

\bibitem{}
	Roberts MS. 1957. {\it Publ. Astron. Soc. Pac.} 69:59--64

\bibitem{}
	Rodr\'{i}guez LF, Poveda A, Lizano S, Allen C. 2005a. 
	{\it Ap. J. Lett.} 627:L65--68

\bibitem{}
	Rodr\'{i}guez LF, Garay G, Brooks KJ, Mardones D. 
	2005b. {\it Ap. J.} 626:953--58 

\bibitem{}
	Rubio M, Barb\'{a} RH, Walborn NR, Probst RG, 
	Garc\'{i}a J, Roth MR.  1998. {\it Astron. J.} 116:1708--18

\bibitem{}
	Salpeter EE. 1955. {\it Ap. J.} 121:161--67

\bibitem{}
	Sana H, Rauw G, Gosset E. 2005. In {\it Massive Stars 
	and High-Energy Emission in OB Associations, 
	Proc. Workshop JENAM,} ed. G Rauw, Y Naz\'{e}, 
	R Blomme, E Gosset, pp. 107--10 

\bibitem{}
	Sandage A. 1986. {\it Astron. Astrophys.} 161:89--101

\bibitem{}
	Sanz-Forcada J, Franciosini E, Pallavicini R. 2004. 
	{\it Astron. Astrophys.} 421:715--27

\bibitem{}
	Scalo JM. 1998. In {\it The Stellar Initial Mass 
	Function,} ed. G Gilmore, D Howell. {\it ASP Conf. 
	Ser.} 142:201--36

\bibitem{}
	Schaerer D, de Koter A. 1997. {\it Astron. Astrophys.} 
	322:598--614

\bibitem{}
	Schertl D, Balega YY, Preibisch T, Weigelt G. 2003. 
	{\it Astron. Astrophys.} 402:267--75

\bibitem{}
	Schmeja S, Klessen RS. 2004. {\it Astron. Astrophys.} 
	419:405--17

\bibitem{}
	Schmid-Burgk J, Guesten R, Mauersberger R, 
	Schulz A, Wilson TL. 1990. {\it Ap. J. Lett.} 
	362:L25--28

\bibitem{}
	Schneider N, Bontemps S, Simon R, Jakob H, Motte 
	F, et al. 2006. {\it Astron. Astrophys.} 458:855--71

\bibitem{}
	Schreyer K, Stecklum B, Linz H, Henning T. 2003. 
	{\it Ap. J.} 599:335--41

\bibitem{}
	Schreyer K, Semenov D, Henning T, Forbrich J. 
	2006. {\it Ap. J. Lett.} 637:L129--32

\bibitem{}
	Schulz NS, Berghoefer TW, Zinnecker H. 1997. 
	{\it Astron. Astrophys.} 325:1001--12

\bibitem{}
	Schwartz PR, Thronson HA Jr., Odenwald SF, 
	Glaccum W, Loewenstein RF, Wolf G. 1985. 
	{\it Ap. J.} 292:231--37

\bibitem{}
	Schwarzschild M, H\"arm R. 1959. {\it Ap. J.} 129:637--46

\bibitem{}
	Schweickhardt J, Schmutz W, Stahl O, Szeifert T, 
	Wolf B. 1999. {\it Astron. Astrophys.} 347:127--36

\bibitem{}
	Selman FJ, Melnick J. 2005. {\it Astron. Astrophys.} 
	443:851--61

\bibitem{}
	Sewilo M, Churchwell E, Kurtz S, Goss WM, 
	Hofner P. 2004. {\it Ap. J.} 605:285--99

\bibitem{}
	Shepherd D. 2005. See Cesaroni et al. 2005a, pp. 
	237--46

\bibitem{}
	Shepherd DS, Churchwell E. 1996a. {\it Ap. J.} 472:225--39

\bibitem{}
	Shepherd DS, Churchwell E. 1996b. {\it Ap. J.} 457:267--76

\bibitem{}
	Shepherd DS, Claussen MJ, Kurtz SE. 2002. See 
	Crowther 2002, pp. 415--16

\bibitem{}
	Shirley YL, Evans NJ II, Young KE, Knez C,
	Jaffe DT. 2003. {\it Ap. J. Suppl.} 149:375--403

\bibitem{}
	Shu FH. 1977. {\it Ap. J.} 214:488--97

\bibitem{}
	Shu FH, Adams FC, Lizano S. 1987. {\it Annu. Rev. 
	Astron. Astrophys.} 25:23--81

\bibitem{}
	Shu FH, Najita J, Ostriker E, Wilkin F, Ruden S, 
	Lizano S.  1994. {\it Ap. J.} 429:781--96

\bibitem{}
	Shu FH, Najita J, Ostriker EC, Shang H. 1995. {\it Ap. J. 
	Lett.} 459:L155--8

\bibitem{}
	Silk J. 1978. In {\it Protostars and Planets I,} ed. T 
	Gehrels, p. 172. Tucson: Univ. Ariz. Press

\bibitem{}
	Silk J. 1995. {\it Ap. J. Lett.} 438:L41--44

\bibitem{}
	Silk J. 1997. {\it Ap. J.} 481:703--09

\bibitem{}
	Silk J. 2005.  See Corbelli et al. 2005, pp. 439--47

\bibitem{}
	Simon R, Jackson JM, Rathborne JM, Chambers ET. 
	2006. {\it Ap. J.} 639:227--36

\bibitem{}
	Sirianni M, Nota A, Leitherer C, De Marchi G, 
	Clampin M. 2000. {\it Ap. J.} 533:203--14

\bibitem{}
	Smith J, Bentley A, Castelaz M, Gehrz RD, 
	Grasdalen GL, Hackwell JA.  1985. {\it Ap. J.} 
	291:571--80

\bibitem{}
	Smith LF, Mezger PG, Biermann P. 1978. {\it Astron. 
	Astrophys.} 66:65--76

\bibitem{}
	Smith N, Bally J, Shuping RY, Morris M, Hayward 
	TL. 2004. {\it Ap. J. Lett.} 610:L117--20

\bibitem{}
	Smith N, Owocki SP. 2006. {\it Ap. J. Lett.} 645:L45--48

\bibitem{}
	Smith N, Brooks KJ. 2007. {\it MNRAS.} In press 
	(arXiv:0705.3053, astro-ph)

\bibitem{}
	Sridharan TK, Williams SJ, Fuller GA. 2005. {\it Ap. J. 
	Lett.} 631:L73--76

\bibitem{}
	Stahl O, Kaufer A, Rivinius T, Szeifert T, Wolf B, et 
	al. 1996. {\it Astron. Astrophys.} 312:539--48

\bibitem{}
	Stahler SW, Palla F, Ho PTP. 2000. See Mannings et 
	al. 2000, pp. 327--51

\bibitem{}
	Stecklum B, Brandl B, Henning T, Pascucci I, 
	Hayward TL, Wilson JC.  2002. {\it Astron. 
	Astrophys.} 392:1025--29

\bibitem{}
	Stolte A, Brandner W, Brandl B, Zinnecker H. 2006. 
	{\it Astron. J.} 132:253--70

\bibitem{}
	Stone JM, Ostriker EC, Gammie CF. 1998. {\it Ap. J. 
	Lett.} 508:L99--102

\bibitem{}
	Stothers RB, Chin C-W. 1993. {\it Ap. J. Lett.} 408:L85--88

\bibitem{}
	Sung H, Bessell MS. 2004. {\it Astron. J.} 127:1014--28
	
\bibitem{}
	Suttner G, Yorke HW. 2001. {\it Ap. J.} 551:461--77

\bibitem{}
	Tamura M, Yamashita T. 1992. {\it Ap. J.} 391:710--18

\bibitem{}
	Tan JC. 2004. {\it Ap. J. Lett.} 607:L47--50

\bibitem{}
	Tan JC. 2005. See Cesaroni et al. 2005a, pp. 318--27

\bibitem{}
	Tan JC, Krumholz MR, McKee CF. 2006. {\it Ap. J. 
	Lett.} 641:L121--24

\bibitem{}
	Taresch G, Kudritzki RP, Hurwitz M, Bowyer S, 
	Pauldrach AWA, et al. 1997. {\it Astron. Astrophys.} 
	321:531--48

\bibitem{}
	Terebey S, Shu FH, Cassen P. 1984. {\it Ap. J.} 286:529--51

\bibitem{}
	Terquem CEJMLJ. 2001. See Zinnecker \& Mathieu 
	2001, pp. 406--09

\bibitem{}
	Testi L, Palla F, Prusti T, Natta A, Maltagliati S. 
	1997. {\it Astron. Astrophys.} 320:159--66

\bibitem{}
	Testi L, Palla F, Natta A. 1999. {\it Astron. Astrophys.} 
	342:515--23

\bibitem{}
	Tohline JE, Durisen RH. 2001. See Zinnecker \& 
	Mathieu 2001, pp. 40--44

\bibitem{}
	Trinidad MA, Curiel S, Cant\'{o} J, D'Alessio P, 
	Rodr\'{i}guez LF. 2003. {\it Ap. J.} 589:386--96

\bibitem{}
	Truelove JK, Klein RI, McKee CF, Holliman JH II, 
	Howell LH, Greenough JA.  1997. {\it Ap. J. Lett.} 
	489:L179--83

\bibitem{}
	Turner JL, Beck SC, Crosthwaite LP, Larkin JE, 
	McLean IS, Meier DS.  2003. {\it Nature} 423:621--23

\bibitem{}
	Turner NJ, Blaes OM, Socrates A, Begelman MC, 
	Davis SW. 2005. {\it Ap. J.} 624:267--88

\bibitem{}
	Turner NJ, Quataert E, Yorke HW. 2007. {\it Ap. J.} In 
	press (astro-ph/0701800) 

\bibitem{}
	Vall\'{e}e JP, MacLeod JM. 1994. {\it Astron. J.} 108:998--1001

\bibitem{}
	van Albada TS. 1968. {\it Bull. Astron. Inst. Neth.} 
	20:57--68

\bibitem{}
	van Altena WF, Lee JT, Lee J-F, Lu PK, Upgren 
	AR. 1988. {\it Astron. J.} 95:1744--54 

\bibitem{}
	van Bever J, Vanbeveren D. 1998. {\it Astron. Astrophys.} 
	334:21--28

\bibitem{}
	van der Tak FFS, Menten KM. 2005. {\it Astron. 
	Astrophys.} 437:947--56

\bibitem{}
	van der Tak FFS, Walmsley CM, Herpin F, 
	Ceccarelli C. 2006. {\it Astron. Astrophys.} 447:1011--25

\bibitem{}
	Vanhala HAT, Boss AP, Cameron AGW, Foster PN. 
	1998. {\it 29th Annu. Lunar Planet. Sci. Conf., 
	March 16-20, 1998,} Houston. Abstr. 1470 

\bibitem{}
	Vanhala HAT, Cameron AGW. 1998. {\it Ap. J.} 
	508:291--307

\bibitem{}
	V\'{a}zquez-Semadeni E, Kim J, Shadmehri M, 
	Ballesteros-Paredes J. 2005. {\it Ap. J.} 618:344--59

\bibitem{}
	Verbunt F. 1993. {\it Annu. Rev. Astron. Astrophys.} 
	31:93--127

\bibitem{}
	von Hoerner S. 1968. In {\it Interstellar Ionized 
	hydrogen, Proc. Symp. HII Regions,} ed. Y 
	Terzian, pp. 101--69. New York: Benjamin

\bibitem{}
	Walborn NR. 2003. In {\it A Massive Star Odyssey: 
	From Main Sequence to Supernova,} ed. K van 
	der Hucht, A Herrero, E C\'{e}sar. {\it IAU Symp.} 
	212:13--21. San Francisco: ASP 

\bibitem{}
	Walborn NR. 2007. In {\it Massive Stars: From Pop III 
	and GRBs to the Milky Way,} ed. M Livio, E 
	Villaver. In press (astro-ph/0701573) 

\bibitem{}
	Walborn NR, Parker JW. 1992. {\it Ap. J. Lett.} 
	399:L87--89

\bibitem{}
	Walborn NR, Drissen L, Parker JW, Saha A, 
	MacKenty JW, White RL.  1999. {\it Astron. J.} 
	118:1684--99

\bibitem{}
	Walborn NR, Howarth ID, Lennon DJ, Massey P, 
	Oey MS, et al. 2002. {\it Astron. J.} 123:2754--71

\bibitem{}
	Walborn NR, Ma\'{i}z-Apell\'{a}niz J, Barb\'{a} RH. 2002. 
	{\it Astron. J.} 124:1601--24

\bibitem{}
	Walborn NR, Howarth ID, Herrero A, Lennon DJ. 
	2003. {\it Ap. J.} 588:1025--38

\bibitem{}
	Walsh AJ, Burton MG, Hyland AR, Robinson G. 
	1998. {\it MNRAS} 301:640--98

\bibitem{}
	Weidenschilling SJ, Ruzmaikina TV. 1994. {\it Ap. J.} 
	430:713--26

\bibitem{}
	Weidner C, Kroupa P. 2004. {\it MNRAS} 348:187--91

\bibitem{}
	Weigelt G, Baier G. 1985. {\it Astron. Astrophys. Lett.} 
	150:L18--20

\bibitem{}
	Whitworth AP, Bhattal AS, Chapman SJ, Disney 
	MJ, Turner JA. 1994. {\it MNRAS} 268:291--98

\bibitem{}
	Whitworth AP, Boffin HMJ, Francis N. 1998. 
	{\it MNRAS} 299:554--61

\bibitem{}
	Williams JP, Blitz L, McKee CF. 2000. See 
	Mannings et al. 2000, pp. 97--120

\bibitem{}
	Winkler K-H, Newman M. 1980. {\it Ap. J.} 236:201--11

\bibitem{}
	Wolff SC, Strom SE, Dror D, Lanz L, Venn K. 
	2006. {\it Astron. J.} 132:749--55

\bibitem{}
	Wolfire MG, Cassinelli JP. 1987. {\it Ap. J.} 319:850--67

\bibitem{}
	Woodward PR. 1978. {\it Annu. Rev. Astron. Astrophys.} 
	16:555--84

\bibitem{}
	Wu Y, Zhang Q, Chen H, Yang C, Wei Y, Ho PTP.  
	2005. {\it Astron. J.} 129:330--47

\bibitem{}
	Wynn-Williams CG. 1982. {\it Annu. Rev. Astron. 
	Astrophys.} 20:587--618

\bibitem{}
	Wyrowski F, Walmsley CM, Goss WM, Tielens 
	AGGM. 2000. {\it Ap. J.} 543:245--56

\bibitem{}
	Yorke HW. 1986. {\it Annu. Rev. Astron. Astrophys.} 
	24:48--87

\bibitem{}
	Yorke HW. 1988. In {\it Dust in the Universe, Proc. 
	Conf.,} ed. ME Bailey, DA Williams, pp. 355--72.
	Cambridge, UK: Cambridge Univ. Press

\bibitem{}
	Yorke HW. 2002. See Crowther 2002, pp. 165--78

\bibitem{}
	Yorke HW, Kr\"ugel E. 1977. {\it Astron. Astrophys.} 
	54:183--94

\bibitem{}
	Yorke HW, Welz A. 1996. {\it Astron. Astrophys.} 
	315:555--64

\bibitem{}
	Yorke HW, Bodenheimer P. 1999. {\it Ap. J.} 525:330--42

\bibitem{}
	Yorke HW, Sonnhalter C. 2002. {\it Ap. J.} 569:846--62

\bibitem{}
	Zavagno A, Deharveng L, Brand J, Massi F, Caplan 
	J, et al. 2005. See Cesaroni et al. 2005a, pp. 346--51

\bibitem{}
	Zhang Q. 2005. See Cesaroni et al. 2005a, pp. 135--44

\bibitem{}
	Ziebarth K. 1970. {\it Ap. J.} 162:947--62

\bibitem{}
	Zinnecker H. 1982. {\it NY Acad. Sci.} 395:226--35

\bibitem{}
	Zinnecker H. 1985. In {\it Birth and Infancy of Stars. 
	Proc. Les Houches Summer School,} ed. R Lucas, 
	A Omont, R Stora, pp. 473--75. NATO: 
	Knudsen.  

\bibitem{}
	Zinnecker H. 1986. In {\it Luminous Stars \& 
	Associations in Galaxies,} ed. CWH de Loore, AJ 
	Willis, P Laskarides,  {\it IAU Symp.} 116:271--73. 
	Dordrecht: Kluwer Academic Publishers

\bibitem{}
	Zinnecker H. 1991. In {\it Fragmentation of Molecular 
	Clouds and Star Formation,} ed. E Falgarone, F 
	Boulanger, G Duvert, {\it IAU Symp.} 147:526--32. 
	Dordrecht: Kluwer Academic Publishers

\bibitem{}
	Zinnecker H. 1996. In {\it The Interplay Between 
	Massive Star Formation, the ISM and Galaxy 
	Evolution. Proc. 11th IAP Astrophys. Meet.,} ed. 
	D Kunth, B Guiderdoni, M Heydari-Malayeri, 
	TX Thuan, pp. 249--58. Gif-sur-Yvette: Editions 
	Frontieres. 

\bibitem{}
	Zinnecker H. 2002. In {\it The Origin of 
	Stars and Planets: The VLT View,} ed. JF Alves, 
	MJ McCaughrean, pp. 179--86. 
	Berlin/Heidelberg: Springer-Verlag

\bibitem{}
	Zinnecker H. 2003. In {\it A Massive Star Odyssey: 
	From Main Sequence to Supernova,} ed. K van 
	der Hucht, A Herrero, E C\'{e}sar. {\it IAU Symp.} 
	212:80--90. San Francisco: ASP 

\bibitem{}
	Zinnecker H. 2004. In {\it The Formation and Evolution 
	of Massive Young Star Clusters,} ed. HJGLM 
	Lamers, LJ Smith, A Nota, {\it ASP Conf. Ser.} 
	322:349--58 

\bibitem{}
	Zinnecker H. 2006a. In {\it The Scientific Requirements 
	for Extremely Large Telescopes,} ed. PA 
	Whitelock, M Dennefeld, B Leibundgut. {\it IAU 
	Symp.} 232:324--27. Cambridge: Cambridge 
	University Press 

\bibitem{}
	Zinnecker H. 2006b. In {\it Stellar Evolution at Low 
	Metallicity: Mass Loss, Explosions, Cosmology,} 
	ed. HJGLM Lamers, N Langer, T Nugis, K 
	Annuk, {\it ASP Conf. Ser.}  353:339--47

\bibitem{}
	Zinnecker H, McCaughrean MJ, Wilking BA. 1993. 
	In {\it Protostars and Planets III,} ed. EH Levy, JI 
	Lunine, pp. 429--95. Tucson: Univ. Ariz. Press

\bibitem{}
	Zinnecker H, McCaughrean MJ, Rayner JT. 1998. 
	{\it Nature} 394:862--65

\bibitem{}
	Zinnecker H, Mathieu RD, ed. 2001. {\it The Formation 
	of Binary Stars, IAU Symp. 200.} San Francisco: 
	ASP

\bibitem{}
	Zinnecker H, Bate MR. 2002. See Crowther 2002, 
	pp. 209--18

\end{thebibliography}
\end{document}